\documentclass{jpp}
\usepackage{graphicx}

\usepackage[utf8]{inputenc}
\usepackage[T1]{fontenc}
\usepackage{amsmath,amssymb}
\usepackage{mathtools}
\usepackage{aas_macros}
\usepackage{layouts}
\usepackage{xcolor}
\usepackage{natbib}

\ifCUPmtlplainloaded \else
  \checkfont{eurm10}
  \iffontfound
    \IfFileExists{upmath.sty}
      {\typeout{^^JFound AMS Euler Roman fonts on the system,
                   using the 'upmath' package.^^J}%
       \usepackage{upmath}}
      {\typeout{^^JFound AMS Euler Roman fonts on the system, but you
                   dont seem to have the}%
       \typeout{'upmath' package installed. JPP.cls can take advantage
                 of these fonts, if you use 'upmath' package.^^J}%
       \providecommand\upi{\pi}%
      }
  \else
    \providecommand\upi{\pi}%
  \fi
\fi

\ifCUPmtlplainloaded \else
  \checkfont{msam10}
  \iffontfound
    \IfFileExists{amssymb.sty}
      {\typeout{^^JFound AMS Symbol fonts on the system, using the
                'amssymb' package.^^J}%
       \usepackage{amssymb}%
       \let\le=\leqslant  
       \let\ge=\geqslant  
      }{}
  \fi
\fi

\ifCUPmtlplainloaded \else
  \IfFileExists{amsbsy.sty}
    {\typeout{^^JFound the 'amsbsy' package on the system, using it.^^J}%
     \usepackage{amsbsy}}
    {\providecommand\boldsymbol[1]{\mbox{\boldmath $##1$}}}
\fi

\newcommand{\pD}[2]{\frac{\partial #2}{\partial #1}}

\newcommand{\bigD}[2]{\frac{{\rm D} #2}{{\rm D} #1}}

\newcommand{\pDD}[2]{\frac{\partial^2 #2}{\partial #1^2}}
\newcommand{\D}[2]{\frac{{\rm d} #2}{{\rm d} #1}}

\newcommand\bb[1]{\mbox{\boldmath{$#1$}}}
\newcommand\grad{\bb{\nabla}}
\renewcommand\bcdot{\,\bb{\cdot}\,}
\newcommand\btimes{\,\bb{\times}\,}
\newcommand{\mc}[1]{\mathcal{#1}}

\newcommand{\msb}[1]{\mathsfbi{#1}}
\newcommand{\lop}[1]{\mathcal{#1}}
\newcommand{\imag}{{\rm i}}
\newcommand{\rme}{{\rm e}}
\newcommand{\rmd}{{\rm d}}

\newcommand{\ex}{\hat{\bb{x}}}
\newcommand{\ey}{\hat{\bb{y}}}
\newcommand{\ez}{\hat{\bb{z}}}
\newcommand{\eb}{\hat{\bb{b}}}
\newcommand{\const}{{\rm const}}
\newcommand{\gammaw}{\gamma_{\text{w}}}

\newcommand{\nueff}{\nu_{\rm eff}}

\newcommand{\nuce}{\nu_{\rm CE}}

\newcommand{\nuql}{\nu_{\rm QL}}

\newcommand{\numodel}{\nu_{\rm model}}

\newcommand{\nufp}{\nu_{\rm FP}}
\newcommand{\nupic}{\nu_{\rm PIC}}
\newcommand{\C}{{\rm C}}
\newcommand{\CCE}{C_{\rm CE}}
\newcommand{\CQL}{C_{\rm QL}}

\newcommand{\qpare}{q_{\parallel e}}

\newcommand{\Var}{{\rm Var}}

\newcommand{\tac}{\tau_{\text{ac}}}
\newcommand{\taclin}{\tau_{\text{ac}}^{\text{lin}}}
\newcommand{\ts}{t_{\rm s}}
\newcommand{\te}{t_{\rm e}}
\newcommand{\kpar}{k_{\parallel}}
\newcommand{\kperp}{k_{\perp}}
\newcommand{\vpar}{v_{\parallel}}
\newcommand{\wpar}{w_\parallel}
\newcommand{\vperp}{v_{\perp}}

\newcommand{\wb}{\omega_{\rm b}}

\newcommand{\vw}{v_{\text{w}}}
\newcommand{\rmw}{\text{w}}
\newcommand{\fMe}{f_{{\rm M}e}}
\newcommand{\vth}{v_{\text{th}}}

\newcommand{\vthe}{v_{\text{th}e}}
\newcommand{\vthi}{v_{\text{th}i}}
\newcommand{\vtheo}{v_{\text{th}e,0}}

\newcommand{\mfp}{\lambda_{\text{mfp}}}
\newcommand{\mfpe}{\lambda_{\text{mfp,e}}}

\newcommand{\mfpeff}{\lambda_{\text{mfp,eff}}}
\newcommand{\betaeo}{\beta_{e,0}}
\newcommand{\rhoeo}{\rho_{e,0}}
\newcommand{\wce}{\Omega_e}
\newcommand{\wceo}{\Omega_{e,0}}

\newcommand{\Th}{T_{\rm h}}

\newcommand{\Tc}{T_{\rm c}}

\newcommand{\sa}[1]{#1_{\alpha}}

\newcommand{\unit}[1]{\hat{\boldsymbol{#1}}}

\newcommand{\trv}{t,\bb{r},\bb{v}}

\newcommand{\tv}{t,\bb{v}}
\newcommand{\tr}{t,\bb{r}}
\newcommand{\tvxi}{t,v,\xi}
\newcommand{\gradpz}{\grad p_0=0}
\newcommand{\gradpg}{\grad p_0=\rho_0\bb{g}}
\newcommand{\xicrit}{\xi_{\rm crit}}

\newcommand{\avg}[1]{\langle #1\rangle}

\newcommand{\primed}[1]{#1^{\prime}}
\newcommand{\sign}{{\rm sign}}

\def\apjs{Astrophys.~J.~Supp.~Ser.}
\def\apj{Astrophys.~J.}
\def\apjl{Astrophys.~J.~Lett.}
\def\aap{Astron.~Astrophys.}
\def\araa{Ann.~Rev.~Astron.~Astrophys.}
\def\mnras{Mon.~Not.~Roy.~Astron.~Soc.}

\def\prl{Phys.~Rev.~Lett.}
\def\jgr{J.~Geophys.~Res.}

\def\grl{Geophys.~Res.~Lett.}
\def\physrep{Phys.~Rep.}

\def\pof{Phys.~Fluids}

\def\jpp{J.~Plasma Phys.}

\newcommand{\besselj}[1]{{\rm J}_{#1}}

\defcitealias{Komarov2018}{K18}
\defcitealias{Roberg-Clark2018a}{RC18}

\shorttitle{Whistler Transport}
\shortauthor{E.~L.~Yerger, M.~W.~Kunz, A.~F.~A.~Bott and A.~Spitkovsky}

\title{Collisionless conduction in a high-beta plasma: a collision operator for whistler turbulence}

\author{Evan L.~Yerger\aff{1,2,3}
  \corresp{\email{evan.yerger@unh.edu}},
  Matthew W.~Kunz\aff{1,2},
  Archie F.~A.~Bott\aff{1,4},
 \and Anatoly Spitkovsky\aff{1}}

\affiliation{\aff{1}Department of Astrophysical Sciences, Princeton University, Peyton Hall, Princeton NJ, 08544, USA
\aff{2}Princeton Plasma Physics Laboratory, PO Box 451, Princeton NJ, 08543, USA
\aff{3} Space Science Center, University of New Hampshire, Durham, NH 03824\\
\aff{4}Department of Physics, University of Oxford, Clarendon Laboratory, Oxford, OX1 3PU}

\begin{document}

\maketitle

\begin{abstract}
The regulation of electron heat transport in high-$\beta$, weakly collisional, magnetized plasma is investigated. A temperature gradient oriented along a mean magnetic field can induce a kinetic heat-flux-driven whistler instability (HWI), which back-reacts on the transport by scattering electrons and impeding their flow. Previous analytical and numerical studies have shown that the heat flux for the saturated HWI scales as $\beta_e^{-1}$. These numerical studies, however, had limited scale separation and consequently large fluctuation amplitudes, which calls into question their relevance at astrophysical scales. To this end, we perform a series of particle-in-cell simulations of the HWI across a range of $\beta_e$ and temperature-gradient length scales under two different physical setups. The saturated heat flux in all of our simulations follows the expected $\beta_e^{-1}$ scaling, supporting the robustness of the result. We also use our simulation results to develop and implement several methods to construct an effective collision operator for whistler turbulence. The results point to an issue with the standard quasi-linear explanation of HWI saturation, which is analogous to the well-known $90$-degree scattering problem in the cosmic ray community. Despite this limitation, the methods developed here can serve as a blueprint for future work seeking to characterize the effective collisionality caused by kinetic instabilities.
\end{abstract}

\section{Introduction}\label{sec:intro}

\subsection{Physical motivation}

\subsubsection{The intracluster medium of galaxy clusters}

Galaxy clusters are the largest virialized structures in the Universe, spanning more than a megaparsec in diameter. Clusters are bound by dark matter, which forms a deep gravitational well. Dark matter accounts for the majority of cluster mass, ${\approx}84\%$. The remaining mass is attributed to baryonic matter in the hundreds to thousands of galaxies scattered across the cluster (${\approx}3\%$) or the relatively hot, diffuse plasma between them (${\approx}13\%$) known as the intracluster medium (ICM) [see \citet{Peterson&Fabian2006} for a review]. Studies of temperature gradients in various clusters \citep[e.g.,][]{Binney&Cowie1981,Ettori&Fabian2000,Vikhlinin2001,Markevitch2003,Zakamska2003} have suggested that heat conduction must be suppressed by a multiplicative factor of ${\sim}0.3$ to ${\sim}10^{-2}$ relative to the Spitzer value \citep{Fabian2002}, which, because of its strong temperature dependence ${\propto}T^{7/2}$ \citep{Spitzer1962}, would otherwise isothermalize the ICM rapidly or fail to prevent runaway cooling \citep[e.g.,][]{BregmanDavid1988,Kim2003,Conroy&Ostriker2008}.

\subsubsection{The solar wind}

The modern understanding of the solar wind has its origins in seminal work by \citet{Parker1958}: the corona, which is heated to temperatures upwards of $10^6~{\rm K}$ [e.g., by  Alfv{\'e}n-wave turbulence  \citep{Chandran_2009,Squire_2022,Chen_2022} or perhaps interchange reconnection \citep{Raouafi_2023,Drake_2023}], provides much of the free energy and mass required to accelerate the plasma to velocities exceeding ${\approx}400~{\rm km~s}^{-1}$ \citep[see][for a review]{Verscharen_Review2019}. The observed radial electron temperature profile inside a few au does not match predictions from a purely adiabatic wind \citep{Richardson2003}; heat fluxes are therefore thought to play an important role in the thermodynamic evolution of the solar wind.
A statistical study of \textit{Wind} spacecraft measurements taken from the $\beta\gtrsim 1$, weakly collisional slow wind \citep{Bale2013} revealed that the scaling of the electron heat flux with the Coulomb-collisional mean free path $\mfpe$ depends on the local temperature-gradient length scale $L_T$: for $L_T\gtrsim\mfpe$ the measured heat flux matched collisional predictions  \citep{Spitzer1962}, whereas for $L_T/\mfpe \lesssim 3$ the heat flux was found to be constant with $L_T$. Similar results were also found in data from the {\em Parker Solar Probe} \citep{Halekas2021}.

\subsubsection{Low-luminosity black-hole accretion flows}

Accretion flows onto supermassive black holes are often significantly underluminous when compared to predictions from classical thin-disc theory. For example, if plasma were accreted at the \citet{Bondi1952} rate onto the ${\approx}4\times 10^6~{\rm M}_\odot$ black hole at the Galactic centre, Sgr~A$^\ast$, via a geometrically thin, optically thick accretion disc \citep{ss73}, then the luminosity would be ${\sim}10^5$ times larger than that observed \citep[see reviews by][]{Quataert2003,Yuan2015}. Such radiatively inefficient accretion flows (RIAFs) can be explained by a combination of substantially sub-Bondi accretion, because much of the inflowing plasma is gravitationally unbound and lost to a magnetically driven wind \citep[e.g.,][]{Blandford1999,Hawley2002}, and low radiative efficiency, because the liberated gravitational potential energy is stored as thermal energy primarily in the poorly radiating ion population \citep[e.g.,][]{Narayan1994}. The result is a geometrically thick accretion flow, one in which the thermodynamics of the putative collisionless, high-$\beta$ plasma should play an important role, whether by instigating hydrodynamic \citep{Narayan2000,Quataert2000} or magnetothermal \citep{Balbus2001} convective transport, or by being modified directly by conductive transport \citep{Tanaka2006,Johnson2007,Ressler2015}.

\subsection{General transport considerations and history}

Given the potential importance of conductive and convective heat transport in the ICM, solar wind, and RIAFs, it is not surprising that it remains an active area of research within each of the associated communities. Obtaining definitive physical models for this transport, however, is made difficult by the extreme scale separations characterizing these plasmas and by the complexities such multi-scale physics brings. Namely, all of these plasmas are weakly collisional, with the ratio of Coulomb mean free path and macroscopic temperature-gradient length scales ranging from ${\sim}10^{-2}$ to ${\sim}1$ in the ICM, in the solar wind, and around Sgr~A$^\ast$ near the Bondi radius (increasing to values ${\gg}1$ towards the event horizon). They are also highly magnetized, with the ratio of electron gyroradius $\rho_e$ to $L_T$ being (very roughly) ${\sim}10^{-15}$ in the ICM, ${\sim}10^{-8}$ in the solar wind, and ${\sim}10^{-12}$ at the Bondi radius of Sgr~A$^\ast$. As a result, the motions of charged particles are tightly bound to magnetic-field lines and the transport of both momentum and heat perpendicular to the field is highly suppressed relative to the parallel transport. Thus, both the angle of the magnetic field with respect to the local temperature gradient \citep[e.g., in insulated cluster cold fronts;][]{Vikhlinin2001,Markevitch&Vikhlinin2007} and the extent to which the field lines are tangled by turbulence \citep[e.g.,][]{Chandran&Cowley1998,Narayan2001} have a significant effect on heat transport. If the temperature gradient is aligned or anti-aligned with a confining gravitational field, even energetically weak magnetic fields can drive buoyancy instabilities like the magnetothermal instability \citep{Balbus2000,Balbus2001} or heat-flux buoyancy instability \citep{Quataert2008}, provided that the conductive heat transport between fluid elements is rapid and restricted along field lines [see \citet{Kunz2011b} and \citet{Xu&Kunz2016} for detailed treatments of these instabilities for weakly collisional and collisionless plasmas, respectively].

Kinetic instabilities can also strongly affect transport in weakly collisional, magnetized plasma. The transport of heat or momentum implies a distortion of the distribution function \citep[e.g.,][]{Braginskii1965}, which can be a source of free energy for kinetic instabilities \citep[e.g.,][]{Bott_2024}. If the plasma beta $\beta=8\upi p/B^2$ (the ratio of thermal pressure $p$ and magnetic pressure $B^2/8\upi$) is large, small departures from local thermodynamic equilibrium, and thus small amounts of transport implied by the departure, are enough to grow Larmor-scale distortions in the energetically weak magnetic field. Such distortions can scatter or trap particles, pushing the particles' velocity distribution function back towards isotropy and thereby limiting transport.

One instability of note in this context is the heat-flux-driven whistler instability (HWI), which was first investigated in the case of a weakly collisional plasma by \citet{Levinson_Eichler1992}. Those authors derived a growth rate for the case of a wavevector oriented parallel to the local magnetic field; however, they considered the case where the HWI is saturated by wave-mode coupling and found that the saturated heat flux was independent of the quasi-linear scattering rate. A few years later, \citet{Pistinner_Eichler1998} found that oblique whistler waves, which are elliptically (rather than circularly) polarized, could diffusively scatter heat-flux-carrying electrons via cyclotron resonance and limit the heat flux $\propto\beta_e^{-1}$. Following these analytical results has been a relatively recent flurry of numerical works, namely particle-in-cell (PIC) simulations of the HWI. The first of these were 1D simulations \citep{Roberg-Clark2016}, which found little reduction of the heat flux by parallel whistlers, confirming the need for oblique waves. It was not until 2D PIC simulations were performed that the predicted ${\sim}\beta_e^{-1}$ scaling was shown empirically. These runs were performed concurrently by two independent groups: \citet{Komarov2018} (hereafter \citetalias{Komarov2018}) and \citet{Roberg-Clark2018a} (hereafter \citetalias{Roberg-Clark2018a}).

The literature for heat-flux instabilities in the solar wind evolved largely independently from the more astrophysics-focused work discussed in the preceding paragraph. This is likely due to the emphasis on high-$\beta_e$ plasma for the latter and the multiple electron populations encountered in the former. The `whistler heat flux instability', or WHFI as it is called in the solar wind literature, was first investigated in \citet{Gary1975} for the case of two bi-Maxwellian electron populations representing the core and the halo. The instability was presented as a viable mechanism for heat-flux reduction and was supported by a semi-empirical model by \citet{Gary1994}, who used data from the \textit{Ulysses} spacecraft to show that the parallel heat flux is suppressed by a factor of $\beta_{\parallel, {\rm c}}^{-0.9}$, where $\beta_{\parallel, {\rm c}}$ is the parallel beta of the electron core population. This scaling was later supported by analytic calculations in the limit of high $\beta_{\parallel, {\rm c}}$ \citep{Gary&Li2000}. A statistical study by \citet{Scime1994} also reported that the WHFI best explained the heat flux measured by \textit{Ulysses}. A later study using data from the \textit{Artemis} mission \citep{Tong2019} revealed correlations between the properties of whistler waves seen in the magnetic-field data and the macroscopic plasma parameters, in a way that is consistent with their excitation by the HWI.

\subsection{Purpose and organization}

Previous numerical work on the HWI suffers from a number of limitations. First, the equilibria adopted in these works -- isobaric in the case of \citepalias{Komarov2018} and collisionless in the case of \citepalias{Roberg-Clark2018a} -- are  unlikely to be found in actual galaxy clusters. Secondly, the scale separations used in the numerical simulations are limited, with the ratio of temperature-gradient length scale to electron gyroradius being no more than 256. As a result, the saturated fluctuation amplitude of the whistlers, predicted to satisfy $\delta B/B_0 \sim (\beta_e \rho_e/L_T)^{1/2}$, approached the strength of the background field. On these topics, the aim of this paper is to assess the extent to which prior results on the whistler-mediated heat flux by \citetalias{Komarov2018} and \citetalias{Roberg-Clark2018a} are robust with respect to physical setup and scale separation. For numerical results of the HWI to be extrapolated reliably to astrophysical scale separations and more realistic astrophysical environments, it ought to be checked that the underlying physics of the saturated instability converges at large scale separation and that the HWI is robust to different physical setups. We therefore perform electromagnetic PIC simulations similar to those in \citetalias{Roberg-Clark2018a} and \citetalias{Komarov2018}, but with an additional, physically motivated equilibrium (i.e., in which thermal stratification is associated with hydrostatic equilibrium in a gravitational field) and a focus on studying the largest scale separations available to us numerically.

We then use the results of our simulations to obtain a numerical effective collision operator for the saturated HWI using three distinct methods: the first leverages a Chapman--Enskog expansion to calculate a pitch-angle scattering frequency from the electron distribution function, the second uses a quasi-linear operator to obtain a pitch-angle scattering rate from the magnetic spectrum, and the third utilizes a Fokker--Planck method to obtain a pitch-angle scattering operator from tracked particle data. The results of each of these methods will be discussed and synthesized in the context of building a model effective collision operator for the HWI. Our model operator has inherent limitations associated with it, which we show to be a manifestation of a longstanding issue with quasi-linear analyses. Despite our inability to construct a fully self-consistent model operator, we hope the present work will advance the methods available and that our experience will provide a useful road map so that future studies of the saturation of kinetic instabilities may be more successful.

The paper is organized as follows. In section~\ref{sec:background}, we provide a brief quantitative introduction to the HWI. We then detail our numerical methods and simulation diagnostics in section~\ref{sec:methods}. Section~\ref{sec:results1} contains simulation results that largely confirm past work -- namely, the dependence of the saturated heat flux and wave amplitude on $\beta_e$ and the temperature-gradient length scale. For those already familiar with the instability, the heat flux and wave amplitude dependencies are given by \eqref{eqn:qb_v_beta} and are consistent with an effective collision frequency given by \eqref{eqn:nueff_sat}. All of these relationships hold regardless of the equilibrium state and agree with previous work. Section~\ref{sec:results2} contains the bulk of our original analysis. Here, we detail three methods for obtaining model collision operators from our simulations, namely: leveraging a Chapman--Enskog expansion (\S\ref{sec:CE}), a quasi-linear method (\S\ref{sec:QL}), and a Fokker--Planck method (\S\ref{sec:FP}). In Section~\ref{sec:results3}, we present a physically motivated model collision operator (\S\ref{sec:model_CO}), discuss its shortcomings in explaining the heat flux observed in our simulations (\S\ref{sec:IHF}), and end the section by exploring various ways to deal with those shortcomings (\S\ref{sec:scattering_gap}). Finally, we present our conclusions in Section~\ref{sec:conclusion}.

\section{Background}\label{sec:background}

Given the existing body of work on the HWI, we present here only a brief overview of the instability and refer the reader to the works cited in the Introduction (\S\ref{sec:intro}) for more detail. In \S\ref{sec:HWI_background}, we provide a simplified description of the HWI, focusing mostly on its qualitative features. It should be noted that the quantitative details used to support this description differ slightly from those obtained from the fully self-consistent, collisionless simulations presented in subsequent sections. A reader already familiar with the HWI can skip to section~\ref{sec:methods} for our numerical methods and setup, section~\ref{sec:results1} for a summary of our simulation results, or section~\ref{sec:results2} for our model collision operator results.

\subsection{The heat-flux-driven whistler instability (HWI)}\label{sec:HWI_background}

The HWI is a destabilization of the whistler branch of the plasma dispersion relation in the presence of a heat flux. To see this, consider a plasma threaded by a magnetic field $\bb{B}$. If the plasma is magnetized, meaning that the electron gyroradius $\rho_e$ is much smaller than some characteristic macroscale length scale $L$ in the plasma, then the transport of heat and momentum will be highly anisotropic, with transport along (`$\parallel$') the magnetic field occurring much more rapidly than transport across (`$\perp$') the field. For example, if a macroscopic electron temperature gradient $\grad T_e$ has a component oriented along the magnetic-field direction $\unit{b}=\bb{B}/|\bb{B}|$ (i.e.,~$\nabla_\parallel T_e \ne 0$), then electrons will flow to produce a heat flux $\bb{q}_e$ that is oriented predominantly along the field, {\em viz.}~$\bb{q}_e\simeq q_{\parallel,e}\unit{b}$. If, furthermore, the collisional mean free path $\mfp$ is much smaller than the temperature gradient length scale along the field, $L_T=(-\nabla_\parallel \ln T_e)^{-1}$, then the distortion in the velocity distribution function of the electrons associated with the heat flux will be small, ${\sim}\mfp/L_T$, relative to the equilibrium (Maxwellian) distribution. Assuming that electrons are pitch-angle scattered at some velocity-independent rate $\nu$, these considerations suggest a steady-state electron distribution function given by
\begin{equation}\label{eqn:fe_model}
    f_e(v,\xi)=\frac{n_e}{\upi^{3/2}\vthe^3}\,\rme^{-v^2/\vthe^2}\bigg[1-\frac{\vthe}{\nu L_T}\, \xi\,\frac{v}{\vthe} \bigg(\frac{v^2}{\vthe^2}-\frac{5}{2}\bigg)\bigg],
\end{equation}
where $\xi\doteq \vpar/v$ is the cosine of the pitch angle, $n_e$ is the local electron number density, and $\vthe$ is the local electron thermal speed \citep[e.g.,][]{Braginskii1965}. For a derivation of a similar expression, see Appendix~\ref{apx:CEO}.

Parallel-propagating whistler waves in a plasma with an electron distribution function given by \eqref{eqn:fe_model} are unstable, with a growth rate dependent upon the product of the mean free path normalized by the temperature-gradient length scale and the electron plasma beta parameter $\beta_e=8\upi p_e/B^2$ -- the ratio of scalar electron thermal pressure $p_e$ to magnetic pressure $B^2/8\upi$. An analytic expression for the growth rate of whistlers with parallel wavenumber $\kpar=\bb{k}\bcdot \unit{b}>0$ may be obtained in the asymptotic limit $\vthe\beta_e/\nu L_T\ll 1$ and $\kpar\rho_e/\beta_e^{1/2}\ll 1$ \citep[see \S 3.3.1 of][]{Bott_2024}; it is
\begin{equation}\label{eqn:HWI_growth_rate}
    \frac{\gamma}{\wce}\simeq\frac{\sqrt{\upi}}{\kpar^2\rho_e^2}\bigg(\frac{\vthe}{2\nu L_T}-\frac{\kpar^3\rho_e^3}{\beta_e}\bigg)\exp\bigg(-\frac{1}{\kpar^2\rho_e^2}\bigg),
\end{equation}
where $\wce=\vthe/\rho_e$ is the (positive) electron gyrofrequency. (The qualitative features of \eqref{eqn:HWI_growth_rate} do not change outside the limit $\vthe\beta_e/\nu L_T\ll 1$; the same is not true for the $\kpar\rho_e/\beta_e^{1/2}\ll 1$ limit.) Note, the growth rate is only positive when $L_T>0$, i.e. for whistlers moving down the temperature gradient. The corresponding dispersion relation for the (real) wave frequency $\omega$ is
\begin{equation}\label{eqn:cpwdr}
    \frac{\omega}{\Omega_e} \simeq \frac{k\kpar \rho_e^2}{\beta_e}.
\end{equation}
Coincidentally, this is the same as the cold plasma dispersion relation for whistler waves, $\omega/\Omega_e \simeq kk_\parallel d^2_e$, where $d_e$ is the electron skin depth.

\begin{figure}
    \centering
    \includegraphics[width=0.85\textwidth]{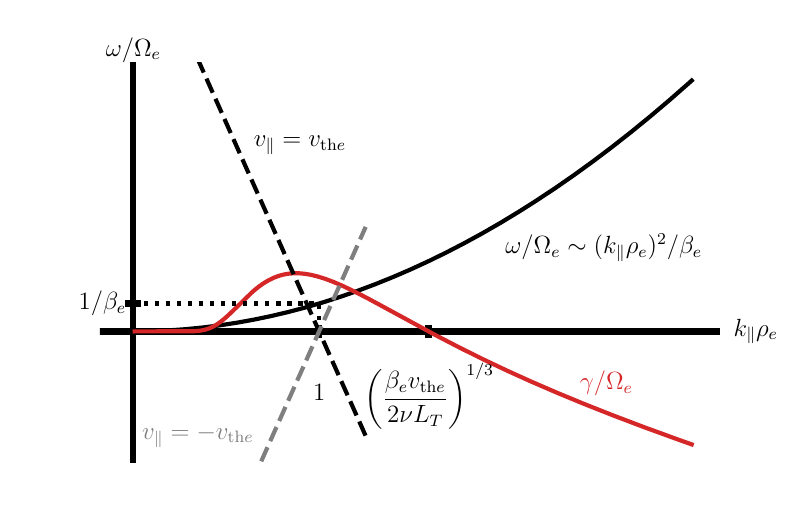}
    \caption{HWI growth rate (red line), whistler dispersion relation (solid black line), and cyclotron resonances (dashed lines) plotted as functions of $\kpar\rho_e$. For thermal electrons, only the cyclotron resonances intersect the whistler dispersion relation at values of $\kpar\rho_e$ corresponding to maximum growth rate.}
    \label{fig:whistler_dispersion}
\end{figure}

The growth rate \eqref{eqn:HWI_growth_rate} and dispersion relation \eqref{eqn:cpwdr}
are plotted schematically in figure~\ref{fig:whistler_dispersion}. The growth rate has a maximum near $\kpar\rho_e\sim 1$ (while this statement does not strictly follow from \eqref{eqn:HWI_growth_rate} because the expression was derived in a certain asymptotic limit, the statement is rigorous in the correct limit; see \citet{Bott_2024} for specific details). For this wave vector, the whistler phase velocity,
\begin{equation}\label{eqn:vw}
    \vw\sim\frac{\omega}{\kpar}\sim \frac{\vthe}{\beta_e}
\end{equation}
is very sub-thermal. Resonant wave-particle interactions occur when the frequency of the wave vanishes in the frame co-moving with a particle having parallel velocity $\vpar$, i.e.
\begin{equation}\label{eqn:res_condition}
    \omega-\kpar\vpar+n\Omega_e=0,
\end{equation}
where $n$ is an integer. Whistler waves in the HWI, therefore, are only Landau $(n=0)$ resonant with electrons that are buried deep in the core of the distribution and do not contribute to the instability. Instead, the principal cyclotron $(n=\pm 1)$ resonances, which couple the waves with thermal $(\vpar\sim\vthe)$ electrons in the unstable region of the distribution function \eqref{eqn:fe_model}, mediate the instability. This is represented schematically in figure~\ref{fig:whistler_dispersion}, where the cyclotron resonances corresponding to thermal electrons intersect the whistler dispersion relation at the $\kpar\rho_e$ corresponding to maximum growth rate. The decline in growth rate at small $\kpar\rho_e$ can be identified with electrons that have superthermal parallel velocities, of which there are exponentially few [thus the multiplicative factor of $\exp(-1/\kpar^2\rho_e^2$)]. The growth rate has a zero at $\kpar\rho_e=(\beta_e\vthe/2\nu L_T)^{1/3}$; any waves with larger $\kpar$ are cyclotron damped on subthermal electrons in the core of the distribution.

\subsection{HWI saturation}\label{sec:HWI_sat}
As the whistlers grow to finite amplitude, resonant wave-particle interactions satisfying \eqref{eqn:res_condition} begin to diffuse electrons, which pushes the distribution function toward isotropy. This in turn reduces the whistler growth rate. The system eventually reaches saturation, a steady state in which the waves have large enough amplitudes to scatter electrons and thereby hold the distribution function at a marginally stable state in which the growth rate is zero. Denoting the rate at which the nonlinear whistlers diffuse electrons as $\nueff$, an `effective collisionality', and substituting $\nueff$ for $\nu$ in \eqref{eqn:HWI_growth_rate} with $\kpar\rho_e\sim 1$, we find that the HWI should saturate when
\begin{equation}\label{eqn:nueff_sat}
    \nueff \sim \frac{\beta_e\vthe}{L_T} .
\end{equation}
The saturated heat flux in the saturated state should then scale as
\begin{equation}\label{eqn:q_scaling}
    q^{\rm sat}_{\parallel, e }\sim m_en_e\frac{\vthe^4}{\nueff L_T}\sim m_en_e\frac{\vthe^3}{\beta_e}.
\end{equation}
If whistlers do indeed scatter electrons diffusively, the effective scattering frequency can be related to the whistler fluctuation energy via
\begin{equation}\label{eqn:nueff_diff}
    \nueff\sim \wce\frac{\delta B^2}{B_0^2} .
\end{equation}
Matching this expression for $\nueff$ with \eqref{eqn:nueff_sat} implies that
\begin{equation}\label{eqn:dB2_prediction}
    \frac{\delta B^2}{B_0^2}\sim \frac{\beta_e}{L_T/\rho_e}
\end{equation}
in saturation.

The scaling \eqref{eqn:dB2_prediction} was predicted by \citetalias{Komarov2018} and obtained empirically using bespoke kinetic PIC simulations. Those authors then argued that whistlers pitch-angle scatter electrons in the frame of the background flow. \citetalias{Roberg-Clark2018a} argued instead that whistlers advect thermal energy at their phase velocity, so that
\begin{equation}\label{eqn:q_RC18}
    q_{\parallel,e}\sim n_eT_e\avg{\vpar}\sim n_eT_e\vw \sim m_en_e\frac{\vthe^3}{\beta_e}.
\end{equation}
Both arguments reproduce the scaling \eqref{eqn:q_scaling} for the saturated heat flux. This was recognized by \citet{Drake2021}, who concluded that whistlers should act as pitch-angle scatterers in a frame moving at the wave phase velocity. The argument for electromagnetic waves scattering electrons in this way goes back to \citet{Kennel&Engelmann1966}, who derived a quasi-linear wave-particle diffusion operator and found that, for particles with $\vpar \gg \omega/\kpar$, the gyroresonant scattering is dominated by scattering in pitch angle due to the parallel electric field vanishing in the frame of the particle; only particles with $\vpar \sim \omega/\kpar$ have strong velocity diffusion.

Another aspect agreed upon by \citetalias{Komarov2018}, \citetalias{Roberg-Clark2018a}, and \citet{Drake2021} is the need for oblique -- and thus elliptically polarized -- whistler waves to achieve the heat flux scaling $\qpare\sim \beta_e^{-1}$. This requirement, initially discovered by \citet{Pistinner_Eichler1998}, is due to both the right-handedness of purely parallel whistler waves and the fact that the HWI only drives unstable whistlers traveling down the temperature gradient. Any electron traveling down the temperature gradient faster than the whistler phase speed will therefore never be in cyclotron resonance with the wave, which is Doppler-shifted to a left-hand polarization in that electron's frame. In contrast, the elliptical polarization of oblique waves can be decomposed into left- and right-handed components, which can be in cyclotron resonance with electrons traveling in either direction along the temperature gradient. What seems to have gone unappreciated, however, is that the ratio of energy in left- to right-handed waves is significantly below unity for all but the most oblique wave vectors. Using the cold plasma whistler dispersion \citep{Stix1992}, the ratio of energies in the left-handed ($-$) and right-handed ($+$) components is
\begin{equation}\label{eqn:pol_frac}
    \frac{\mathcal{E}^-}{\mathcal{E}^+}\simeq \bigg(\frac{\cos\theta -1}{\cos\theta+1}\bigg)^2 .
\end{equation}
Even for oblique waves with $\theta = 45^\circ$, $\mathcal{E}^-/\mathcal{E}^+\simeq 0.03$. (Finite-temperature effects can increase this ratio.) Given \eqref{eqn:nueff_diff}, it is possible an asymmetry will arise in the scattering rate with respect to pitch angle; as far as we are aware, this effect has not been commented on in the literature thus far.

The ion species can also affect the saturation of the HWI. At high $\beta_e$, ions may be resonant with oblique whistler waves. For example, when $\beta_e=(m_i/m_e)^{1/2}$, the ion thermal speed is concurrent with the cold-plasma whistler phase speed, {\em viz.}~$\vthi=(m_i/m_e)^{-1/2}\vthe=\vthe/\beta_e\sim \vw$, and the waves can Landau damp on the ions. For $\beta_e=m_i/m_e$, the whistler wave frequency is of order the ion gyrofrequency, {\em viz.}~$\omega\sim\Omega_e/\beta_e=\Omega_i$ and the ions can cyclotron damp the waves. Such interactions would heat the ion species at the expense of the strength of the turbulent fluctuations and therefore might increase the saturated electron heat flux in the HWI.

\subsection{Constructing collision operators}
In this work, we explore three methods to construct a collision operator from our kinetic simulations describing how the electrons interact with the HWI fluctuations: Chapman--Enskog, quasi-linear, and Fokker--Planck. In section~\ref{sec:CE}, we leverage measurements of the distribution function in the simulations to solve the Chapman--Enskog problem for a postulated pitch-angle scattering operator and obtain the associated scattering frequency $\nuce$ as a function of $v$. This method is very powerful in that it solves for the effective collision frequency that explains the observed heat flux for a given scattering model, thereby reproducing the observed heat flux by definition. The downside is that one must assume a scattering model, so the approach is not very flexible. The quasi-linear operator (\S\ref{sec:QL}) computes a collision frequency from the observed spectrum of waves, assuming resonant wave-particle scattering. As we will show, the assumption of purely resonant interactions does not hold because of the large fluctuation amplitudes in our runs (though may be better justified at increasingly large scale separations). Finally, in section \ref{sec:FP}, we construct a Fokker--Planck operator by explicitly computing the short-time velocity and pitch-angle jump moments from ${\sim}500{\rm k}$ tracked electrons. This operator is by far the most general, but can be easily misinterpreted.

\section{Numerical methods and diagnostics}\label{sec:methods}

\subsection{Initial equilibrium states}\label{sec:initial_conditions}

We initialize each run with a thermally stratified ion-electron plasma and a constant background magnetic field $\bb{B}_0=B_0\unit{x}$. We parameterize the temperature profile of each species as a function of $x$ by a factor $N>1$ using
\begin{equation}\label{eqn:temperature_profile}
    T(x) = \Th\bigg(1-\frac{x}{NL_x}\bigg) ,
\end{equation}
where $\Th$ is a `hot' temperature at the left wall, $x=0$. The `cold' temperature at right wall, $x=L_x$, is then given by
\begin{equation}\label{eqn:temperature_ratio}
    \Tc=\Th\bigg( \frac{N-1}{N}\bigg).
\end{equation}
The temperature-gradient length scale $L_T\doteq-(\nabla_{\parallel}\ln T)^{-1}$
is therefore $L_T = NL_x$.

We take the scalar pressure of species $\alpha$, $\sa{p}$, to be related to its density $\sa{n}$ and temperature $\sa{T}$ through an ideal equation of state, $\sa{p}=\sa{n}\sa{T}$. We then construct two different equilibria depending on whether a gravitational field $\bb{g}=-g\ex$ is present. For $g=0$, hydrostatic equilibrium is simply $\rmd p_\alpha/\rmd x = 0$. Using~\eqref{eqn:temperature_profile} for $T_\alpha(x)$, we find that
\begin{equation}\label{eqn:k18_density}
    n(x) = n_0 \biggl(1-\frac{x}{NL_x}\biggr)^{-1} \qquad{\rm when}~g=0,
\end{equation}
which is independent of species. This is the initial condition used in~\citetalias{Komarov2018}. While this setup is simple, it is difficult to imagine a physically relevant scenario in which the gradients of temperature and density have equal and opposite signs, $\rmd\ln T_\alpha/\rmd x = -\rmd\ln n_\alpha/\rmd x$. We therefore consider a complementary setup in which gravity balances a non-zero pressure gradient and the gradients of temperature and density have the same sign. For $g\ne 0$, hydrostatic equilibrium is then given by
\begin{equation}\label{eqn:hydro_force_balance}
    \D{x}{\sa{p}} - q_\alpha n_\alpha E_0 + m_\alpha n_\alpha g = 0 ,
\end{equation}
an equation that must hold for each species individually. We take $T_i=T_e\doteq T$ in the equilibrium state, in which case $E_0$ is the equilibrium electric field that is required to maintain quasi-neutrality with a gravitational force that is greater on the ions than on the electrons. To determine this electric field, one multiplies~\eqref{eqn:hydro_force_balance} by $q_\alpha/m_\alpha$, sums over particle species, and uses $\sum_\alpha q_\alpha n_\alpha = 0$ to find that
\begin{equation}
    E_{0}= \left.\sa{\sum}\frac{\sa{q}}{\sa{m}}\frac{d\sa{p}}{dx}\middle/ \sa{\sum}\frac{\sa{q}^2\sa{n}}{\sa{m}}\right. = - \frac{1}{en}\bigg(\frac{m_i-m_e}{m_i+m_e}\bigg) \D{x}{p} ,
\end{equation}
where in the second equality we have taken $p_i = p_e \doteq p$, $q_i = -q_e = e$, and $n_i=n_e\doteq n$. Directly summing~\eqref{eqn:hydro_force_balance} over species yields the standard hydrostatic equilibrium,
\begin{equation}\label{eqn:hydroeq}
    \sum_\alpha \D{x}{p_\alpha} = 2\D{x}{p} = -\varrho g,
\end{equation}
where $\varrho\doteq(m_i+m_e)n$. Using~\eqref{eqn:temperature_profile} for the temperature profile, equation~\eqref{eqn:hydroeq} can be straightforwardly integrated to find that
\begin{equation}\label{eqn:grav_density}
    n(x) = n_0 \biggl( 1 - \frac{x}{NL_x}\biggr)^{N-1} \quad{\rm when}~g = \frac{2\Th}{(m_i+m_e)L_x} .
\end{equation}
In this case, both the temperature and the density decrease with `height' ($x$).

Before proceeding to discuss our numerical approach for evolving these equilibria, it is worth addressing an issue with the $g\ne 0$ case, which may have caught the attention of erudite students of astrophysical fluid dynamics. The situation in which a thermally conducting magnetic field is aligned with a temperature gradient that points in the direction of gravity is known to be linearly overstable to $g$-mode perturbations when the profiles satisfy the inequalities $1 < (1-1/\gamma)(\rmd\ln p/\rmd\ln T) < 2$, where $\gamma$ is the effective adiabatic index of the plasma \citep{Balbus&Reynolds2008}.\footnote{This configuration is linearly stable to the magnetothermal instability \citep{Balbus2000,Balbus2001}, the heat-flux-driven buoyancy instability \citep{Quataert2008}, and their weakly collisional and collisionless generalizations \citep{Kunz2011b,Xu&Kunz2016}.} Using the profiles \eqref{eqn:temperature_profile} and \eqref{eqn:grav_density}, this criterion requires that $N$ satisfies $1 < (1-1/\gamma)N < 2$. Without specifying $\gamma$, which can differ from the standard monatomic $5/3$ when the plasma in question is collisionless and magnetized, we can nevertheless state that this overstability plays no role in our simulations, because its maximum growth rate is a fraction (typically ${\sim}0.1$) of the characteristic buoyancy frequency, ${\sim}(g/L_x)^{1/2} \simeq \Omega_e (m_e/m_i)^{1/2}(\rho_e/L_x)$ \citep[][\S 4.3]{Kunz2011b}. With our use of $m_i/m_e=1600$ and $L_x/\rho_e\ge 125$ (see \S\ref{sec:simuation_approach}), none of our simulations are run for long enough to realize such slow growth. In fact, with the whistler instability rapidly producing an effective collisionality $\nu_{\rm eff} \sim \beta_e (\vthe/L_T)$, one can show that the fastest-growing $g$-mode has a parallel wavelength larger than the scale height (and therefore the box length).

\subsection{Simulation approach and choice of free parameters}\label{sec:simuation_approach}

The initial conditions described in \S\ref{sec:initial_conditions} are advanced forward in time by the Vlasov--Maxwell set of equations using the particle-in-cell (PIC) method implemented in {\tt TRISTAN-MP}~\citep{Spitkovsky2019}. All runs are 2.5D: the spatial simulation domain is an elongated, two-dimensional grid of size $L_x \times L_y$ with $L_x\gg L_y$, while particle velocities are fully three-dimensional. We choose the `hot' temperature at the left wall to be commensurate with an electron thermal velocity $\vtheo\doteq \vthe(x=0)=\sqrt{2\Th/m_e}=c/5$, where $c$ is the speed of light. We take the same temperature ratio, $\Th/\Tc=2$, across all of our runs. Thus $N=2$ in \eqref{eqn:temperature_profile}--\eqref{eqn:k18_density} and \eqref{eqn:grav_density}. The heat flux driven through the box in the absence of collisions is
\begin{equation}\label{eqn:reference_q}
    \begin{split}
    q_{\parallel,e 0}
    &\doteq m_e n_e(x=0) \vthe^{3}(x=0) - m_e n_e(x=L_x)\vthe^3(x=L_x)\\
    &= \begin{cases}(1-2^{-1/2}) m_en_{e0}\vtheo^{3},&\text{ for } \gradpz\\
    \frac{1}{2}m_en_{e0}\vtheo^{3},&\text{ for } \gradpg
    \end{cases}
    \end{split}
\end{equation}
which we use to normalize the computed parallel heat flux in all of our runs.
We set the density at the left wall to be
\begin{equation}
    n_{e0}\doteq {\rm ppc_0}=\begin{cases}
        175~{\rm ppc},&\text{ for }\gradpz\\
        350~{\rm ppc},&\text{ for }\gradpg
    \end{cases}
\end{equation}
where ${\rm ppc}$ is particles per cell; in both cases the electron density at the center of the domain is ${\simeq} 262$. Electrons are first sampled in the $x$-$y$ plane according to the appropriate density distribution, whereupon their temperature is calculated following~\eqref{eqn:temperature_profile}. An ion is deposited at the same spatial coordinates as each electron so that the initial condition is exactly charge neutral. Following \citet{Zenitani2015}, the particle speed is sampled from a Maxwell--J{\"u}ttner distribution with that temperature and projected onto a uniform sphere to obtain an isotropic velocity distribution. We set the electron inertial length $d_e(x=0)=7.5\Delta x$, where $\Delta x$ is the grid spacing, across all runs. The electron Debye length, $\lambda_{{\rm D},e}(x=0)=1.5\Delta x$, is large enough with respect to the grid to minimize spurious particle heating.

The strength of the background field $B_0$ is controlled by the initial electron plasma beta parameter $\betaeo$ measured at the left boundary of the domain, $x=0$:
\begin{equation}
    \betaeo=\frac{8\upi p_{e,0}}{B_0^2},
\end{equation}
where $p_{e,0}=p_e(x=0)=n_0\Th$. The size of the domain is given in units of the electron gyroradius measured at $x=0$, $\rhoeo=\vtheo/\wceo$, with $L_y/\rhoeo = 25$. We vary $L_x/\rhoeo$ to change the temperature-gradient length scale $L_T$. This ensures that the expected effective whistler mean free path,
\begin{equation}
    \mfpeff\sim \frac{L_T}
    {\betaeo}\sim \frac{2}{\betaeo}L_x
\end{equation}
is much smaller than $L_x$ for $\betaeo\gg 1$.

\renewcommand{\arraystretch}{1.25}
\renewcommand{\tabcolsep}{6pt}
\begin{table}
    \centering
    \begin{tabular}{ l l r r r r c c c }
		Name & Equilibrium & $\beta_{e,0}$ & $\dfrac{m_i}{m_e}$
        & $\dfrac{L_x}{\rho_{e,0}}$ & $\dfrac{L_T}{\rho_{e,0}}$ & $T_{\text{run}}\Omega_{e,0}$ & $\beta_{e,0}\dfrac{\vw}{\vtheo}$ & ($\ts - \te)\Omega_{e,0}$
        \\ \hline
		b10     & $\gradpz$   & 10    & 1600	& 125	& 250   & 6150 &
        $0.20$ & $4500-6000$\\
		b10g    & $\gradpg$   & 10    & 1600	& 125	& 250   & 6125 &
        --   & -- \\

		b25     & $\gradpz$   & 25    & 1600	& 125	& 250   & 7476 &
        $0.19$   & $5000-7476$ \\
		b25g    & $\gradpg$   & 25    & 1600	& 125	& 250   & 2430 &
        --   & -- \\

		b40     & $\gradpz$   & 40    & 1600	& 125	& 250   & 7630 &
        $0.21$   & $5500-7500$\\
		b40g    & $\gradpg$   & 40	  & 1600	& 125	& 250   & 3125 &
        --  & -- \\
		b40x2   & $\gradpz$   & 40	  & 1600	& 250	& 500   & 6988 &
        $0.25$  & $4500-6988$ \\
		b40gx2  & $\gradpg$   & 40	  & 1600	& 250	& 500   & 3255 &
        --  & -- \\
		b40x4   & $\gradpz$   & 40    & 1600	& 500	& 1000  & 7975 &
        $0.25$  & $6500-7500$ \\
        b40x8   & $\gradpz$   & 40    & 1600	& 1000	& 2000  & 7805 &
        0.25  & $6000-7800$\\

		b50     & $\gradpz$   & 50    & 1600	& 125	& 250   & 3992 &
        --  & -- \\
		b50g    & $\gradpg$   & 50    & 1600	& 125	& 250   & 3015 &
        --  & -- \\

		b100    & $\gradpz$   & 100   & 1600  & 125   & 250   & 4600 &
        $0.28$  & $3000-4500$ \\
		b100m   & $\gradpz$   & 100   & 100   & 125   & 250   & 4000 &
        --  & -- \\
    \end{tabular}
    \caption{For all runs performed, from left to right: run name, equilibrium initial condition, initial electron plasma beta parameter, proton-to-electron mass ratio $m_i/m_e$, length of the box along the temperature gradient $L_x$, temperature gradient length scale $L_T$, and run time $T_{\rm run}$. For $\gradpz$ runs used to compute an effective collision operator, we include the whistler phase speed $\vw$ measured from spectrograms and the time interval over which our Fokker--Planck coefficients are calculated, $\ts-\te$.}
    \label{tab:runs}
\end{table}

We conducted a number of simulations varying $\betaeo$, $L_T/\rhoeo$, and the equilibrium state; these runs are summarized in Table \ref{tab:runs}. To determine how the heat flux scales with $\betaeo$, we conduct simulations at $\betaeo=\{10,25,40,50,100\}$ using the \citetalias{Komarov2018} equilibrium and $\betaeo=\{10,25,40,50\}$ for the gravitational equilibrium, keeping $L_T=250\rhoeo$ fixed throughout. Similarly, we fix $\betaeo=40$ while varying $L_T/\rhoeo=\{250,500,1000,2000\}$ and $L_T/\rhoeo=\{250,500\}$ for the \citetalias{Komarov2018} and gravitational equilibria, respectively, to quantify how the heat flux scales with the temperature gradient length scale. All but one of the aforementioned runs were conducted using $m_i/m_e=1600$; there is one run with $m_i/m_e=100$ and $\betaeo=100$, designed to test the effect of the ion gyroresonance on the instability.

\subsection{Particle boundary conditions and temperature enforcement} \label{sec:particle_bcs}

In the $y$-direction perpendicular to $\bb{B}_0$, the boundary conditions for the particles are periodic. In order to maintain the temperature gradient across the domain and thereby drive the system to a steady state, the boundaries in the $x$-direction fix the temperatures at $x=0$ and $x=L_x$ to $\Th$ and $\Tc$, respectively. This is achieved by placing particle barriers on the edges of the simulation domain and re-sampling any particle that crosses them from a half-shell distribution with the appropriate temperature. The algorithm is as follows. If, after a single particle push a particle crosses a barrier, it is pushed back to the point of collision with the wall and the current generated by the push back is added to the current density on the grid. The particle speed is then sampled from a Maxwell--J{\"u}ttner distribution in the same manner as the simulation domain was initialized in (\S\ref{sec:simuation_approach}); however, instead of projecting the speed onto a uniform sphere, we project onto a half-shell distribution that pushes the particle back into the active domain. The returning particle therefore acts as though it has been thermalized by a heat bath at the appropriate temperature. When this particle boundary condition is enforced, any currents deposited by a particle being rewound to the domain boundary or subsequently re-injected are summed into the total current density for that time step. That total current density is then used to calculate the fields in the next time step, ensuring that the algorithm is physically consistent and conserves charge.

\subsection{Field boundary conditions and wave absorption}\label{sec:field_bcs}

The electromagnetic fields, following the particles, are taken to be periodic across the $y$ boundaries. In the $x$-direction, we employ the masking method for absorbing boundary conditions detailed in \citet{Umeda2001}. The electric and magnetic fields are multiplied by a masking function $M(x,L_D,r)$, which is a function of $x$, the length of the damping region $L_D$, and the masking parameter $r$ $(0<r\le 1)$ given by
\begin{equation}
    M(x,L_D,r) = \begin{cases}
        1 - \bigg(r\,\dfrac{x-L_D}{L_D}\bigg)^2, & x\le L_D \\
        1, & L_D < x < L_x-L_D \\
        1 - \bigg(r\,\dfrac{x-L_x+L_D}{L_D}\bigg)^2, & x\ge L_x-L_D
        \end{cases}.
\end{equation}
This mask is multiplied to the right-hand-side of the discretized Amp{\`e}re's and Faraday's laws. This has the effect of strongly suppressing the fluctuations near the $x$-domain boundaries. We set the length of the damping region $L_D=\lambda_{\rmw}/\upi\sim 2\rhoeo$, assuming a characteristic wavenumber for the whistlers satisfying $\kpar\rho_e\sim 1$, and the masking parameter $r=0.25$. The ratio of the incident wave intensity to the reflected intensity using this scheme is proportional to amount of time the wave spends in the buffer.

\section{Results, I.~Whistler regulation of the heat flux}\label{sec:results1}

To help organize the presentation, our results are divided into two separate sections. The first (this section) focuses on how the imposed temperature gradient drives a heat flux unstable to the whistler instability, which subsequently scatters electrons and thereby regulates that heat flux to a value that is marginally unstable. This section generally serves to confirm previous work -- in particular the dependence of saturated heat flux and whistler amplitude on $\betaeo$ and $L_T$ (\S\ref{sec:sat_scaling}). We therefore keep this section brief, referring the reader to both \citetalias{Komarov2018} and \citetalias{Roberg-Clark2018a} for more detail. We do, however, comment on the prolonged secular phase of the instability we observe (\S\ref{sec:inst_evolution}), directly measure the whistler dispersion relation (\S\ref{sec:spect}), and note a significant reduction in heat flux by ion-cyclotron-resonant damping (\S\ref{sec:q_vs_mratio}) -- all of which are novel contributions. The second section, \S\ref{sec:results2}, documents our efforts to derive, using a combination of analytical arguments and simulation data, an effective collision operator describing how electrons interact with the whistler fluctuations. The latter is the principal contribution of this work.

\subsection{Evolution of the instability}\label{sec:inst_evolution}

\begin{figure}
    \centering
    \includegraphics[width=1.0\textwidth]{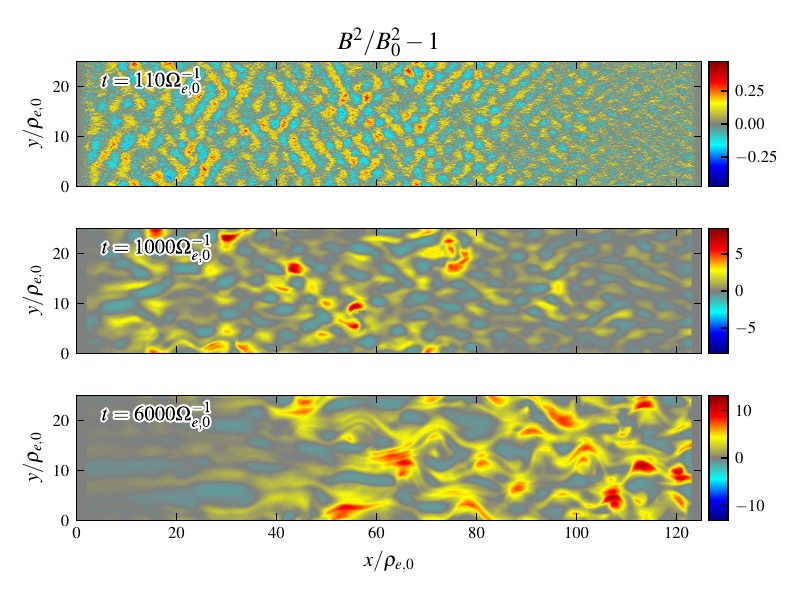}
    \caption{Perturbed magnetic energy in run b40 at the beginning of the exponential phase (top), end of exponential phase (second from top),  and in the saturated state (bottom).}
    \label{fig:b40_snapshots}
\end{figure}

Each of our runs follow a qualitatively similar evolution, going through four distinct phases: (i) establishment of an unstable heat flux, (ii) exponential growth of whistler waves, (iii) sustained power-law (`secular') growth, and (iv) complete saturation of the HWI. Qualitatively, these four stages are manifest in figure~\ref{fig:b40_snapshots}, which shows 2D plots of the magnetic-field energy in run~b40 ($\betaeo=40$ and $L_T=250\rhoeo$) during the latter three stages in the top, middle, and bottom panels, respectively. In the exponential phase (ii), the top panel of figure \ref{fig:b40_snapshots} shows oblique, $k\rhoeo\sim 1$, whistlers throughout the domain. The gray bars at the edges of the domain correspond the masked regions described in section~\ref{sec:field_bcs}. In the secular phase (middle panel), we see a shift towards smaller wavenumbers and stronger field fluctuations, a trend that continues until saturation (third panel).

\begin{figure}
    \centering
    \mbox{\large \hspace{1em}$(a)$\hspace{0.95\textwidth}}
	\includegraphics[width=1.0\textwidth,clip]{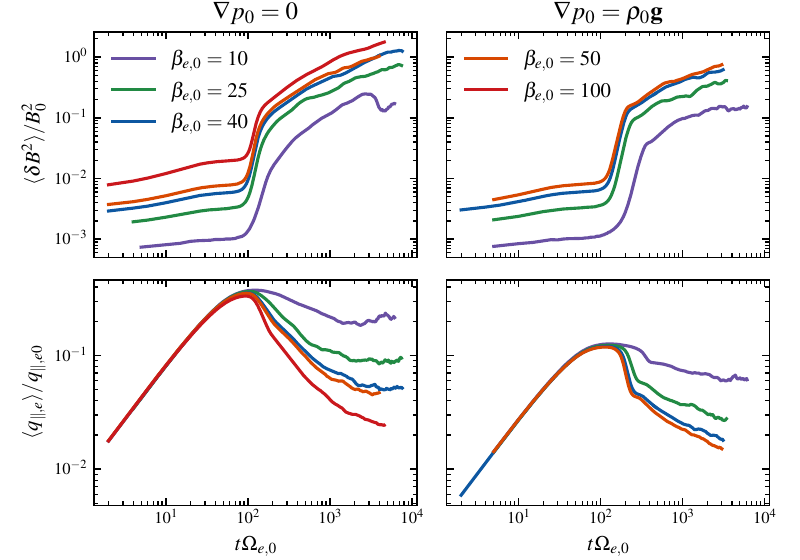}
    \mbox{\large \hspace{1em}$(b)$\hspace{0.95\textwidth}}
    \includegraphics[width=1.0\textwidth,clip]{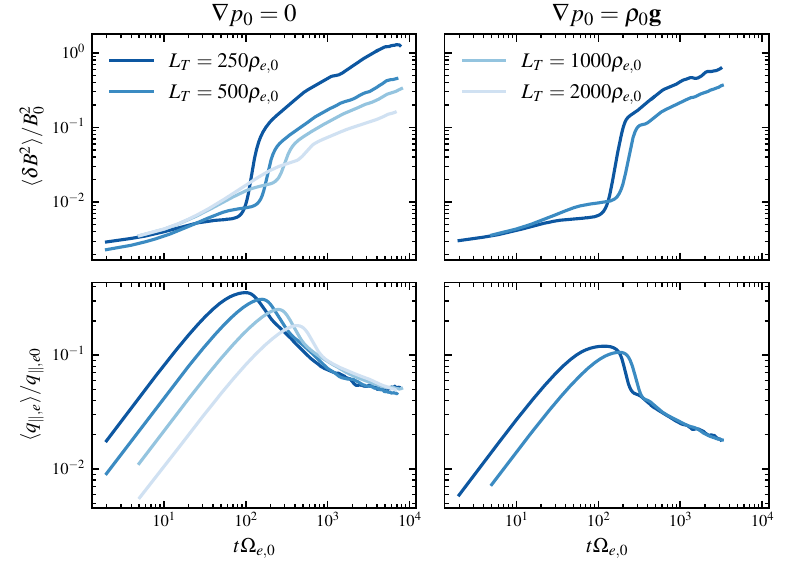}
    \caption{Time evolution of $\delta B^2$ and $q_{\parallel}$ for both the $\gradpz$ (left panels) and $\gradpg$ (right panels) equilibria ($a$) as a function of electron plasma beta $\betaeo$ for $L_T=250\rhoeo$ and ($b$) as a function of temperature-gradient length scale $L_T$ for $\beta_e=40$.}
    \label{fig:HWI_growth}
\end{figure}

\begin{figure}
    \centering
    \includegraphics{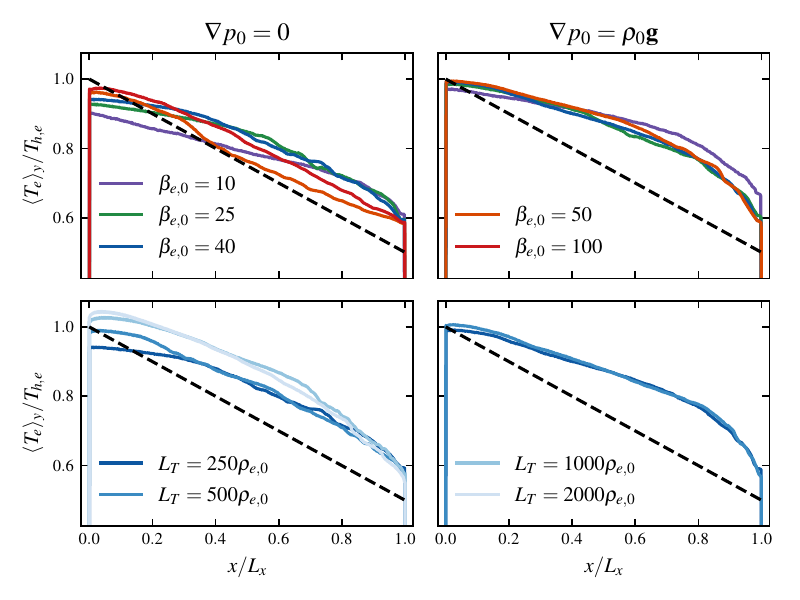}
    \caption{Average temperature as a function of distance along the temperature gradient. The initial temperature profile is shown by the black dotted line. In saturation, the $\grad p_0=0$ runs (save b10) support a temperature gradient close to the initial gradient near the centre of the simulation domain. The runs with a gravitationally supported pressure gradient, however, only support a temperature gradient that is ${\approx}60\%$ of its initial value.
    }
    \label{fig:T_yave}
\end{figure}

\begin{figure}
    \centering
    \includegraphics{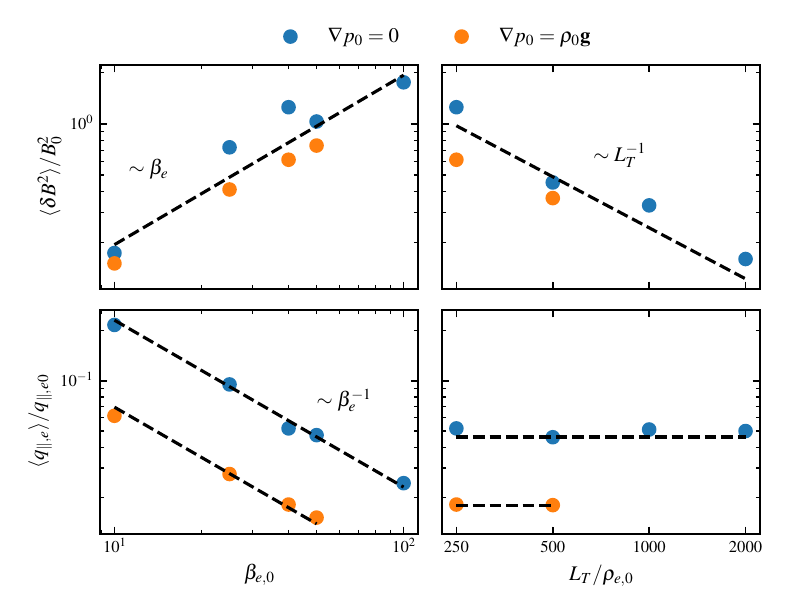}
    \caption{Normalized magnetic fluctuation amplitude (top) and heat flux (bottom) vs. $\betaeo$ (left) and $L_T\rhoeo$ (right) for the K18 setup (blue) and gravity setup (orange). The heat flux and fluctuation amplitude show good agreement with \eqref{eqn:q_v_beta} and \eqref{eqn:b_v_beta}, respectively.
    }
    \label{fig:qb_vs_beta}
\end{figure}

More quantitatively, in figure~\ref{fig:HWI_growth} we plot the box-averaged magnetic-field fluctuations and heat flux as functions of time, with panel ($a$) corresponding to a scan in $\betaeo$ at fixed $L_T=250\rhoeo$ and panel ($b$) corresponding to a scan in $L_T$ at fixed $\betaeo=40$. For the first ${\sim}100-500\wceo^{-1}$, depending on $L_T/\rhoeo$, the whistler instability is inactive. This is a consequence of our initial conditions: we initialize the domain with a temperature gradient, but with no heat flux. It therefore takes some fraction of the sound crossing time $T_c=L_x/\vtheo$ for the heat flux -- and the associated anisotropy in the distribution function -- to grow large enough for the whistlers to overcome cyclotron damping.

Once the heat flux is sufficiently strong, the instability enters an exponential phase, which is predicted by the whistler growth rate \eqref{eqn:HWI_growth_rate}. At $\kpar\rho_e= 1$,
\begin{equation*}
    \frac{\gammaw}{\Omega_e} = \frac{C}{\beta_e}\bigg(\frac{\vthe\beta_e}{2\nu L_T}-1\bigg) ,
\end{equation*}
where $C=\sqrt{\upi}\rme^{-1}$. The time rate of change of the whistler fluctuation energy at this scale is therefore
\begin{equation}\label{eqn:whistler_db_growth}
    \pD{t}{}\delta B^2=2\gammaw \delta B^2=\frac{2C}{\beta_e} \bigg(\frac{\beta_e \vthe}{2L_T \nu}-1\bigg)\Omega_e\,\delta B^2.
\end{equation}
In the analysis of \citet{Bott_2024}, $\nu$ is the scattering rate associated with a weakly collisional background and which results in the whistler-unstable steady-state distribution function \eqref{eqn:fe_model}. We argue that we, too, have a (very) weakly collisional background in the form of finite-particle-number noise, which scatters electrons at a rate $\nupic$. We additionally have effective scattering from whistler fluctuations at a rate $\nueff$ [see \eqref{eqn:nueff_diff}], so we take $\nu=\nupic + \nueff$. Exponential growth (ii) occurs in the limit $\nueff \ll \nupic\ll \beta_e\vthe/L_T$, in which the $-1$ term in \eqref{eqn:whistler_db_growth} can be ignored and the scattering rate is approximately independent of the magnetic field. This rapid growth is sustained only for a short time -- approximately $10-100\Omega_e^{-1}$, depending on $\betaeo$ and $L_T$ -- until the effective scattering rate by the magnetic-field fluctuations, $\nueff$, becomes comparable to $\nupic$ and the whistler fluctuations begin to feed back on themselves. In the limit where $\nu\simeq \nueff\sim \Omega_e(\delta B/B_0)^2$, the instability grows linearly in time following
\begin{equation}
    \pD{t}{}\delta B^2\sim \frac{C\vthe}{L_T} B_0^2 \qquad\Longrightarrow\qquad
    \frac{\delta B^2(t)}{B^2_0} - \frac{\delta B^2(t_{\rm sec})}{B^2_0} \sim \frac{C\vthe}{L_T} (t-t_{\rm sec}),
\end{equation}
where $t_{\rm sec}$ marks the start of the secular phase.
For all runs at $L_T/\rhoeo=250$, $t_{\rm sec}\sim100\wceo^{-1}$; this value increases up to $\sim500\wceo^{-1}$ at $L_T/\rhoeo=2000$. The secular phase lasts~${\sim}10^{3-4}\wceo^{-1}$, depending on the simulation,
until the magnetic energy saturates at a value that is dependent upon $\beta_e$ and $L_T$. This phase (iii) was not reported in earlier numerical work \citepalias[][]{Komarov2018,Roberg-Clark2018a}.

\subsection{Saturated heat-flux scaling with $\beta_e$ and $L_T$}\label{sec:sat_scaling}

Once the fields are saturated, they are able to maintain a temperature gradient throughout the box, which we show in figure \ref{fig:T_yave}. For the $\gradpz$ setup, the saturated temperature gradient in the centre of the domain is close to the initialized temperature gradient. However, for the $\gradpg$ setup, the gradient is ${\sim}60\%$ less, likely due to the stronger equilibrium heat flux. In figure \ref{fig:qb_vs_beta}, we plot the saturated box-averaged heat flux (bottom panels) and magnetic-field perturbation energy (top panels) versus $\betaeo$ (left panels) and $L_T/\rhoeo$ (right panels). These results are consistent with the scalings
\begin{subequations}\label{eqn:qb_v_beta}
\begin{align}
    \frac{\langle q_{\parallel,e}\rangle}{q_{\parallel,e0}} &\sim \frac{1}{\beta_e},\label{eqn:q_v_beta}\\
    \frac{\langle\delta B^2\rangle}{B_0^2} &\sim\frac{\beta_e}{L_T/\rho_e}\label{eqn:b_v_beta},
\end{align}
\end{subequations}
and are independent of the equilibrium setup. Our result (\ref{eqn:qb_v_beta}) not only confirms the work of both \citetalias{Komarov2018} and \citetalias{Roberg-Clark2018a}, but extends the validity of the scaling to larger scale separations and to an equilibrium set by gravity. The empirical scalings \eqref{eqn:q_v_beta} and \eqref{eqn:b_v_beta}, along with the analysis in section~\ref{sec:HWI_sat}, strongly suggest that the relevant dimensionless parameter in the saturation of the HWI is $\betaeo\rhoeo/L_T$. For the remainder of the paper, therefore, we normalize the magnetic energy $\avg{\delta B^2}/B_0^2$ and all effective collision frequencies using $(\betaeo\rhoeo/L_T)\Omega_{e0}$.

\subsection{Magnetic-field spectra and spectrograms}\label{sec:spect}

\begin{figure}
    \centering
    \includegraphics[]{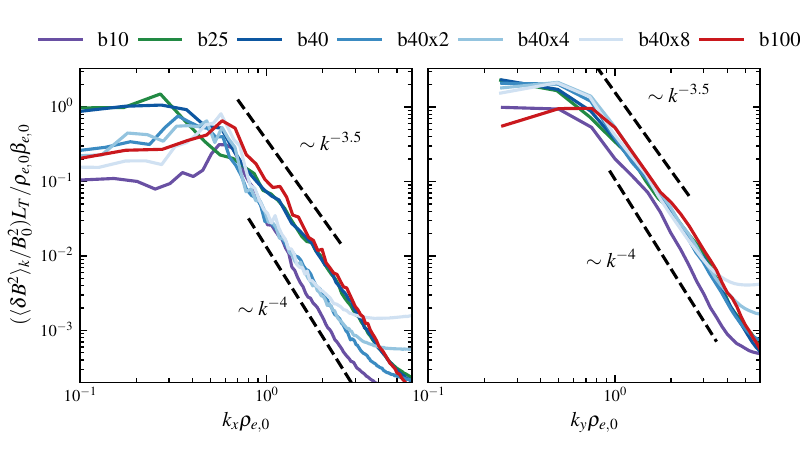}
    \caption{One-dimensional spectra of $\delta B^2$ for all runs versus $k_x$ (left column) and $k_y$ (right column), computed by integrating along the free dimension. For either direction, the spectrum is consistent with a power law index of $-4$ for $k\rho_e\ge 1$.}
    \label{fig:power_spectra}
\end{figure}

We compute the time-averaged magnetic energy spectrum for each run over the final $10^3\wceo$. Results are plotted in figure \ref{fig:power_spectra}. The spectra are qualitatively similar between runs and are characterized by a sharp break around $k\rhoeo\sim0.6-0.7$. At smaller $k$, the spectral index is approximately zero, while the index (or indices) at larger $k$ are more difficult to interpret due to the noise in the spectrum. For $k\rhoeo>1$, the spectra can be characterized by an index of $-4$, which was also reported in \citet{Zhdankin2017} for PIC simulations of driven turbulence in magnetized, collisionless, relativistic pair plasma. The two spectra that differ quantitatively are runs b10 and b40x8, which have a pronounced peak at $k\rhoeo\sim 0.6$.

\begin{figure}
    \centering
    \includegraphics{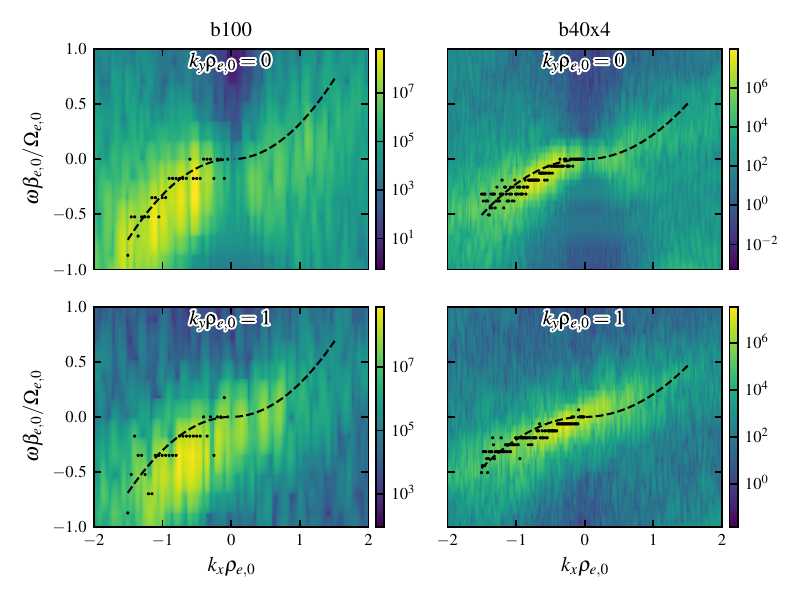}
    \caption{Spectrograms of $B_y+{\rm i}B_z$ for runs b100 (left column) and b40x4 (right column) for $k_y\rho_e=0$ (top row) and $k_y\rho_e=1$ (bottom row). Energy is concentrated in right-hand polarization ($\omega<0$) for parallel modes, but energy does go into left-hand polarization for oblique modes, as expected with whistler waves. Black dots represent the frequency $\omega$ with the highest Fourier amplitude at each $\kpar$; black dashed lines are the best fits to these points. The best fit lines differ from the cold plasma dispersion by a factor of $0.32-0.17$, with an average value of $0.23$.}
    \label{fig:spectrogram}
\end{figure}

We determine the whistler dispersion relation in our simulations by calculating the spectrogram of the combination $B_y+\imag B_z$, which reveals information about the polarization of the waves: positive wave frequencies $(\omega > 0)$ correspond to left-hand-polarized waves, while $\omega < 0$ corresponds to right-hand-polarized waves. We show results for runs b100 and b40x4 in figure~\ref{fig:spectrogram}; spectrograms from the other runs were qualitatively similar. Data was taken over the final ${\sim}4000\Omega^{-1}_e$ of each run to obtain sufficient resolution at low frequencies. We also applied Gaussian windows in $x$ and $t$ with standard deviations $1/4$ and $1/6$ of the total window widths, respectively, to ensure non-periodic edge effects were minimized. The black dashed lines in figure~\ref{fig:spectrogram} are best fits for the wave dispersion relation multiplied by a free parameter $A$, i.e.
\begin{equation}\label{eqn:cpwdr_fit}
    \frac{\omega}{\Omega_e}\simeq A\frac{(\kpar \rho_e)(k\rho_e)}{\beta_e},
\end{equation}
which implies a phase speed
\begin{equation}\label{eqn:cpwdr_fit_vw}
    \vw \simeq A\frac{\vthe}{\beta_e}\kpar\rho_e.
\end{equation}
The best fit value of $A$, expressed as a multiplicative constant of $\betaeo\vw/\vtheo$ with $\kpar\rhoeo=1$, is reported for each $\gradpz$ run in table~\ref{tab:runs}. In general, $A$ increases with $\betaeo$: from ${\simeq}0.20$ at $\betaeo=10$ to ${\simeq}0.28$ at $\betaeo=100$. Our analysis indicates that the whistler phase speed is slower than that of the cold-plasma dispersion prediction at $\kpar\rho_e=1$ (cf.~equation~\eqref{eqn:cpwdr}). For the remainder of the work, we treat $\vw$ as a constant, given by the $\kpar\rhoeo=1$ values reported in table~\ref{tab:runs}. Given the complexity of the analysis in the following sections, we view this as an expedient assumption and leave the effects of a more general treatment to future work.

\subsection{Effect of ion-Landau and ion-cyclotron resonances}\label{sec:q_vs_mratio}

\begin{figure}
    \centering
    \includegraphics{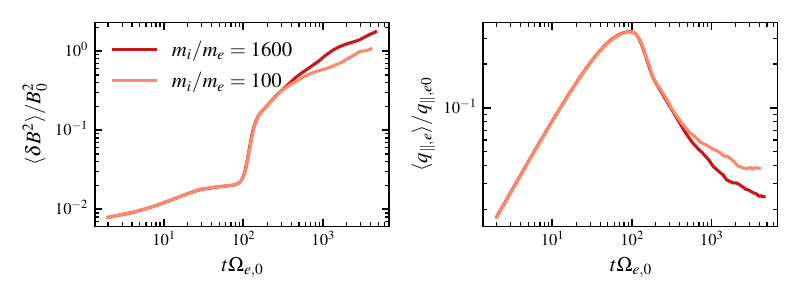}
    \caption{Box-averaged magnetic perturbation (left panel) and box-averaged heat flux (right panel) vs. time for two $\betaeo=100$, $L_T=250\rhoeo$ runs with $m_i/m_e=1600$ and $m_i/m_e=100$. When $\beta_e\sim m_i/m_e$, ions are in cyclotron resonance with the whistler waves. This results in both a reduction of $\delta B^2/B_0^2$ and an increase in the saturated heat flux by a factor of 2.}
    \label{fig:qb_vs_mratio}
\end{figure}

Runs b40 and b40g, with $\betaeo=40$ and $m_i/m_e=1600$, allow us to assess any impact from the ion-Landau resonance. Run b100m, with $\betaeo=100$ and $m_i/m_e=100$, tests the effect of the ion-cyclotron resonance. As explained at the end of \S\ref{sec:HWI_sat}, ions may be Landau-resonant with whistler waves when the phase velocity of the former is of order the thermal speed of the latter, i.e., when $\beta_e\sim (m_i/m_e)^{1/2}$. The ions can also be cyclotron-resonant with whistlers if the wave frequency equals the ion cyclotron frequency -- this occurs when $\beta_e\sim m_i/m_e$. Re-examining the left column of figure~\ref{fig:qb_vs_beta}, the runs at $\betaeo=40$ do not stray appreciably from the $q\sim\beta_e^{-1}$ scaling and so we conclude that effects due to the ion-Landau resonance are negligible. \citet{Komarov2018} report a $20\%$ increase in saturated heat flux and $15\%$ decrease in magnetic field fluctuation energy. Their conclusion comes from comparing a $\beta_e=15$, $m_i/m_e=225$ run to one at the same $\beta_e$, but at $m_i/m_e=\infty$. In our tests we saw changes in saturation levels with immobile ions at $\beta=10$, so it is possible that their conclusion is affected by comparison to a run with immobile ions instead of a different mass ratio, resulting in a deviation higher than ours.

In contrast to \citetalias[][]{Komarov2018}, we find that the ion-cyclotron resonance affects the saturated state of the HWI significantly. In figure~\ref{fig:qb_vs_mratio}, we plot the box-averaged magnetic-field fluctuation amplitude and heat flux as functions of time for two $\betaeo=100$ runs, one with $m_i/m_e=1600$ and the other with $m_i/m_e=100$. While the heat flux accumulation and exponential stages are nearly identical between runs, the cyclotron-resonant run saturates much sooner than does the run with mass ratio $1600$, with a $25\%$ smaller fluctuation amplitude and $57\%$ higher heat flux. Our result that the ion-cyclotron resonance significantly alters the saturated heat flux relative to the Landau resonance is consistent with our simple explanation of the instability in \S\ref{sec:background}. Evidently, cyclotron-resonant ions can strongly damp the oblique whistlers that are necessary for scattering heat-flux-carrying electrons. We also conjecture that Landau-resonant ions preferentially damp the parallel whistlers and have a diminishing effect on the heat flux. An in-depth analysis of the interaction between whistlers and ions, however, is outside the scope of this work.

\section{Results, II.~An effective collision operator for heat-flux-driven whistler turbulence}\label{sec:results2}

In this section, we present detailed methods and results for three different means of obtaining a collision operator from our simulations. Section~\ref{sec:CE} is devoted to a method based upon the Chapman--Enskog expansion. After stipulating a model collision operator that captures pitch-angle scattering in the frame of the whistlers, we derive the implied `Chapman--Enskog' scattering frequency $\nuce$ in section~\ref{sec:CE_der}, characterize the equilibrium electron distribution function in section~\ref{sec:CE_f}, and present our numerical findings in section~\ref{sec:CE_res}. Section~\ref{sec:QL} focuses on the quasi-linear operator; we define the operator in section~\ref{sec:QL_def} and give our numerical results in section~\ref{sec:QL_res}. Finally, we detail the Fokker--Planck method in section~\ref{sec:FP}. We define the resulting operator in section~\ref{sec:FP_def} and discuss the relevant timescales, particle statistics, and limitations in section~\ref{sec:FP_times}. In sections~\ref{sec:FP_v} and \ref{sec:FP_xi} we present our numerical results for the velocity-space Fokker--Planck operator in $(v,\xi$) coordinates.

For all of these methods, we leverage a key finding from section~\ref{sec:results1}: that the energy of magnetic-field fluctuations in the saturated state of the HWI and the effective collisionality implied by the steady-state heat flux both scale proportionally with $\beta_e (\rho_e/L_T)$. We organize the various features of our model collision operators according to this dimensionless free parameter.

\subsection{Chapman--Enskog pitch-angle scattering operator}\label{sec:CE}

Our first method to obtain a collision operator for whistler turbulence relies on a Chapman--Enskog expansion of the electron kinetic equation. In the usual
expansion, one solves for the collisional transport resulting from the free-energy gradients (usually velocity or temperature) in a fluid with a known scattering operator. We instead solve for the collision operator that self-consistently describes the transport and fluid gradients measured in our simulations. Using this method, one assumes the mathematical form of the collision operator and solves for the implied diffusion coefficient as a function of phase-space variables. This is simultaneously a strength and a limitation, as the diffusion coefficient is designed to explain all the observed transport given the model.

\subsubsection{Derivation}\label{sec:CE_der}

For the remainder of section~\ref{sec:CE}, we assume that whistlers pitch-angle scatter electrons in a frame moving at the whistler phase speed, $\vw$. This assumption is consistent with the quasi-linear operator for slow electromagnetic modes \citep[see \S\ref{sec:QL} of][]{Kennel&Engelmann1966} and is the same as chosen by \citet{Drake2021}. What follows is a brief outline of the full calculation; see Appendix~\ref{apx:CEO} for more details. The model collision operator is given by
\begin{equation}\label{eqn:pa_operator}
	\CCE[f]=\pD{\primed{\xi}}{}\biggl[\frac{1-\xi^{\prime 2}}{2}\nuce(\primed{v},\primed{\xi})\pD{\primed{\xi}}{f}\biggr],
\end{equation}
where the prime denotes quantities evaluated in the reference frame of the whistlers; note that $C_{\rm CE}[f]$ is independent of the gyro-angle, $\phi$. We take $\vw$, $\beta_e^{-1}$, the electron gyroradius, and the effective electron mean free path to all be small quantities of the same order $\epsilon$, {\em viz.}
\begin{equation*}
	\epsilon\sim\frac{\vw}{\vthe} \sim \frac{1}{\beta_e}\sim
    \frac{\rho_e}{L_T}\sim\frac{\mfpe}{L_T}\ll 1.
\end{equation*}
Expanding $\CCE[f]$ in $\epsilon$, we find that the electron kinetic equation evaluated to leading order in $\epsilon$ is
\begin{equation}\label{eqn:CE_ep0}
	0 = \Omega_e\pD{\phi}{f_{e0}}+
	\pD{\xi}{}\biggl[\frac{1-\xi^{2}}{2}\nuce(w,\xi)\pD{\xi}{f_{e0}}\biggr],
\end{equation}
where $\bb{w}=\bb{v}-\bb{u}_e$ is the velocity of the electrons in the frame of any bulk flow $\bb{u}_e$. This constraint is satisfied trivially by an isotropic $f_{e0}=f_{e0}(v)$. To next order, after gyroaveraging, we obtain the correction equation for parallel transport,
\begin{equation}\label{eqn:CE_correction}
        \wpar\nabla_{\parallel} f_{e0}+\frac{\nabla_{\parallel}		 p_e}{m_en_e} \frac{w_\parallel}{w}\D{w}{f_{e0}}
		=\pD{\xi}{}\biggl[\frac{1-\xi^2}{2}\nuce(w,\xi)\bigg(\pD{\xi}{\avg{f_{e1}}_{\phi}} + \vw\D{w}{f_{e0}}\bigg) \biggr].
\end{equation}
The term proportional to $\vw$ arises as a consequence of the ordering of the whistler phase speed and ensures that the scattering occurs in the wave frame. The correction equation \eqref{eqn:CE_correction} can be solved for $\nuce$ by using $\wpar=w\xi$ and integrating both sides with respect to $\xi$. Choosing the integration constant to keep ${\p \avg{f_{e1}}_\phi/\p \xi}$ finite, we find that
\begin{equation}\label{eqn:CEnuw_CEsec}
	\nuce(w,\xi)= \left.
	-\bigg(w\nabla_\parallel f_{e0}+\frac{\nabla_{\parallel}p_e}{m_en_e}\D{w}{f_{e0}}\bigg)
	\middle/ \bigg(\pD{\xi}{\avg{f_{e1}}_{\phi}} + v_w\D{w}{f_{e0}}\bigg) \right. .
\end{equation}

\subsubsection{Calculating the distribution function}\label{sec:CE_f}

In order to evaluate \eqref{eqn:CEnuw_CEsec}, we require expressions for $f_{e0}$ and $\avg{f_{e1}}_\phi$. We calculate these quantities from our simulations by binning individual particle data on a finite phase-space grid. To do so, we break up the central $60\%$ of the simulation domain into $4\%$ intervals and integrate over the $y$-direction; this ensures that we sample a distribution function with nearly uniform temperature in each interval. In velocity space we bin using a $60\times 60$ grid in $(v,\xi)$ coordinates, from $v=0$ to $v=3\vthe$ (where $v_{{\rm th}e}$ is the electron thermal speed local to that interval) and from $\xi=-1$ to $\xi=1$. Because the magnitude of the fluid velocity $\bb{u}_e$ in our simulations is at most a few percent of the whistler phase speed $\vw$, we do not distinguish between $v$ and $w$ in what follows; in real astrophysical contexts with bulk flows, $\bb{u}_e$ may not be small compared to $\vw$. A consequence of our choice to ignore the $\phi$ coordinate is that our measured distributions only contain parallel transport, naturally precluding any transport perpendicular to the background field.

The algorithm is as follows. For every electron at a certain time step, we calculate the bin indices that correspond to its location in velocity space coordinates ($v,\xi$) and add $1$ to that bin. After normalizing, this procedure yields $f_e(v,\xi)$ for each $4\%$ interval of the simulation domain. We then determine $f_{e0}$ by requiring that all fluid quantities ($n_e$, $\bb{u}_e$, $T_e$) are contained in $f_{e0}$, so that the transport is determined solely by $f_{e1}$ \citep{Krommes2018}, {\em viz.}
\begin{equation}\label{eqn:f0_constraints}
    \big(n_e,\bb{u}_e,p_e\big)\equiv\int \rmd^3\bb{v}\,\bigg(1,\frac{\bb{v}}{n_e},\frac{m_e}{3}w^2\bigg)f_e=\int \rmd^3\bb{v}\,\bigg(1,\frac{\bb{v}}{n_e},\frac{m_e}{3}w^2\bigg)f_{e0},
\end{equation}
where $p_e$ is the scalar electron pressure associated with $T_e$. As a consequence of this procedure, $f_{e0}$ is naturally isotropic, as required by the Chapman--Enskog expansion at zeroth order in $\epsilon$ (equation~\eqref{eqn:CE_ep0}). Finally, we calculate the first-order deviation in $f_e$ from the definition
\begin{equation}\label{eqn:f1_def}
	f_{e1}=f_e - f_{e0} ,
\end{equation}
so that
\begin{equation}\label{eqn:f1_constraints}
    \int \rmd^3\bb{v}\, (1,\bb{v},v^2)f_{e1}=0 ,
\end{equation}
i.e., $f_{e1}$ is fully kinetic.

The only subtlety in the above procedure is the choice of $f_{e0}$, since \eqref{eqn:CE_ep0} demands only that $f_{e0}$ is isotropic. In what follows, we choose $f_{e0}$ to be an isotropic Maxwellian: $f_{e0}(v,\xi)=\fMe(v)$. Alternatively, we could have chosen $f_{e0}=\avg{f_e}_{\phi,\xi}$, which puts fewer restrictions on $f_{e0}$. Thankfully, the exact choice of $f_{e0}$ does not substantially change the following analysis or the resulting computed collision frequency. To determine $\nu_{\rm CE}(v,\xi)$, we first normalize each $4\%$ domain segment to a single $\vthe$ and then average both $f_{e0}$ and $f_{e1}$ across all segments. Given $f_{e0}(v)$ and $\nabla_\parallel\ln T_e$ from each simulation, we then need only to determine $\vw$ and $\partial\avg{f_{e1}}_{\phi}/\partial\xi$. The former is straightforwardly computed for each simulation using the method explained in \S\ref{sec:spect}. Unfortunately, the latter presents some difficulties, which are addressed in the next subsection.

\subsubsection{Numerical results}\label{sec:CE_res}

\begin{figure}
    \centering
    \includegraphics{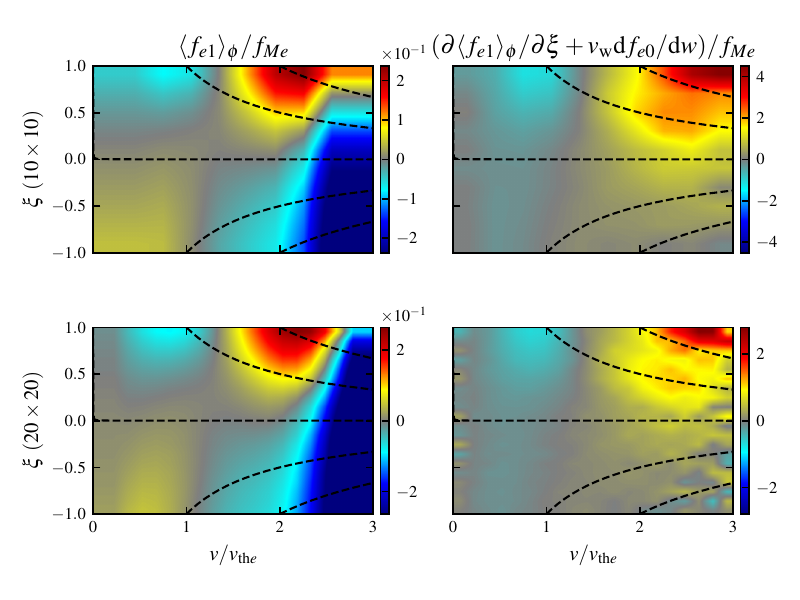}
    \caption{Two-dimensional plots of $\avg{f_{e1}}_\phi/f_{e0}$ (left column) and $(\partial \avg{f_{e1}}/\partial \xi + \vw \rmd f_{e0}/\rmd w)/f_{e0}$ (right column) for run b40x4 at grid resolution $10\times 10$ (top row) and $20\times 20$ (bottom row). The noise and negative values for $v/\vthe\gtrsim 2$ only get worse at increasing resolution and imply a $\nuce(v,\xi)$ that is noisy and contains unphysical negative values. Dashed lines correspond to constant $\vpar$. }
    \label{fig:CE_denom}
\end{figure}

In principle, the CE method can resolve the full velocity-space dependence of the collision frequency \eqref{eqn:CEnuw_CEsec}, at least within the model operator. However, in practice we find that spurious zeros in ${\partial \avg{f_{e1}}_\phi/\partial\xi}$ caused by statistical noise in the binned distribution function as well as shallow gradients in $f_{e1}$ result in a noisy calculation of $\nuce(v,\xi)$, which can have unphysical (i.e., negative) values in certain regions of phase space. See figure~\ref{fig:CE_denom} for plots of $\avg{f_{e1}}_\phi/f_{e0}$ and $(\partial \avg{f_{e1}}_\phi/\partial \xi + \vw \rmd f_{e0}/\rmd w)/f_{e0}$ for run b40x4 at reduced resolutions of $10\times 10$ and $20\times 20$, which exhibit noise and negative values for $v/\vthe\gtrsim 2$. Increasing the grid resolution only increases and worsens the artifacts already present in figure~\ref{fig:CE_denom}. It is possible that even lower resolutions might produce favorable numbers; however, much fewer than $10\times 10$ grid cells runs the risk of not resolving the structure of $\nuce$ in both $v$ and $\xi$. To circumvent these issues, we average $\partial f_{e1}/\partial\xi$ over pitch angle to find a $v$-dependent collision frequency
\begin{equation}\label{eqn:CEnu_xiavg}
    \nuce(v)= \left.
	-\bigg(v\nabla_\parallel f_{e0}+\frac{\nabla_{\parallel}p_e}{m_en_e}\D{v}{f_{e0}}\bigg)
	\middle/ \bigg(\bigg\langle\pD{\xi}{f_{e1}}\bigg\rangle_{\!\xi,\phi} + v_w\D{v}{f_{e0}}\bigg) \right. .
\end{equation}
A particularly unfortunate consequence of having to average over $\xi$ is that any resonant structure in $\nu_{\rm CE}(v,\xi)$ (e.g., due to Landau-cyclotron resonances occurring along lines of constant $v_\parallel$) is lost. A distinct benefit, however, is that we are able to  resolve precisely the velocity dependence of the scattering frequency.

\begin{figure}
    \centering
    \includegraphics{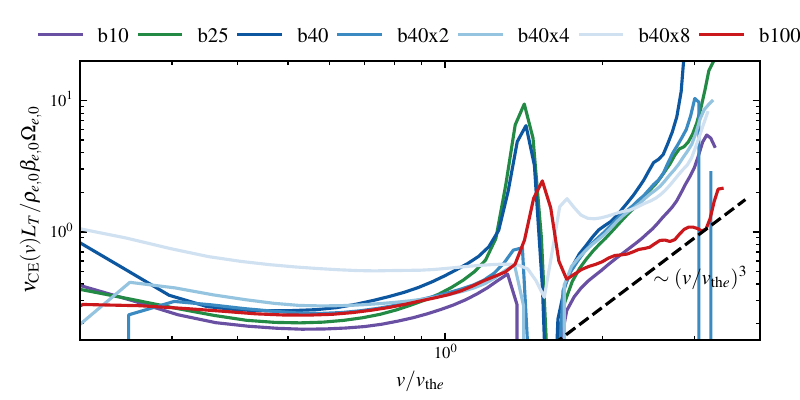}
    \caption{Chapman--Enskog whistler scattering frequency $\nuce$ \eqref{eqn:CEnu_xiavg} as a function of speed $v/\vthe$. A Gaussian filter with standard deviation of one cell ($\vthe/20$ and $1/30$ in $v$ and $\xi$, respectively) is applied for smoothing. The distribution function used to calculate $\nuce$ was taken at the end of each of the runs (see table~\ref{tab:runs}). To guide the eye, we include a black dashed line ${\propto}(v/\vthe)^3$.
    }
    \label{fig:CE_lineouts}
\end{figure}

In figure \ref{fig:CE_lineouts}, we plot \eqref{eqn:CEnu_xiavg} as a function of speed $v$ for each of the $\gradpz$ runs. The discontinuities and spikes in the range $1\lesssim v/\vthe \lesssim 2$ are the result of zeros in the denominator not exactly canceling with the zero in the numerator at $v/\vthe=\sqrt{5/2}\simeq 1.6$. Within this model, sub-thermal electrons are scattered at a rate that is weakly dependent upon speed, with only a slight increase as $v$ becomes very sub-thermal. Super-thermal electrons, on the other hand, have a pitch-angle-scattering rate that increases steeply with $v$, approximately as $(v/\vth)^{3}$ (black dashed line) though with some significant spread. (Note that this is rather different than in Coulomb-collisional plasmas, for which the scattering rate {\em decreases} with $v$ at super-thermal speeds.) While the model operator is unable to resolve the pitch-angle dependence of the scattering operator, it does serve as an important data point for comparison with the other methods used in sections~\ref{sec:QL} and \ref{sec:FP}.

\subsection{Quasi-linear operator}\label{sec:QL}
Our next model collision operator for the saturated HWI is a quasi-linear operator that depends explicitly on the energy spectrum of the electromagnetic fluctuations. This dependence arises from averaging the electron kinetic equation over a few wave periods in time and wavelengths in space and separating the electron distribution function into its mean and fluctuating parts; one then finds that the time rate of change of the mean distribution function is due to the mean of the interaction of the perturbed fields with the perturbed distribution function. If one further assumes that the perturbed distribution function is the linear response to the fields (hence the qualifier `quasi-linear'), one obtains an operator that depends only on the fields resonantly interacting with particles. A benefit of this method is a rich insight into the physics of the saturation mechanism. Its main disadvantage is that the quasi-linear operator contains a number of strict assumptions that are often suspect. For our results in particular, wave-particle resonances are likely to be broadened.

\subsubsection{Definition}\label{sec:QL_def}
We adopt the electromagnetic quasi-linear operator given in \citet{Stix1992} (also given by \eqref{eqn:fpstix}--\eqref{eqn:Lstix}) under two simplifying assumptions: (i) that the turbulence is only two-dimensional (as in our simulations) and (ii) that the phase speed of the scattering whistlers relative to the electron thermal velocity is small and order $\epsilon$, i.e.
\begin{equation}
    \epsilon \sim \frac{\vw}{\vthe} \sim \frac{1}{\beta_e}\sim \frac{\rho_e}{L}\sim\frac{\mfpe}{L}\ll 1.
\end{equation}
The first assumption allows us to write the original quasi-linear diffusion operator \eqref{eqn:fpstix}, which is expressed in terms of the electric field, in terms of the magnetic and parallel-electric fields (see \eqref{eqn:Ek}). The second assumption allows us to expand the electron kinetic equation in $\epsilon\ll 1$ (cf.~\S\ref{sec:CE_der}). At zeroth order, we find that the quasi-linear operator is simply a pitch-angle scattering operator associated with the magnetic-field fluctuations. At this order, there is no Landau damping from the parallel electric field, but there is transit-time damping as the guiding centres of Landau-resonant particles surf the mirror force associated with low-frequency fluctuations in the magnetic-field strength \citep{Stix1992,Barnes1966}. At first order in $\epsilon$, we find a correction equation in which the distortion in the distribution function caused by the parallel temperature gradient is balanced by two terms: pitch-angle scattering of $f_{e1}$ and a term proportional to $\vw$. Except for the nature of the collision frequency, this correction equation is identical to that associated with a collision operator that pitch-angle scatters in a frame co-moving with whistler waves at $\vw$ [cf. \eqref{eqn:CE_correction}]. We refer the reader to appendix~\ref{apx:QL} for a more detailed derivation.

Keeping only relevant terms, the simplified operator is
\begin{equation}\label{eqn:QL_op}
    \CQL[f]=\pD{\xi}{} \left[\frac{1-\xi^2}{2}\nuql(v,\xi) \left(\pD{\xi}{f} + \vw\pD{v}{f}\right)\right] ,
\end{equation}
in which the quasi-linear collision frequency is given by
\begin{subequations}\label{eqn:nu_QLall}
\begin{equation}\label{eqn:nu_QL}
    \nuql(v,\xi)\simeq
    2\upi \frac{\Omega_e}{N_xN_y}
    \sum_n
    \sum_{k_x}\sum_{k_y}\delta\bigg(\frac{\omega(\kpar\rho_e)}{\wce}-\kpar\rho_e\frac{\vpar}{\vthe}+n
    \bigg) \biggl|\frac{\Psi_{n,k}}{B_0}\biggr|^2 .
\end{equation}
The magnetic field enters this expression via
\begin{equation}\label{eqn:psi_nk}
    \Psi_{n,k} \simeq     B_k^-    \besselj{n-1}(z)+    B_k^+    \besselj{n+1}(z) ,
\end{equation}
with the argument of the Bessel functions $\besselj{n}$ being $z=k_y\vperp /\Omega_e$, and the complex quantities
\begin{equation}\label{eqn:B_pol}
    B_k^\pm =\frac{B_{kz}\pm \imag B_{ky}}{2}
\end{equation}
\end{subequations}
are the Fourier-transformed magnetic fields corresponding to right-handed ($+$) and left-handed ($-$) polarizations. For parallel waves, $k_y=k_z=0$ and the cyclotron resonances correspond to purely right-handed modes for $n=-1$ and left-handed modes for $n=1$. For $n=0$, $\Psi_{0,k}=\imag B_{ky}\besselj{1}(z)=
-\imag (k_x/k_y)B_{kx} \besselj{1}(z)$; at long wavelengths such that $z\ll 1$ and $\besselj{1}(z)\approx z/2$, we have that $2\Omega_e|\Psi_{0,k}/B_0|^2 \approx (v^2_\perp/2\Omega_e) k_x^2 |B_{kx}/B_0|^2$, which corresponds to transit-time damping.

\subsubsection{Resonance condition}\label{sec:resonance_condition}
The argument of the delta function in \eqref{eqn:nu_QL}, commonly called the resonance condition, defines the parallel wavenumber that resonantly interacts with an electron at a given parallel velocity. Using the whistler dispersion relation \eqref{eqn:cpwdr_fit}, the resonance condition is
\begin{equation}\label{eqn:specific_res_condition}
    -\sign(n)\frac{0.23}{\beta_e}(\kpar\rho_e)(k\rho_e)-\kpar\rho_e\frac{\vpar}{\vthe}+n=0.
\end{equation}
The factor of $\sign(n)$ in front of the whistler dispersion relation ensures that \eqref{eqn:specific_res_condition} has the correct solution for electrons with $\vpar/\vthe>0$ in resonance with oblique whistler waves; we take $\sign(0)=-1$ . The general solution to \eqref{eqn:specific_res_condition} for quasi-parallel ($\kperp/\kpar\ll 1$) waves is
\begin{equation}\label{eqn:res_general_solution}
    \kpar\rho_e=-\sign(n)\frac{1}{2}\frac{\beta_e}{0.23}\frac{\vpar}{\vthe}\Biggl[1\pm \sqrt{1+4\bigg(\frac{\beta_e}{0.23}\frac{\vpar}{\vthe}\bigg)^{-2}\bigg(\frac{\beta_e}{0.23}|n|+(\kperp\rho_e)^2\bigg)}\Biggr].
\end{equation}
For $n=0$, exactly parallel whistlers have their resonance at
\begin{equation}\label{eqn:kpar_res_0}
    \kpar\rho_e=\frac{\beta_e}{0.23}\frac{\vpar}{\vthe}\qquad (n=0),
\end{equation}
excluding $\kpar\rho_e=0$. For $n\ne 0$, we expand \eqref{eqn:res_general_solution} in $\beta_e\gg 1$ to find the resonance conditions
\begin{equation}\label{eqn:kpar_res_n}
    \kpar\rho_e\approx n\bigg(\frac{\vpar}{\vthe}\bigg)^{-1}\qquad (n\ne 0) ,
\end{equation}
as well as
\begin{equation}\label{eqn:kpar_res_n_unphysical}
    \kpar\rho_e=-\sign(n)\frac{\beta_e}{0.23}\frac{\vpar}{\vthe} + n\frac{\vthe}{\vpar}\qquad (n\ne 0).
\end{equation}
We omit the solution \eqref{eqn:kpar_res_n_unphysical} for the reason that, in the $\beta_e\gg 1$ limit, it yields a negative $\kpar$ for $|\vpar|/\vthe\sim 1$ electrons and a proper choice of $n$; such wavevectors are stable to the HWI. Keeping only solutions \eqref{eqn:kpar_res_0} and \eqref{eqn:kpar_res_n} and performing some simple manipulations of the delta function, we have, for purely parallel waves,
\begin{equation}\label{eqn:resonant_delta}
\delta\bigg(\frac{\omega(\kpar\rho_e)}{\Omega_e}-\kpar\rho_e\frac{\vpar}{\vthe}+n\bigg)=\bigg|\frac{\vthe}{\vpar}\bigg|
\begin{cases}
\delta\bigg(\kpar\rho_e-\dfrac{\beta_e}{.23}\dfrac{\vpar}{\vthe}\bigg),& n=0\\
\delta\bigg(\kpar\rho_e-n\dfrac{\vthe}{\vpar}\bigg),& n\ne 0
\end{cases}.
\end{equation}
For the numerical results provided in \S\ref{sec:QL_res}, we solve the the full oblique resonance condition \eqref{eqn:specific_res_condition} numerically.

\subsubsection{Numerical results}\label{sec:QL_res}

\begin{figure}
    \centering
    \includegraphics{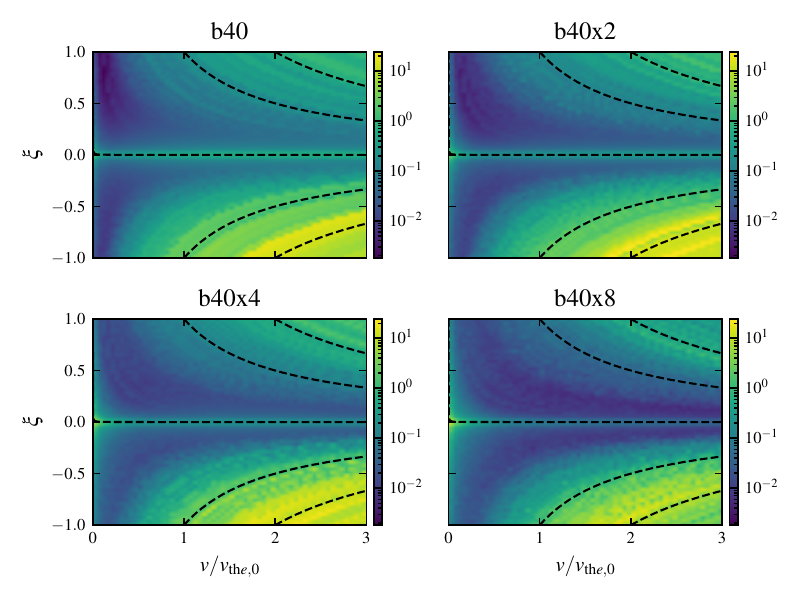}
    \caption{Two-dimensional plots of the quasi-linear pitch-angle collision frequency $\nuql(v,\xi)$ \eqref{eqn:nu_QLall} for all $\betaeo=40$ runs normalized to $\betaeo\rhoeo/L_T\wceo$. Dashed lines correspond to contours of constant $\vpar$.}
    \label{fig:QL_2d}
\end{figure}
\begin{figure}
    \centering
    \includegraphics{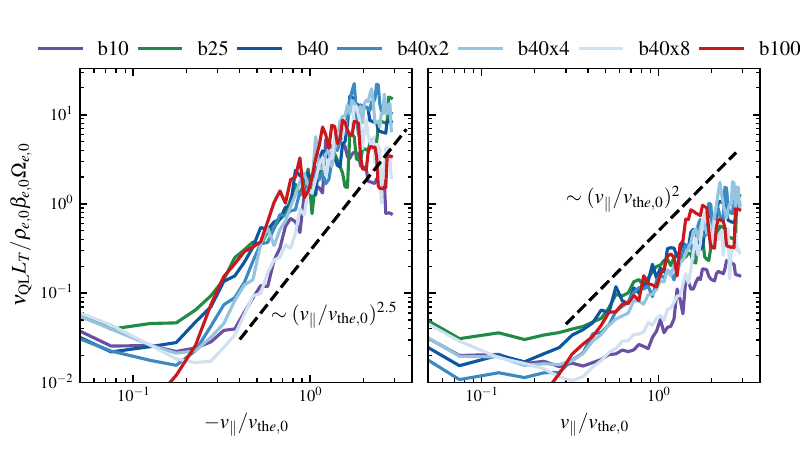}
    \caption{Quasi-linear collision frequency calculated for each of the runs as a function of $\vpar$. Electrons with $\xi<0$ are plotted on the left and those with $\xi>0$ are on the right; the former are scattered at a rate ${\sim}10$ times faster than the latter due to the relative amounts of energy in the right- and left-hand-polarized components of the wave spectrum. We include lines with power laws $(\vpar/\vtheo)^{2.5}$ and $(\vpar/\vtheo)^{2}$ on the left and right plots, respectively.}
    \label{fig:QL_vpar}
\end{figure}

In figure~\ref{fig:QL_2d}, we plot $\nuql(v,\xi)L_T/(\rhoeo\betaeo\wceo)$ from \eqref{eqn:nu_QLall} for simulations b40 and b40x4. One property of the operator that is readily apparent is the asymmetry about $\xi=0$. The scattering frequency for $\xi<0$ is an order of magnitude higher than that for $\xi>0$, and is consistent with the fraction of energy in the left-hand-polarized component of an elliptically polarized oblique whistler wave (cf.~\eqref{eqn:pol_frac}). As discussed in section~\ref{sec:HWI_sat}, whistler waves  appear to be left-hand polarized to any electron with $\vpar>\vw$, so no gyroresonance can occur. Resonance can only be facilitated by the left-hand component of oblique whistlers, which will appear right-handed in the frame of these electrons. This asymmetry is prominent and a significant refinement over the results in the previous section.

\begin{figure}
    \centering
    \includegraphics{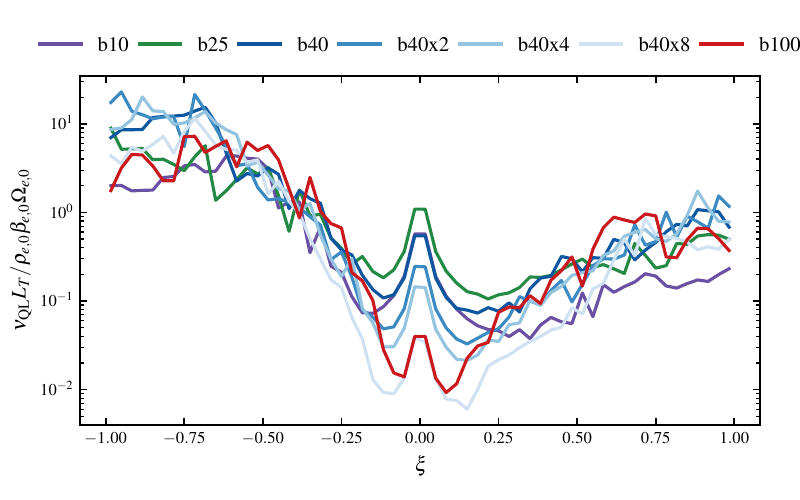}
    \includegraphics{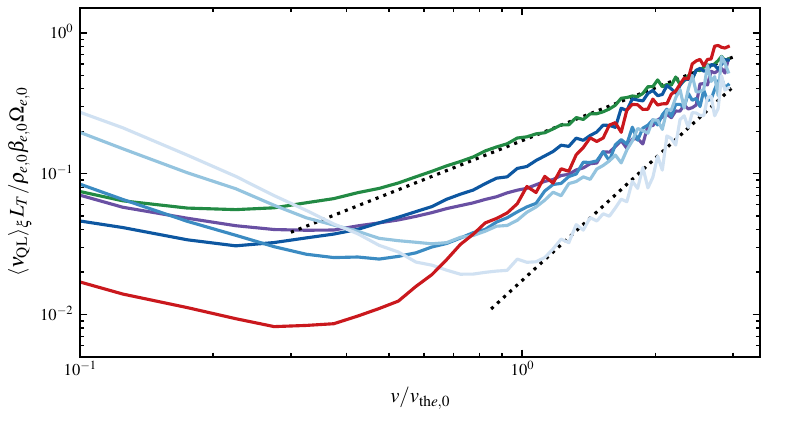}
    \caption{Quasi-linear scattering frequency averaged over the range $v/\vthe=[2,3]$ as a function of pitch-angle $\xi$ (top) and averaged over pitch angle as a function of $v/\vthe$ (bottom), the latter of which exhibits power laws in $v$ with indices from $1.2$ to $2.9$ (dotted lines).}
    \label{fig:QL_leftover}
\end{figure}

It is also clear from figure~\ref{fig:QL_2d} that the collision frequency is approximately constant along lines of constant $\vpar$ (dashed lines), a functional dependence that is expected given the form of \eqref{eqn:nu_QLall}. Suppose the magnetic energy scales with the parallel wavenumber as ${\delta B^2/B_0^2\propto \kpar^{-\alpha}}$; ignore any $k_y$-dependence for simplicity. This scaling implies a resonant scattering frequency $\nuql\propto|\vpar|^{\alpha-1}$. In figure~\ref{fig:QL_vpar}, we plot $\nuql(v,\xi)$ as a function of $\vpar$ for electrons with $\xi<0$ (left) and $\xi>0$ (right). Regardless of the sign of $\xi$, the scattering frequency grows as a power law in $|\vpar|$. For electrons with $\vpar < 0$, $\nuql \sim (\vpar/\vtheo)^{2.5}$, implying a field spectrum ${\sim}\kpar^{-3.5}$ that is compatible with the results shown in figure~\ref{fig:power_spectra}. There is a break in this power law at $|\vpar|/\vthe\sim 2$, which corresponds to the break in magnetic energy spectrum around $k_x\rhoeo\sim 0.6$ for the $n=\pm 1$ principal resonances. There is another break near $|\vpar|/\vthe\sim 0.3$ which is due to contributions from both the $(n=\pm 1)$ cyclotron resonances and the $(n=0)$ resonance. The contribution from the principal cyclotron resonances is the smallest of the two and corresponds to the transition of the spectrum from being dominated by the $k_x\rhoeo>1$ cascade to being dominated by noise with a spectral index of $0$. The dominant contribution is from the $n=0$ resonance and is the result of transit-time damping. The consequence is scattering at very small parallel velocities $\vpar/\vthe\sim\vw/\vthe\le 0.1$ at a rate generally much smaller than that at thermal velocities; however, the effect is most pronounced for positive parallel velocities.

In general we find rough agreement with $\nu_{\rm QL}/\Omega_e\propto \betaeo\rhoeo/L_T$, matching the predicted scaling \eqref{eqn:nueff_diff} and the observed scaling using our Chapman--Enskog method in section~\ref{sec:CE}. In figure~\ref{fig:QL_leftover}, we plot $\nuql$ as a function of pitch-angle (top) and pitch-angle-averaged as a function of $v/\vtheo$ (bottom). The latter shows a strong resemblance to the pitch-angle-averaged Chapman--Enskog results in figure~\ref{fig:CE_lineouts}: the scattering frequency shows a clear power law in $v/\vtheo$ for both superthermal and subthermal velocities. The increase in scattering for subthermal velocities is due to transit-time damping and is significantly more pronounced when plotting $\nuql$ as a function of $v$ versus as a function of $\vpar$ because the increased scattering at small parallel velocities applies for electrons with any $\vperp$ and thus any $v$. The scattering frequency for superthermal electrons grows as ${\sim}(v/\vtheo)^{1.2-2.9}$, which is compatible with the power laws observed in the Chapman--Enskog case. It is possible that the superthermal power-law indices have a $\beta_e$ or $L_T$ dependence; however, comparing figures \ref{fig:CE_lineouts} and \ref{fig:QL_leftover} -- and indeed looking ahead to figure~\ref{fig:FP_xi} -- it is not clear that there is a consistent dependence between all three of our methods, and it seems best not to over-interpret these indices.

\subsection{Fokker--Planck method}\label{sec:FP}

The most direct and detailed method for calculating the HWI collision operator is the Fokker--Planck method. This method is particularly powerful because it calculates drag and diffusion directly from the statistics of the particle motion. Moreover, this method requires no assumptions about the physics underlying the drag and diffusive processes. That being said, it does require statistical assumptions;  in particular, the Fokker--Planck method works well when there is a large separation between the auto-correlation time of the forcing and the collisional timescale. For a number of our runs, this scale separation is not especially large and results in some complications that are otherwise avoided by the other methods we have employed.

\subsubsection{Definition}\label{sec:FP_def}

The Fokker--Planck equation describes the time rate of change of a distribution function $f(t,\bb{v})$ undergoing drag, modeled by the vector $\bb{A}(\bb{v})$, and diffusion, modeled by the tensor $\msb{B}(\bb{v})$. These coefficients may be calculated by tracking the Lagrangian changes in velocity,
\begin{equation}\label{eqn:deltav}
    \Delta \bb{v}(t;\Delta t) = \bb{v}(t+\Delta t,\bb{x}(t+\Delta t)) - \bb{v}(t,\bb{x}(t)) ,
\end{equation}
of a large number of particles having the trajectories $\bb{x}(t)$ and computing various `jump moments' given by statistical averages over all of these particles within a time interval $t\in[\ts,\te]$, in which the system is assumed to be in steady-state. The specific interval for each run is given in table \ref{tab:runs}. The Fokker--Planck drag vector and diffusion tensor are given by
\begin{equation}\label{eqn:jumps}
    \bb{A}(\bb{v}) \doteq \lim_{\Delta t \rightarrow \textrm{`0'}} \frac{\langle\Delta\bb{v}\rangle}{\Delta t} \quad{\rm and}\quad\msb{B}(\bb{v}) \doteq \lim_{\Delta t\rightarrow \textrm{`0'}} \frac{\langle\Delta\bb{v}\Delta\bb{v}\rangle}{\Delta t} ,
\end{equation}
respectively, where
\begin{equation}\label{eqn:ens_avg_defn}
    \langle\,\dots\,\rangle \doteq \frac{1}{\te-\ts} \int^{\te}_{\ts} \rmd t \, \int \rmd{\bb{v}} \, \big(\,\dots\,\big) \, f(\tv) .
\end{equation}
The jump interval $\Delta t$ must be taken to be much smaller than the characteristic `collisional' timescale over which $f$ evolves and yet not so small that it is comparable to the auto-correlation time of the (assumed random) kicks experienced by the particles as they interact with the electromagnetic fluctuations, i.e.,
\begin{equation}\label{eqn:Dt_ordering}
    \tac \ll \Delta t\ll \nu^{-1},
\end{equation}
where $\tau_{\rm ac}$ is the auto-correlation time and $\nu$ is the effective collision frequency (thus the notation $\Delta t\rightarrow \textrm{`0'}$ rather than $\Delta t\rightarrow 0$ in \eqref{eqn:jumps}). Assuming that the jumps are small, so that the consequent changes in the distribution function can be approximated by a Taylor expansion in $\Delta\bb{v}$, one can show that the Fokker--Planck operator is given by
\begin{equation}\label{eqn:general_Fokker--Planck}
    C_{\rm FP}[f] = -\pD{\bb{v}}{}\bcdot \bigl[\bb{A}(\bb{v})f(\tv)\bigr] + \frac{1}{2}\frac{\partial^2}{\partial \bb{v}\partial\bb{v}}\,\bb{:}\,\bigl[ \msb{B}(\bb{v})f(\tv)\bigr].
\end{equation}
Provided that $\bb{A}$ and $\msb{B}$ can be reliably computed, equation~\eqref{eqn:general_Fokker--Planck} is the model collision operator for our Fokker--Planck method.

In what follows, we work in spherical velocity coordinates, $(v,\xi,\phi)$. We further assume that the interactions between the electrons and the fluctuations do not push the distribution function sufficiently far from gyrotropy (i.e., $\partial f/\partial\phi \approx 0$) to warrant retaining jump moments in $\phi$. To transform \eqref{eqn:general_Fokker--Planck} into this coordinate system, we follow \citet{Rosenbluth1957} in writing \eqref{eqn:general_Fokker--Planck} as
\begin{equation}
\begin{split}
    C_{\rm FP}[f] &= -\bigl(A^\mu f)_{;\mu} + \frac{1}{2}\bigl( B^{\mu\nu} f \bigr)_{;\mu\nu} \\*
    \mbox{} &= -\frac{1}{\sqrt{g}} \pD{q^\mu}{} \bigl(\sqrt{g} A^\mu f \bigr) + \frac{1}{2\sqrt{g}}\frac{\partial^2}{\partial q^\mu q^\nu} \bigl(\sqrt{g} B^{\mu\nu} f\bigr) + \frac{1}{2\sqrt{g}}\pD{q^\nu}{} \bigl(\sqrt{g} \Gamma^\nu_{\lambda\mu} B^{\mu\lambda} f \bigr) ,
\end{split}
\end{equation}
where the semi-colon denotes the covariant derivative, $g\doteq\det(g_{\mu\nu})$ is the determinant of the metric tensor $g_{\mu\nu}$, $\Gamma^\nu_{\lambda\mu}$ are the associated Christoffel symbols, and $q^\mu$ ranges over the coordinates $(v,\xi,\phi)$. With the metric given by
\begin{equation}
    \rmd s^2 \doteq g_{\mu\nu} \rmd q^\mu \rmd q^\nu = \rmd v^2 + \frac{v^2}{1-\xi^2} \,\rmd \xi^2 + v^2(1-\xi^2)\,\rmd\phi^2 ,
\end{equation}
it is then a straightforward calculation to show that \eqref{eqn:general_Fokker--Planck} becomes
\begin{equation}\label{eqn:vxi_Fokker--Planck}
\begin{split}
    C_{\rm FP}[f] &= \frac{1}{v^2} \pD{v}{} \biggl( - v^2 A^v f + \frac{1}{2} \pD{v}{} v^2 B^{vv} f \biggr) + \pD{\xi}{} \biggl( - A^\xi f + \frac{1}{2} \pD{\xi}{} B^{\xi\xi} f \biggr) \nonumber\\*
    \mbox{} &+ \frac{1}{v^3} \frac{\partial^2}{\partial v\partial\xi} \biggl( v^3 B^{v\xi} f \biggr)  + \frac{1}{2v^2} \biggl( \xi \pD{\xi}{} - v \pD{v}{} \biggr) \biggl( \frac{ v^2 }{1-\xi^2} B^{\xi\xi} f \biggr).
\end{split}
\end{equation}
where the jump moments are given by
\begin{equation}\label{eqn:vxi_jumps}
\begin{gathered}
    A^{v} \doteq \lim_{\Delta t \rightarrow \textrm{`0'}} \frac{\langle\Delta{v}\rangle}{\Delta t}, \quad A^{\xi} \doteq \lim_{\Delta t \rightarrow \textrm{`0'}} \frac{\langle\Delta{\xi}\rangle}{\Delta t}, \\*
    B^{vv} \doteq \lim_{\Delta t\rightarrow \textrm{`0'}} \frac{\langle(\Delta{v})^2\rangle}{\Delta t} , \quad B^{\xi\xi} \doteq \lim_{\Delta t\rightarrow \textrm{`0'}} \frac{\langle(\Delta{\xi})^2\rangle}{\Delta t}, \quad B^{v\xi} \doteq \lim_{\Delta t\rightarrow \textrm{`0'}} \frac{\langle\Delta{v}\Delta\xi\rangle}{\Delta t} .
\end{gathered}
\end{equation}
Note that \eqref{eqn:vxi_Fokker--Planck} can be simplified further if we demand that the operator annihilates a particular form of distribution function. For example, if we demand that $C_{\rm FP}[f] = 0$ for an isotropic Maxwellian having temperature $T=(1/2) m v^2_{\rm th}$ and bulk velocity $\vw\eb$, then equation~\eqref{eqn:vxi_Fokker--Planck} becomes
\begin{equation}\label{eqn:redFP}
    C_{\rm FP}[f] = \frac{1}{2} \pD{\xi}{} B^{\xi\xi}\biggl(\pD{\xi}{f}+\vw \pD{v}{f} \biggr)  + \frac{1}{v^2}\pD{v}{} \frac{v^2}{\vth^2} B^{vv} \biggl[ (v-\vw\xi)  f + \frac{\vth^2}{2} \pD{v}{f} \biggr] .
\end{equation}
Each of the three terms in this equation has a clear physical interpretation. The first term corresponds to perpendicular diffusion at fixed energy, i.e., pitch-angle scattering, occurring in a frame moving at speed $\vw$ with a velocity-dependent rate given by
\begin{equation}\label{eqn:nufp}
    \nufp(v,\xi)=\frac{B_{\xi\xi}}{1-\xi^2}.
\end{equation}
This term then recovers the form of both the Chapman--Enskog \eqref{eqn:pa_operator} and quasi-linear \eqref{eqn:QL_op} pitch-angle scattering operators, {\em viz.}
\begin{equation}\label{eqn:FP_op}
    C_{\rm FP}[f]=\pD{\xi}{} \left[ \frac{1-\xi^2}{2} \nu_{\rm FP}(v,\xi) \biggl(\pD{\xi}{f}+\vw \pD{v}{f} \biggr) \right].
\end{equation}
The second term in \eqref{eqn:redFP} proportional to $(v-\vw\xi)$ corresponds to drag, and the third term proportional to $\partial f/\partial v$ to energy diffusion. The physical consequence of these final two terms are that the distribution function relaxes to a Maxwellian with thermal speed $v_{\rm th}$ and bulk velocity $\vw\eb$ at a rate given by $B^{vv}/\vth^2$. Notwithstanding the relatively pleasant form of \eqref{eqn:redFP}, we caution that the HWI is not guaranteed to push the distribution function towards an isotropic, drifting Maxwellian; indeed, a more natural expectation is that the instability pushes the system towards marginal stability, with a non-zero heat flux and its associated asymmetry in $f(v,\xi)$. In this case, equation~\eqref{eqn:FP_op}, with the jump moments~\eqref{eqn:vxi_jumps} calculated from the particle trajectories, is arguably a more appropriate operator than \eqref{eqn:redFP}.

We calculate the expectation values $\langle \Delta v\rangle$, $\langle (\Delta v)^2\rangle$, $\langle \Delta\xi\rangle$,  $\langle(\Delta\xi)^2\rangle$, and $\langle\Delta v\Delta\xi\rangle$ from the simulated particle data for a wide variety of $\Delta t$. Without {\em a~priori} knowledge of the auto-correlation and collisional timescales, we select values of $\Delta t$ that are logarithmically spaced between $\Omega_{e,0}\Delta t\in [ 0.1,500]$ and use the method described in the next subsection (\S\ref{sec:FP_times}) to choose the $\Delta t$ most consistent with the limit $\Delta t\rightarrow \textrm{`0'}$. As with our calculation of the distribution function, we focus on the central 4\% of the domain, which restricts our sample of particles to an area of nearly uniform temperature that is far removed from the domain boundaries. After computing indices describing the velocity-space location of each particle on a $60\times 60$ grid in $(v,\xi)$, we add the individual Lagrangian jump moment to the grid at the indices that correspond to the particle's initial phase-space location. This ensures that electrons that jump out of the 4\% domain under consideration in any given interval $\Delta t$ are included in the jump moment calculations inside the domain. Doing this for all particles in the sample, we generate an ensemble average. We repeat this process at different times throughout the interval $t\in [\ts,\te]$ (again, reported in table~\ref{tab:runs}) to give a time-averaged operator, which drastically reduces the amount of noise in the computed jump moments. The accuracy of the time-averaged operators relies on the system being in steady state -- i.e., in saturation -- over the measurement interval. While this is not exactly satisfied for all runs, they are all very near saturation: the heat flux varies at most by $10\%$ across any given interval. Given the expected sampling noise, particularly out to $v/\vtheo=3$, we take this amount of error to be acceptable.

\subsubsection{Timescales and hierarchies}\label{sec:FP_times}

The limit $\Delta t\rightarrow \textrm{`0'}$ in the definitions of the jump moments \eqref{eqn:jumps} ensures that the jumps are sufficiently small that the Taylor expansion that relates them to the rate of change of the distribution function converges to the true value. We estimate the autocorrelation time of the HWI fluctuations as the quasi-linear autocorrelation time $\taclin$, the time it takes a resonant particle to interact with a wave packet \citep{Krommes2002}. Given a packet of waves with central wave number $\bar{k}$ and width $\Delta k$,
\begin{equation}
    \taclin\sim (|v_{\rm p}(\bar{k}) - v_{\rm g}(\bar{k})|\Delta k\big)^{-1} ,
\end{equation}
where $v_{\rm p}$ is the wave phase velocity and $v_{\rm g}$ is the group velocity of the packet. From the spectra in our simulations (figure \ref{fig:power_spectra}) we see that whistlers are energetically peaked around $k\rhoeo\sim 1$ and that most of the energy is contained within a few $k\rhoeo$ of the peak. Taking $k \rho_e\sim 1$ as our characteristic wavenumber and $\Delta k/k\sim 1$, we find that
\begin{equation}
    \taclin\Omega_e\sim \big(n+1/\beta_e\big)^{-1}.
\end{equation}
For our simulations in the limit of high $\beta_e$, the cyclotron $(n\ne 0)$ resonances will therefore have an autocorrelation time on the order of an inverse cyclotron frequency:
\begin{equation}
    \taclin(n\ne 0)\Omega_e\sim 1.
\end{equation}
However, the Landau $(n=0)$ resonance can have an arbitrarily long autocorrelation time
\begin{equation*}
    \taclin(n=0)\Omega_e\sim\beta_e
\end{equation*}
in the limit of high $\beta_e$. It is conceivable that in our simulations, there does not exist a $\Delta t$ for electrons with $\vpar\sim\vw$ that satisfies the ordering \eqref{eqn:Dt_ordering}. We therefore rely on a model process, called an Ornstein--Uhlenbeck process, to inform our choice of an appropriate $\Delta t$ with which to calculate our drag and diffusion values.

Consider a one-dimensional Fokker--Planck equation in $v$ with drag to a mean velocity $\bar{v}$ at a rate $\nu$ and a diffusion coefficient $D$:
\begin{equation}\label{eqn:FP1D}
    \pD{t}{f(t,v)}=\pD{v}{}\nu (v-\bar{v})f(t,v)+\frac{D}{2}\pDD{v}{}f(t,v).
\end{equation}
This equation represents a particular statistical process called an Ornstein--Uhlenbeck process \citep{Uhlenbeck&Ornstein1930} and will serve as a prototypical model for understanding our numerical results. The Green's function for the differential equation \eqref{eqn:FP1D} has a well-known solution \citep{Risken1991}, which can be written in terms of the jump moments as a Gaussian:
\begin{equation}\label{eqn:Gausian_PDF}
    G(\Delta v,\Delta t)=\frac{1}{\sqrt{2\upi\Var(\Delta v)}}\exp\bigg[-\frac{(\Delta v-\avg{\Delta v})^2}{2\hspace{2pt}\Var(\Delta v)}\bigg],
\end{equation}
with the first and second jump moments given by
\begin{subequations}\label{eqn:OU_scaling}
\begin{align}
    \avg{\Delta v}& = \big(\bar{v}-v\big) \big(1-e^{-\nu\Delta t}\big), \label{eqn:OU_scaling_drag}\\
    \Var(\Delta v)=\avg{\Delta &v^2} - \avg{\Delta v}^2 = \frac{D}{2\nu}\big(1-e^{-2\nu\Delta t}\big),\label{eqn:OU_scaling_diffn}
\end{align}
\end{subequations}
for stationary solutions.

The expressions for the jump moments \eqref{eqn:OU_scaling}, taken in the limit $\Delta t\rightarrow 0$, recover the drag and diffusion coefficients in \eqref{eqn:FP1D}. In the opposite limit of $\nu\Delta t\gg1$, $\avg{\Delta v}\simeq \bar{v}-v$ and $\Var (\Delta v)\simeq D/\nu\sim \vth^2/2$. On this long timescale, particles drift at the mean velocity $\bar{v}$ and equilibrate to a mean temperature corresponding to $\vth$. If one were to use such a $\Delta t$ when calculating the coefficients, all kinetic information would be lost and the measured coefficients would be incorrect. We must therefore be cautious in choosing an appropriate $\Delta t$ when calculating jump moments. For cases where the jump moments are constant for a range of $\Delta t\gg \tac$, the proper choice of $\Delta t$ is unambiguously in that range and the coefficients can be simply read off from the jump moment at a proper $\Delta t$. This is the case for drag and diffusion in speed, as we show first in \S\ref{sec:FP_v}. If no such interval is present, for instance because $\nu^{-1}\sim \tac$, the choice of an appropriate $\Delta t$ is much more complicated. One must choose the $\Delta t\ge \tac$ that corresponds to the range in which the jump moments are most constant with $\Delta t$, much before the jump moments begin to scale as $\Delta t^{-1}$ and represent thermalized, non-kinetic physics. If the scattering rate is near the autocorrelation time, often the best choice is to take $\Delta t = \tac$. Critically, if the scattering frequency varies in velocity space, one must repeat this process for every point in phase space one wishes to compute coefficients, or risk reporting a drag or diffusion coefficient that corresponds to thermalized physics in a portion of phase space. This is the case for drag and diffusion in pitch-angle in many of our runs; we show in detail the process and results in \S\ref{sec:FP_xi}.

\subsubsection{Fokker--Planck jump moments: velocity}\label{sec:FP_v}
\begin{figure}
    \centering
    \includegraphics{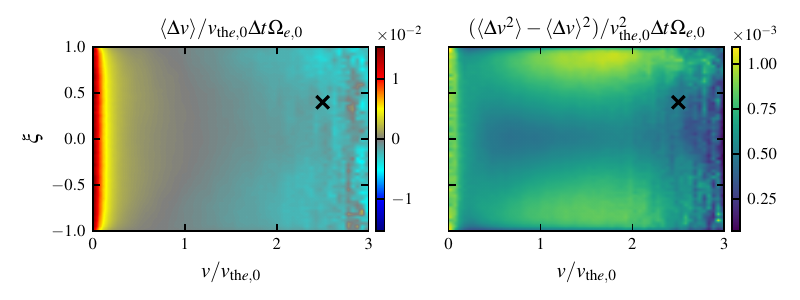}
    \caption{First (left) and second (right) velocity jump moments with $\Delta t\wceo=10$ from simulation b40x4. The cross marks the point in phase space where the jump moments are plotted as functions of $\Delta t$ in figure~\ref{fig:dvdt1d} and where the PDFs of the jump moments are plotted for $\Delta t\wceo=10$ in figure~\ref{fig:Dv_PDFs}.}
    \label{fig:b40x4_dvdt}
\end{figure}
\begin{figure}
    \centering
    \includegraphics{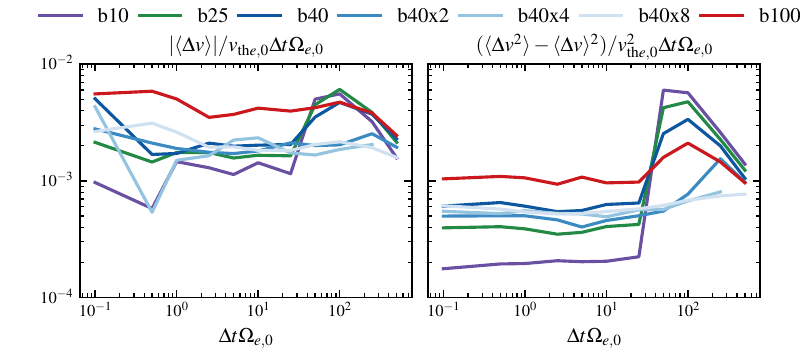}
    \caption{Fokker--Planck drag (left) and diffusion (right) coefficients as a function of $\Delta t$ for all $\gradpz$ simulations at the location in phase space marked by the cross in figure~\ref{fig:b40x4_dvdt}. While the moments are roughly constant for $\Delta t\wceo<25$, as expected for an Ornstein-Uhlenbeck process, there is clearly some non-Markovian behavior for larger $\Delta t$.}
    \label{fig:dvdt1d}
\end{figure}
\begin{figure}
    \centering
    \includegraphics{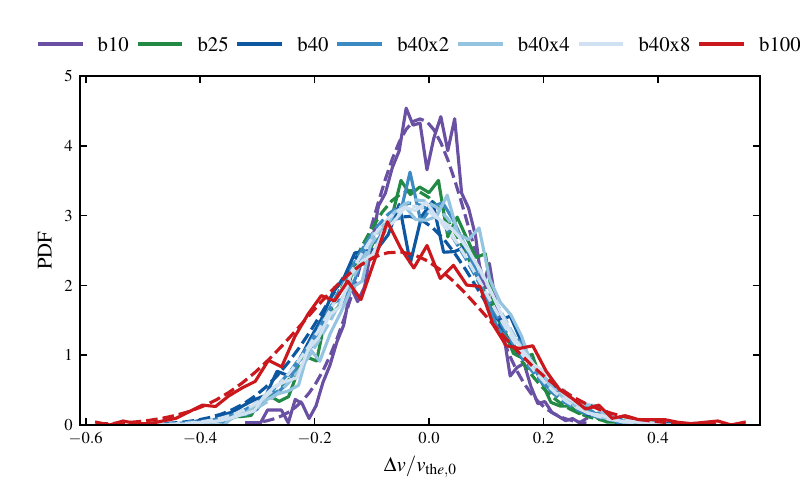}
    \caption{Probability densities for jump in velocity $\Delta v/\vthe$ for $\Delta t/\Omega_e=10$ at the location in phase space denoted by the cross in figure \ref{fig:b40x4_dvdt}. Gaussian PDFs constructed from the moments of the densities are plotted in dashed lines, showing the densities are in fact Gaussian.}
    \label{fig:Dv_PDFs}
\end{figure}

The drag and diffusion coefficients in velocity are remarkably simple to recover from the jump moments. In figure \ref{fig:b40x4_dvdt}, we show two-dimensional plots of drag and diffusion coefficients for run b40x4 at $\Delta t\wceo=10$. In figure \ref{fig:dvdt1d}, we plot the values of these coefficients for all runs as a function of $\Delta t\wceo$ at the point in phase space denoted by the `X' in figure \ref{fig:b40x4_dvdt}, approximately $v/\vtheo=2.5$ and $\xi=0.45$. Across all runs, both the drag and diffusion coefficients are nearly constant as a function of $\Delta t$ for $\Delta t\Omega_e< 25$. Runs at low $\betaeo$ and small $L_T/\rhoeo$ (runs b10, b25, and b40) show a non-Markovian increase in the jump moments for $25 \le \Delta t\wceo\le 500$.

In the following, we assume the underlying process is Ornstein--Uhlenbeck and take as an appropriate jump interval $\Delta t\wceo=10$. This assumption is well justified. The jump interval lies comfortably in the interval $\tac\wceo\sim 1\ll \Delta t\wceo=10\ll (\nu/\wceo)^{-1}\sim 10^3$, and both coefficients are approximately constant with $\Delta t$ in the vicinity of $\Delta t\wceo=10$ for all points in $(v,\xi)$ space. In figure~\ref{fig:Dv_PDFs}, for all runs we plot with solid lines histograms of individual electron jumps in speed for the jump interval $\Delta t\wceo=10$ at the same location in velocity space as in figure~\ref{fig:dvdt1d}. Using dashed lines we overlay  Gaussian distributions with the same mean and variance as each histogram. The correspondence between the data and the Gaussian distributions further suggest that the observed process is Ornstein--Uhlenbeck.

\begin{figure}
    \centering
    \includegraphics{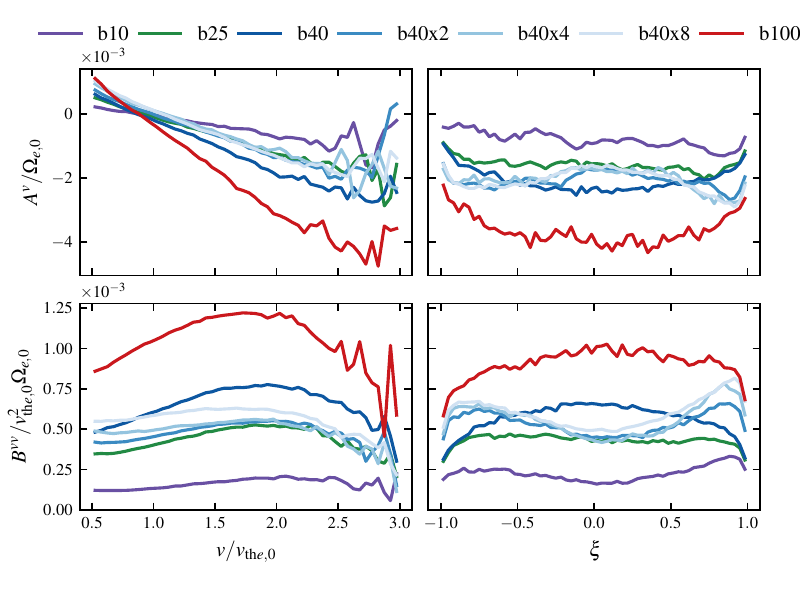}
    \caption{Drag (top) and diffusion (bottom) coefficients in speed as functions of $v/\vtheo$ (left) and $\xi$ (right) for all $\gradpz$ runs. All coefficients are approximately constant except for drag, which linearly decreases with increasing speed.}
    \label{fig:Dv_FP_lineouts}
\end{figure}
\begin{figure}
    \centering
    \includegraphics{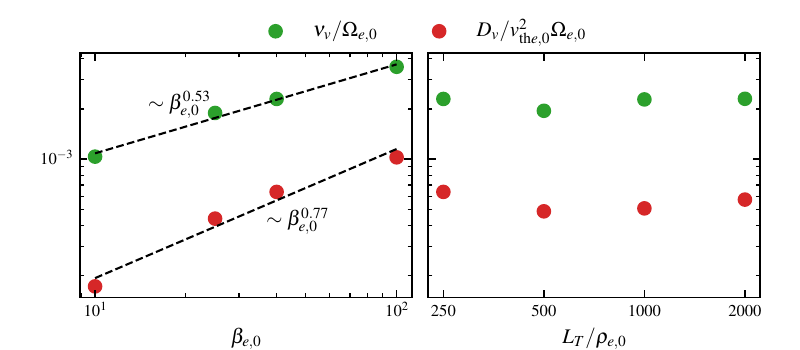}
    \caption{Drag rate $\nu_v$ (green) and diffusion coefficient $D_v$ (red) vs. $\betaeo$ (left) and $L_T/\rhoeo$ (right). Both $\nu_v$ and $D_v$ are nearly constant with $L_T/\rhoeo$, but scale as $\betaeo^{0.53}$ and $\betaeo^{0.77}$, respectively. These values are subdominant to drag and diffusion in pitch angle (see~\S\ref{sec:FP_xi}).}
    \label{fig:FP_v_coefs}
\end{figure}

In figure \ref{fig:Dv_FP_lineouts}, we plot normalized velocity drag $A^v$ and diffusion $B^{vv}$ coefficients as functions of both $v/\vtheo$ and $\xi$. The drag coefficient is linear in $v$, with a zero near $v=\vtheo$. The diffusion coefficient peaks at $v/\vtheo\sim 2$, decreasing at higher and lower speeds. The dependence on $\xi$ of these coefficients, however, is more complicated. For runs~b40 and b100 -- i.e., those with the highest $\avg{\delta B^2}/B_0^2$ -- drag and diffusion are greatest at $\xi=0$ and decrease towards $\xi=\pm 1$. For runs with lower fluctuation amplitudes, the dependence inverts and appears to converge towards a single shape as $L_T/\rhoeo$ increases. This feature is evident in the right-hand panel of figure~\ref{fig:b40x4_dvdt}, where the velocity diffusion coefficient for run b40x4 peaks near $\xi=\pm 1$ for $v/\vtheo\in [1,2]$.

Assuming that the speed jump moments are well approximated by their pitch-angle averages, we find that the resulting $A^v(v)$ and $B^{vv}(v)$ can be accurately modeled by
\begin{equation}\label{eqn:FP_v_model}
    A^v(v) = -\nu_v(\beta_e,L_T)(v-\vthe) \quad\text{and}\quad B^{vv}(v) = D_v(\beta_e,L_T),
\end{equation}
with the constant $\nu_v$ and $D_v$ plotted in figure~\ref{fig:FP_v_coefs} as functions of $\betaeo$ and $L_T/\rhoeo$. These rates appear to be independent of $L_T/\rhoeo$ but exhibit power-law dependencies on $\betaeo$ (shown by the dashed-line fits). In sections~\ref{sec:CE_res}, \ref{sec:QL_res}, and \ref{sec:FP_xi}, we show that these rates are much smaller than those associated with pitch-angle scattering. We therefore neglect $A^v$ and $B^{vv}$ in what follows.

\subsubsection{Fokker--Planck jump moments: pitch angle}\label{sec:FP_xi}

\begin{figure}
    \centering
    \includegraphics{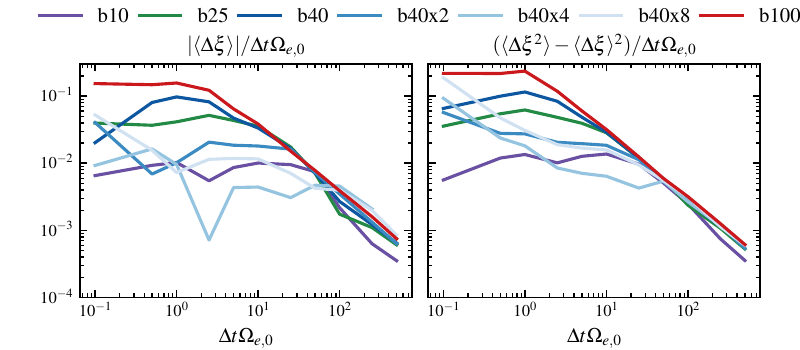}
    \caption{Pitch-angle drag (left) and diffusion (right) as a function of $\Delta t$ for all $\gradpz$ simulations at the location in phase space denoted by the cross in figure~\ref{fig:b40x4_dvdt}. These moments exhibit clear non-Markovian behavior and in general do not conform to an Ornstein-Uhlenbeck process.}
    \label{fig:dxidt1d}
\end{figure}
\begin{figure}
    \centering
    \includegraphics{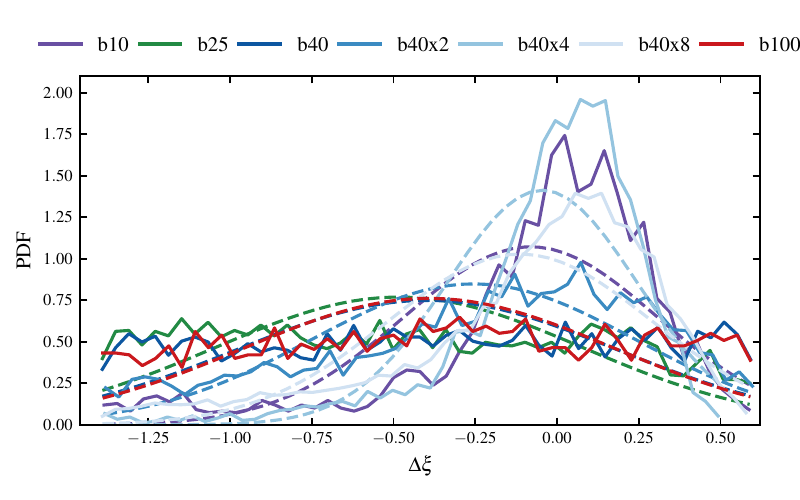}
    \caption{Probability densities for jumps in pitch angle $\Delta \xi$ for $\Delta t\wceo=10$ calculated at $(v,\xi)=(2.5,0.45)$. Gaussian PDFs constructed from the moments of the densities are plotted in dashed lines; the PDFs differ significantly from Gaussian due to strong scattering.}
    \label{fig:Dxi_PDFs}
\end{figure}

Obtaining the Fokker--Planck jump moments in pitch angle, $A^\xi$ and $B^{\xi\xi}$, turns out to be considerably more complicated than for $A^v$ and $B^{vv}$, the principal reason being insufficient lack of scale separation. Analogously to figure~\ref{fig:dvdt1d}, we plot
in figure~\ref{fig:dxidt1d} the values of the drag and diffusion coefficients in pitch angle as  functions of $\Delta t\wceo$ at $(v,\xi)\simeq (2.5\vtheo,0.45)$ for all of our analyzed runs. There is a stark difference between runs with higher $\avg{\delta B^2}/{B_0^2}$, namely b40 and b100, and those where it is lower, like b10 and b40x8. Runs with large fluctuation amplitude have jump moments ${\sim}0.1\wceo$ that are comparable to the inverse autocorrelation time $\tac^{-1}\sim \wceo$. It is impossible to be certain whether the coefficients are constant in $\Delta t$ within such a small range. Likewise, the probability distributions of individual particle pitch-angle jumps $\Delta \xi$ are far from Gaussian; these are shown in figure~\ref{fig:Dxi_PDFs} for the jump interval $\Delta t\wceo=10$ (solid lines) and compared with Gaussians having identical means and variances (dashed lines). At this jump interval, all of the distributions stray from Gaussian, with runs having larger fluctuation amplitudes exhibiting almost completely flat distributions. The distributions do, however, appear to converge towards Gaussian in the limit of low fluctuation amplitude (see b10, b40x4, and b40x8).

\begin{figure}
    \centering
    \includegraphics{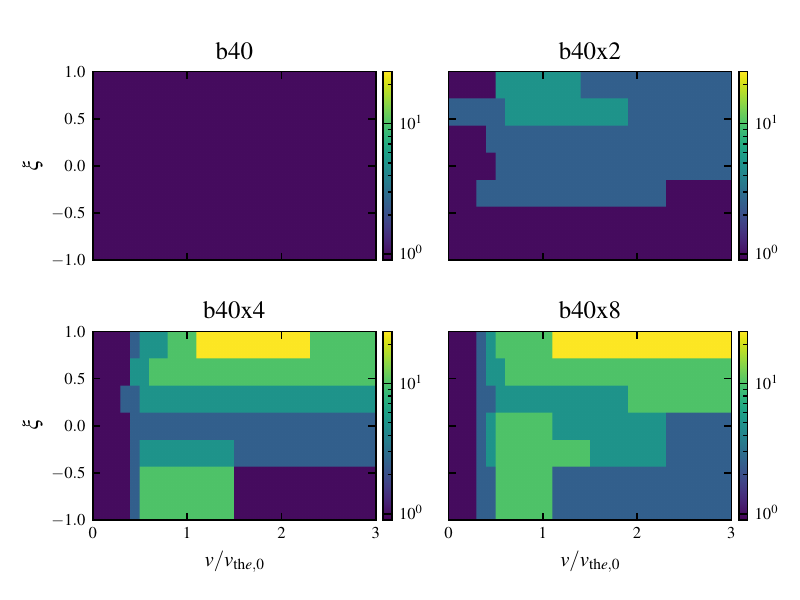}
    \caption{Appropriate jump intervals $\Delta t\wceo$ for a subset of $(v,\xi)$ points for all $\betaeo=40$ runs. Due to a lack of scale separation, all jump intervals for run b40 coincide with the quasi-linear autocorrelation time $\taclin\wceo\sim \Delta t\wceo=1$.}
    \label{fig:Dt2d}
\end{figure}
\begin{figure}
    \centering
    \includegraphics{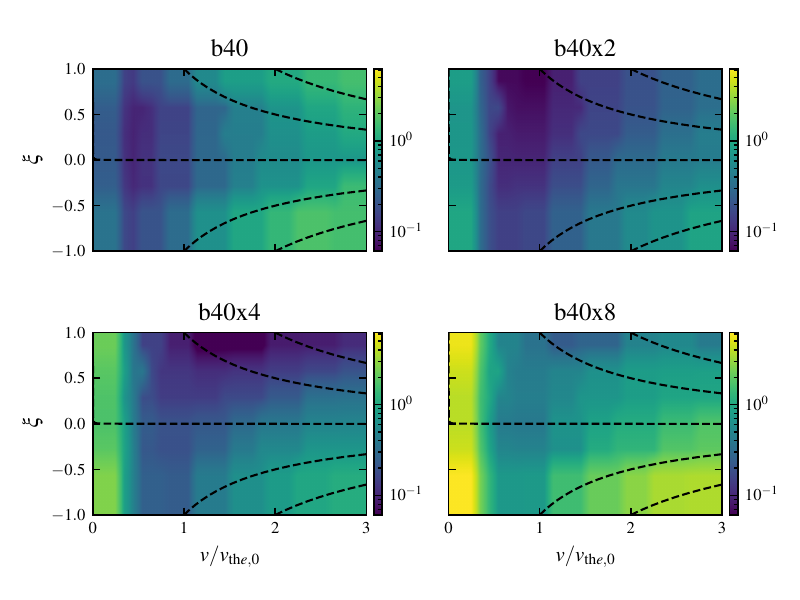}
    \caption{Normalized Fokker--Planck diffusive scattering frequency for all $\betaeo=40$ runs as a function of $v$ and $\xi$, calculated from appropriate jump times $\Delta t$, as shown in figure~\ref{fig:Dt2d}. Dashed lines correspond to contours of constant $\vpar$.}
    \label{fig:xicoefs2d}
\end{figure}
\begin{figure}
    \centering
    \includegraphics{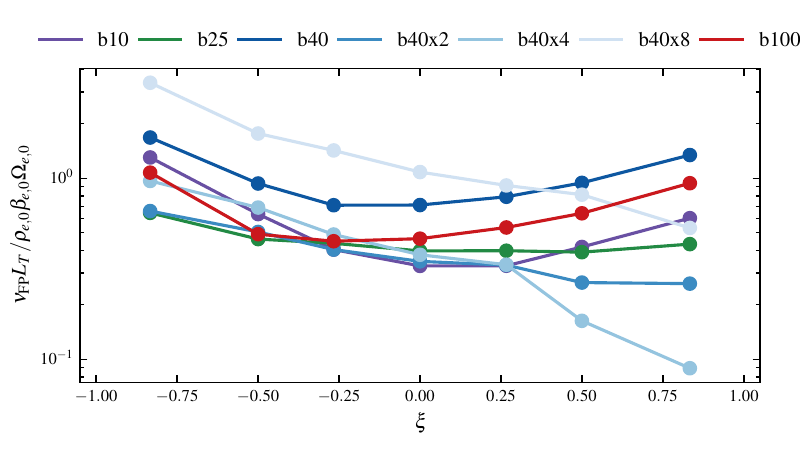}
    \includegraphics{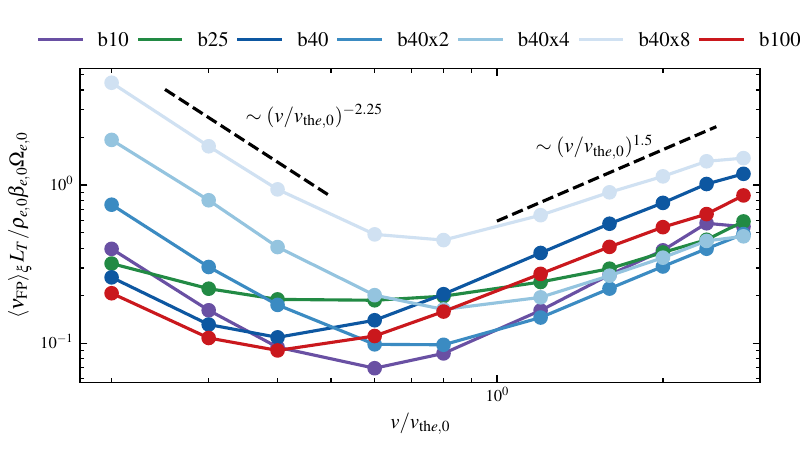}
    \caption{Normalized $\nufp$ for all runs as a function of $\xi$ at $v/\vtheo=2.5$ (top) and pitch-angle-averaged as a function of $v/\vtheo$ (bottom). Runs b40-b40x8 show a transition from nearly symmetric scattering in $\xi$ to electrons with $\vpar<0$ scattering significantly faster than those with $\xi>0$. All runs show a double power law similar to \S\ref{sec:CE_res}, with the normalized scattering frequency of the slow power law increasing with larger $L_T/\rhoeo$.}
    \label{fig:FP_xi}
\end{figure}

Further complicating matters, the drag and diffusion coefficients depend on $v$ and $\xi$, so in most all cases there is no single values of $\Delta t\wceo$ appropriate for use in calculating these coefficients (unlike for the speed jump moments). As explained in \S\ref{sec:FP_times}, we pick an appropriate $\Delta t$ at a subset of locations in phase space and then use this value to calculate $A^\xi$ and $B^{\xi\xi}$ for each $(v,\xi)$ combination. The results of this procedure for runs~b40 and b40x4 are shown in figure~\ref{fig:Dt2d}. For run~b40, the collisional timescale is close to the autocorrelation time across all velocity space. Run~b40x4 does have some separation of timescales, particularly close to $\xi=1$ where the scattering is measured to be weak. The normalized diffusive scattering frequency $\nu_{\rm FP}L_T/\rhoeo\betaeo\wceo$ resulting from the appropriate jump times are shown in figure~\ref{fig:xicoefs2d}. For $v/\vtheo> 1$, there is a clear difference in symmetry about $\xi=0$ between these two runs. This asymmetry shows up in all runs with $L_T/\rhoeo>250$, as shown in the top panel of figure~\ref{fig:FP_xi}. The Fokker--Planck results at large $L_T/\rhoeo$ as a function of pitch angle are structurally similar to the quasi-linear results (cf.~figure~\ref{fig:QL_leftover}). The ratio $\nufp(\xi=-0.83)/\nufp(\xi=0.83)\approx 6$ for run b40x8 is comparable to the quasi-linear result $\nuql(\xi=-1)/\nuql(\xi=1)\sim 10$; however, the scattering rate around $\xi=0$ is much higher for the Fokker--Planck case. The bottom panel of figure~\ref{fig:FP_xi} shows the pitch-angle average of $\nufp$ as a function of $v$. The velocity dependence is a fast and slow power law, similar to \S\ref{sec:CE_res} and \S\ref{sec:QL_res} (cf.~figures~\ref{fig:CE_lineouts} and \ref{fig:QL_leftover}, respectively), which can be attributed to transit-time damping and cyclotron resonances, respectively, per the analysis in \S\ref{sec:QL_res}. Here, as with the other models, the scattering amplitude at small velocities increases with $L_T/\rhoeo$, implying that the relative contribution from transit-time damping also increases with $L_T/\rhoeo$. The velocity dependence of the pitch-angle-average scattering frequency for superthermal particles here is ${\sim}(v/\vtheo)^{1.5}$, which is generally shallower than other models (see figures~\ref{fig:CE_lineouts} and \ref{fig:QL_leftover}).

\subsubsection{Summary of Fokker--Planck results}\label{sec:FP_summary}
In \S\ref{sec:FP}, we calculated an effective Fokker--Planck collision operator for the HWI by constructing drag and diffusion coefficients as functions of velocity-space for both velocity and pitch-angle. The drag and diffusion coefficients were constructed from jump moments of tracked particles in our simulations, which we showed how to construct rigorously using jump intervals informed self-consistently by the particle statistics. The statistics for velocity were found to be consistent with Ornstein--Uhlenbeck statistics while the statistics for pitch-angle were less conclusively Ornstein--Uhlenbeck due to the limited scale separation of our runs. Through a judicious choice of jump interval, we obtained the drag and diffusion coefficients and demonstrated that pitch-angle diffusion clearly dominated drag and diffusion in velocity, as is predicted by the quasi-linear model in \S\ref{sec:QL}. The pitch-angle dependence of the pitch-angle scattering frequency obtained by the Fokker--Planck method, however, is much flatter than the quasi-linear result (\S\ref{sec:QL_res}) and will prove to be a distinct advantage of the former over the latter in \S\ref{sec:results3}.

\section{Results, III. ~A model effective collision operator for saturated HWI}\label{sec:results3}

In section~\ref{sec:results2}, we used our numerical simulation data to calculate three model collision operators describing the drag and diffusion of electrons as they interact with saturated HWI fluctuations. Here we attempt to unify these operators, highlighting their essential ingredients, proposing a model that captures them, and testing whether that model can result in a heat flux that matches, within reason, the heat flux measured directly in the simulations. While doing so, we feature similarities and differences between existing model operators and emphasize what aspects of our model operator could benefit for further refinement.

\subsection{Salient features of a model effective collision operator}\label{sec:salient_features}

First and foremost, all of our model operators show scattering rates that scale as $\beta_e\vthe/L_T$, with the exact rate depending on the model. We also demonstrated that pitch-angle scattering dominates over drag and diffusion in velocity (\S\ref{sec:FP}), and showed in \S\ref{sec:QL_def} that this is expected analytically from an electromagnetic quasi-linear operator for waves with highly subthermal phase velocity. Such an operator includes a drag term proportional to the phase velocity of the scattering waves, which we showed is equivalent to boosting the pitch-angle scattering operator to the wave frame (\S\ref{apx:CEtransform}). Finally, we found that the consequent advection of scattered particles at the phase speed $\vw\sim \vthe/\beta_e$ is of the same order as the diffusive scattering; unsurprisingly then, we show in \S\ref{sec:IHF} that each effect is responsible for approximately half of the observed heat flux.

The pitch-angle average of the pitch-angle scattering frequency derived from each of our methods all show two power laws in $v$: one decreasing as a function of $v$ for subthermal electrons (physically, due to transit-time damping), and the other increasing like ${\sim}(v/\vthe)^3$ for superthermal electrons (physically, due to cyclotron-resonant scattering). The pitch-angle dependence of the operators, however, is more complicated. The quasi-linear operator showed (a) structure that was a function of $\vpar$ rather than of $v$ and $\xi$ separately, and (b) that the scattering frequency for electrons with $\vpar<0$ was an order of magnitude larger than for electrons with $\vpar >0$. For runs with $L_T/\rhoeo=250$, the Fokker--Planck operator showed a largely symmetric structure, but with some evidence that the scattering was a function of $\vpar$. At larger $L_T/\rhoeo$, however, the operator began to show more qualitative similarities to the quasi-linear operator, particularly with resonant structure and a strong asymmetry between electrons with positive versus negative parallel velocities.

\subsection{Basic model collision operator}\label{sec:model_CO}
The preponderance of evidence presented in this work points towards whistlers acting as pitch-angle scatterers in a frame moving at the whistler phase velocity
\begin{subequations}\label{eqn:nu_model_all}
\begin{equation}\label{eqn:Cw}
C[f]=\pD{\xi}{} \left[\frac{1-\xi^2}{2}\nu(v,\xi) \left(\pD{\xi}{f} + \vw\pD{v}{f}\right)\right],
\end{equation}
where we have measured the wave phase velocity in simulation to be
\begin{equation}
\vw \simeq 0.23 \vthe/\beta_e .
\end{equation}
Following \S\ref{sec:salient_features}, we take the model scattering frequency to be given by the quasi-linear value, repeated here, with a few modifications:
\begin{equation}\label{eqn:nu_model}
\frac{\numodel}{\Omega_e}=2\upi
\sum_n \int_0^\infty\rmd (\kpar\rho_e)\,
\delta\bigg(\frac{\omega(\kpar\rho_e)}{\Omega_e}-\kpar\rho_e\frac{\vpar}{\vthe}+n\bigg)\frac{1}{L_x}\bigg|\frac{\Psi_{n,\kpar}}{B_0}\bigg|^2.
\end{equation}
In order to make the full quasi-linear expression more tractable for our model, we remove the explicit dependence of the quasi-linear scattering frequency on $k_\perp$. We accomplish this simplification by assuming that the oblique resonance condition can be approximated by the parallel resonance condition \eqref{eqn:resonant_delta} and that the effect of integrating over oblique modes -- specifically the reduction in resonant power for electrons interacting with the left-handed wave component -- can be encapsulated by a bespoke numerical weighting of the parallel spectrum for each of the principle resonances $n=[-1,0,1]$, {\em viz.}
\begin{equation}\label{eqn:parallel_spectrum_approx}
\begin{split}
\frac{1}{L_x}\bigg|\frac{\Psi_{n,\kpar}}{B_0}\bigg|^2=\frac{\avg{\delta B^2}}{B_0^2}\bigg\{&
\bigg[\bigg(\frac{\kpar\rho_e - 0.5}{0.4}\bigg)^3 + 1\bigg]^{-1}\delta_{n,-1}+
0.1\bigg[\bigg(\frac{\kpar\rho_e - 0.5}{0.25}\bigg)^{2.5} + 1\bigg]^{-1}\delta_{n,0}\\
\mbox{} & + 0.1\bigg[\bigg(\frac{\kpar\rho_e - 0.5}{0.5}\bigg)^3 + 1\bigg]^{-1}\delta_{n,1}
\bigg\}.
\end{split}
\end{equation}
\end{subequations}
The integral in \eqref{eqn:nu_model} has bounds $[0,\infty]$, reflecting the the fact that whistler waves travel only down the temperature gradient ($k_\parallel>0$).

Before we discuss our model spectrum in more detail, the use of the parallel whistler dispersion relation in our model resonance condition requires some justification. The oblique cold plasma dispersion relation \eqref{eqn:cpwdr_fit} differs from the parallel dispersion relation by a factor $1/\cos\theta$, where $\theta$ is again the angle between the wave vector and the magnetic field. For $|\vpar/\vthe|\sim 1$ and $\theta=0$, the wave frequency is a factor $1/\beta_e$ smaller than the (order unity) other terms and $\kpar\rho_e\sim 1$. Only when $\cos\theta\sim 1/\beta_e$ does the obliquity of the wave begin to affect the solution -- and even then only by an order-unity factor. At the same time, $\beta_e\gg 1$ implies that waves are propagating nearly perpendicular to the mean field. Waves in this parameter regime are almost certainly kinetic Alfv\'en waves, not whistler waves, and so the resonance condition in \eqref{eqn:nu_model} is no longer applicable. More realistic (smaller) propagation angles will change the resonant parallel wave number by a factor smaller than unity, thus justifying our use of the parallel dispersion relation.

\begin{figure}
	\centering
	\includegraphics{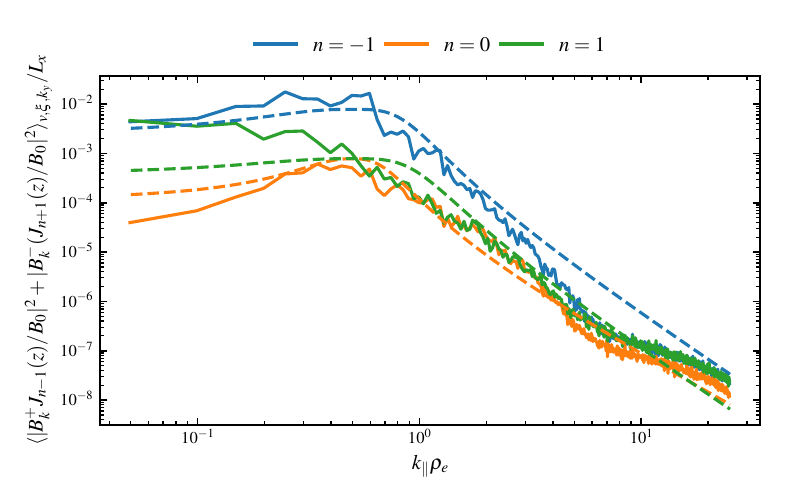}
	\caption{$L_x^{-1}|\Psi_{n ,\kpar}/B_0|^2$ from run b40 using \eqref{eqn:psi_nk} in solid lines and model \eqref{eqn:parallel_spectrum_approx} in dashed lines, for $n=[-1,0,1]$. The spectra calculated from simulation vary from run to run; the model presented here is a good qualitative fit across all runs.}
	\label{fig:b40_psink}
\end{figure}

For each of our runs, we fit \eqref{eqn:parallel_spectrum_approx} to the numerical magnetic-field spectra in saturation; a plot of the fit for run b40 can be found in figure~\ref{fig:b40_psink} along with $L_x^{-1}|\Psi_{n ,\kpar}/B_0|^2$ calculated from the numerical spectrum using \eqref{eqn:psi_nk}. Overall the fit is good, especially so for the $n=-1$ and $n=0$ spectra. However, an issue with our our fit becomes immediately clear: $|\Psi_{n,\kpar}/B_0|^2$ for $n=-1$ and $n=1$ should approach each other at small $\kpar\rho_e$. As it pertains to the heat flux implied by the model, however, this deviation makes little difference as it is the fluctuations at the high-$\kpar\rho_e$ end of the cascade that matter most. One might argue that, in terms of energy, including this high-$\kpar$ part of the cascade makes very little difference to the operator, since the spectra die away very quickly rightward of their peaks. However, the implied heat flux of a collision operator of the form \eqref{eqn:Cw} depends on the inverse of $\nu(v,\xi)$ (see \S\ref{sec:IHF}, specifically \eqref{eqn:f1_nu} and \eqref{eqn:HF_integral}). The heat flux is therefore largely insensitive to regions in velocity space in the vicinity of $\vpar=\vthe$ where particles \textit{are} scattering and is instead dominated by regions where they are \textit{not}. The latter include two populations of particles: those with small enough $\vpar$ that they bounce, and therefore are trapped, in the whistler wave frame; and those that are still passing, but scatter inefficiently because they are resonant with low-energy fluctuations. Proper modelling of both the spectrum at low energies and the effect of resonance broadening and particle trapping is crucial to obtain a scattering frequency that accurately reproduces the observed heat flux.

\begin{figure}
	\centering
	\includegraphics[width=\textwidth]{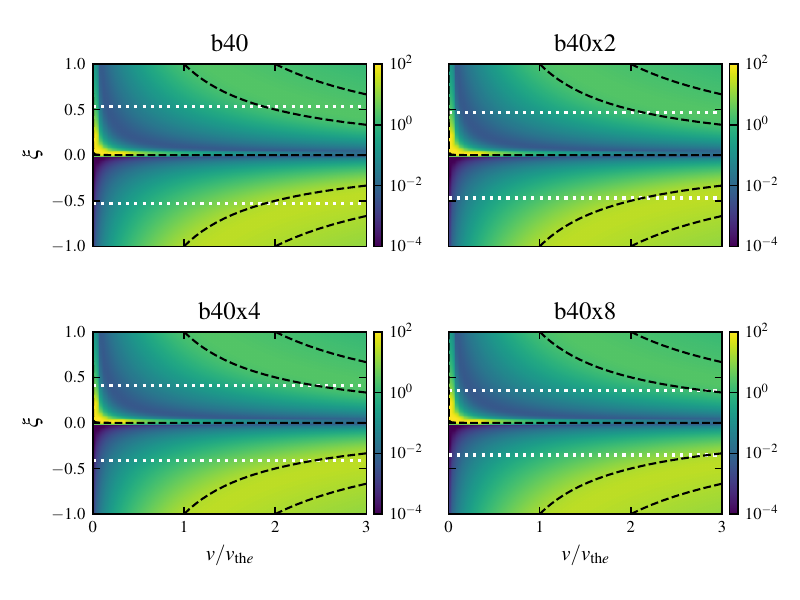}
	\caption{Two-dimensional plots of the model pitch-angle collision frequency $\numodel$ \eqref{eqn:nu_model_all} for all $\betaeo=40$ runs. The scattering frequency is normalized to $\betaeo\rhoeo\wceo/L_T$ and dashed lines correspond to contours of constant $\vpar$. These results are qualitatively similar to the quasi-linear scattering frequency, shown in in figure~\ref{fig:QL_2d}.}
	\label{fig:numodel_delta}
\end{figure}

In figure~\ref{fig:numodel_delta}, we plot the model collision frequency \eqref{eqn:nu_model} normalized by ${\betaeo\rhoeo\wceo/L_T}$ for all ${\betaeo=40}$, $\gradpz$ runs using the model spectrum \eqref{eqn:parallel_spectrum_approx}. Qualitatively, results match those of the quasi-linear operator in figure~\ref{fig:QL_2d}: scattering for electrons traveling up the temperature gradient is higher than that for electrons traveling down the gradient by a factor of ${\sim}10$ and there is a small scattering rate at $\vpar/\vthe\ll 1$ due to transit-time damping. The $n=0$ features in the model, however, are significantly sharper than those in the quasi-linear case, likely a result of using a continuous model for the spectrum rather than a discrete calculation from simulation.

\subsection{Implied heat flux for model collision operators}\label{sec:IHF}
We calculate numerically the heat flux implied by the operators described in \S\ref{sec:QL_res}-\S\ref{sec:FP_xi} by first computing the implied perturbation to the distribution function via
\begin{equation}\label{eqn:f1_nu}
\avg{f_{e1}}_{\phi} = -\int \rmd\xi\, \bigg[\frac{1}{\nu(w,\xi)}\bigg(w\nabla_\parallel f_{e0}+\frac{\nabla_{\parallel}	p_e}{m_en_e}\D{w}{f_{e0}}\bigg)
+v_w\D{w}{f_{e0}}\bigg],
\end{equation}
using $f_{e0}=\fMe$
and $\nu(v,\xi)$ from the appropriate model operator. The first term on the right-hand side of \eqref{eqn:f1_nu} is the perturbation of the distribution function implied by the diffusive flux, which has a scattering frequency $\nu(w,\xi)$, and the last term represents the perturbation implied by advection of particles at the whistler phase velocity $\vw$. The perturbed distribution is then used to compute the parallel electron heat flux according to
\begin{equation}
\begin{split}\label{eqn:HF_integral}
q_{\parallel e}&=\int \rmd^3\bb{w}\,\frac{1}{2}m_ew^2w_\parallel \avg{f_{e1}}_\phi\\
    &=-\frac{m_e}{2}\int_0^\infty \rmd w\,w^5\bigg[\int\rmd\xi\,\xi\bigg(w\nabla_\parallel f_{e0}+\frac{\nabla_{\parallel}	p_e}{m_en_e}\D{w}{f_{e0}}\bigg)\int_0^\xi\frac{\rmd\xi'}{\nu(w,\xi')}
    +\frac{2\vw}{3}\D{w}{f_{e0}} \bigg].
\end{split}
\end{equation}
Before discussing the results of this calculation, we should note that \eqref{eqn:f1_nu} was derived assuming a steady state, i.e. $\partial f_e/\partial t=0$. This assumption may be violated in regions of phase space where our model scattering frequencies are small. Electrons in these regions must experience many collisions to approach a steady state, so $\partial f_e/\partial t\rightarrow 0$ on timescales much longer than $1/\nu(v,\xi)$. Including this effect would serve to reduce the strong variation in $\avg{f_{e1}}_\phi$; however, it likely won't alter the qualitative results that follow.

\begin{figure}
	\centering
	\includegraphics[width=\textwidth]{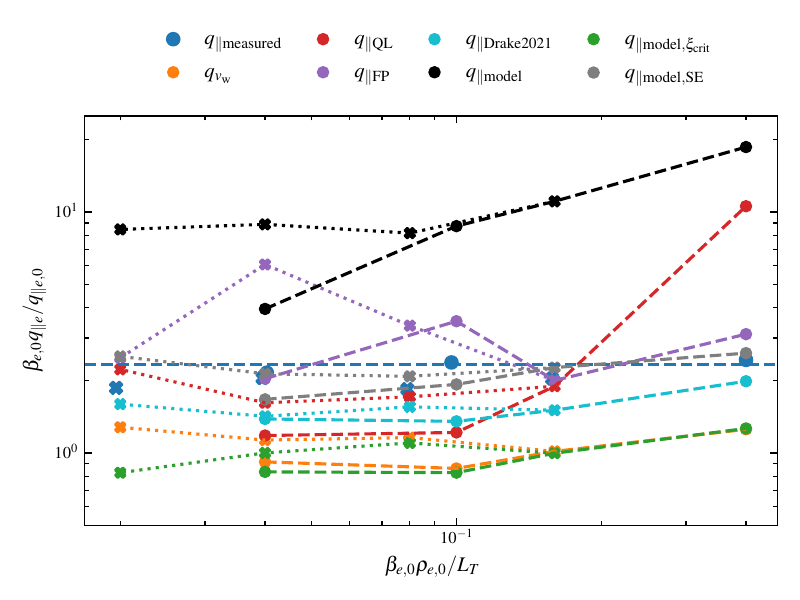}
	\caption{Heat flux measured in simulations ($q_{\parallel{\rm measured}}$) compared to the heat flux implied by advection at the whistler phase speed ($q_{\vw}$), quasi-linear operator ($q_{\parallel{\rm QL}}$), Fokker--Planck operator ($q_{\parallel{\rm FP}}$), Drake et al. (2021) ($q_{\parallel{\rm Drake2021}}$), our model \eqref{eqn:nu_model_all} ($q_{\parallel{\rm model}}$), our model only including passing electrons \eqref{eqn:nuxicrit} assuming the trapped-passing boundary \eqref{eqn:xicrit} ($q_{\parallel{\rm model},\xicrit}$), and our semi-empirical model \eqref{eqn:nuse} ($q_{\parallel {\rm model,SE}}$).
    Points denoted by an `x' are runs with $\betaeo=40$ and are connected by a dotted line. Runs with $L_T=125\rhoeo$ are denoted with a circle (except for run b40) and are connected with dashed lines; the blue dashed line is the average measured heat flux across the runs.}
	\label{fig:IHF}
\end{figure}

In figure~\ref{fig:IHF}, we plot normalized heat fluxes as a function of $\betaeo\rhoeo/L_T$: in blue we show the box-averaged heat flux measured in the $\gradpz$ runs; and in orange, red, purple, and cyan we show the normalized heat flux implied, respectively, by advection at the whistler phase speed, by the quasi-linear operator, by the Fokker--Planck operator, and by the model from \citet{Drake2021} (discussed in \S\ref{sec:model_CO_comp}). In black, green, and gray we show, respectively, heat fluxes implied by our model \eqref{eqn:nu_model_all}, the same model but where we neglect the heat flux contribution of electrons expected to be trapped (see \S\ref{sec:evidence_trapped_electrons} for evidence of trapped electrons in one of our runs, as well as \S\ref{sec:q_passing} for
heat flux calculations), and our semi-empirical model \eqref{eqn:nuse}.
Points marked by an `x' are runs with $\betaeo=40$ and points marked by circles are runs with $L_T=125\rhoeo$; the former are connected with a dotted line and the latter are connected with a dashed line. Run b40, which belongs to both groups, is denoted with an `x'. The blue dashed line is the average of the measured heat flux across all runs. We find that the heat flux implied by advection is on average half of the total measured heat flux, suggesting that the advection and pitch-angle diffusion contribute to the heat flux equally.

Overall, we find that the heat flux implied by our Fokker--Planck operator agrees well with the heat flux measured in our simulations. While there is some deviation of the former from the latter, particularly for $\betaeo\rhoeo/L_T=0.04-0.1$, the deviations appear to be random, i.e. not a function of scale separation. We therefore associate these fluctuations with some error in our Fokker--Planck measurement process, perhaps due to our choice of jump moment or limited velocity-space resolution. In contrast, the heat fluxes implied by both our quasi-linear operator and basic model generally increase with $\betaeo$, as evidenced by the increasing trend of $(q_{\parallel {\rm QL}})$ and $(q_{\parallel {\rm model}})$ for runs with $L_T/\rhoeo=250$ as a function of $\betaeo\rhoeo/L_T$ (all runs with $L_T/\rhoeo=250$ are denoted by dots in figure~\ref{fig:IHF} except for run b40, which is denoted by an `x' at $\betaeo\rhoeo/L_T=0.16$). We attribute this trend to transit-time damping becoming weaker at larger $\betaeo$. For $n=0$, the resonant parallel wavenumber is proportional to $\beta_ev_\parallel/\vthe$ \eqref{eqn:resonant_delta}. At a given parallel velocity, therefore, the resonant $k_\parallel \rho_e$ increases proportionally to $\beta_e$. The result, given the steep spectrum we observe, is that TTD is only active over an ever-narrower region of phase space close to $v_\parallel=0$ as $\beta_e$ increases, the result being an increasingly larger heat flux.

While $q_{\parallel {\rm QL}}$ and $q_{\parallel {\rm model}}$ exhibit the same qualitative trend due to their similar construction, the two imply very different levels of heat flux. In both models, electrons scatter resonantly off magnetic-field fluctuations whose amplitudes scale proportionally to some negative power of $\kpar\rho_e$, resulting in scattering frequencies that decrease rapidly as $|\vpar/\vthe|$ decreases from $|\vpar/\vthe|\sim 1$. Physically, low-$\vpar$ electrons interact via exact resonance only with high-$k_\parallel$ waves, which do not have appreciable power; this implies a large perturbation to the distribution function in that region of velocity space, which in turn contributes to a large heat flux. The critical difference between our quasi-linear operator and our model is that our basic model spectrum does not include the noise floor (due to finite number of PIC particles) present in the actual simulated spectra used to calculate the quasi-linear operator. For a direct comparison of these spectra for run b40, see figure~\ref{fig:b40_psink}; the spectrum measured from our simulations levels out for $\kpar\rho_e\gtrsim 10$ while the model spectrum continues toward $0$. We can therefore expect noise to contribute to the cyclotron resonances of our quasi-linear model, at least in run b40, when $|\vpar/\vthe|\lesssim 0.1$. Referring to parallel-velocity-averages of the quasi-linear operator shown in figure~\ref{fig:QL_vpar}, $|\vpar/\vthe|\sim 0.1$ appears to be a plausible estimate for noise to affect the scattering rate; however, transit-time damping complicates this picture somewhat. The only outlier in figure~\ref{fig:QL_vpar} is run b100, which has the weakest transit-time damping and PIC noise relative to fluctuation amplitude, resulting in the high implied heat flux in figure~\ref{fig:IHF}. Even though our quasi-linear operator does imply heat fluxes close to our simulation measurements for $\betaeo\rhoeo/L_T\lesssim 0.1$, the observed scaling of the implied heat flux with $\beta_e$ attributed to transit-time damping and our difficulties in disambiguating it from the effects of cyclotron scattering and PIC noise call into question the applicability of the basic model to explain the saturated HWI.

There is evidently a physical reality to the Fokker--Planck results that is not captured by the quasi-linear numerical result and our model. The incompatible scaling of transit-time damping rate with $\beta_e$ found from our numerical quasi-linear and simple models also strongly suggest this to be the case. In other words, there must be something scattering electrons with $\vpar/\vthe\ll 1$ that is not captured by resonant physics. This is a manifestation of perhaps the longest-standing issue in quasi-linear theory called the $90$-degree scattering problem; we discuss this issue in detail in \S\ref{sec:scattering_gap}. One physical effect that can alleviate the $90$-degree problem is trapping in large-amplitude waves; see \S\ref{sec:evidence_trapped_electrons} for evidence of trapped electrons in our runs. If we treat these trapped electrons as if they have no contribution to the perturbed distribution function, and therefore to the model diffusive heat flux (see \S\ref{sec:q_passing} for details), then the model produces implied heat fluxes, labeled $q_{\parallel{\rm model,}{\xicrit}}$ and shown in green in figure~\ref{fig:IHF}, which are less than the whistler advection value, i.e. the diffusive heat flux contributions are \textit{negative}. We argue in \S\ref{sec:q_passing} that this is a specific consequence of our model operator, as the same methodology applied to other operators produce reasonable results. It is likely that a more self-consistent treatment of the trapped population is required. In addition to this model where trapped electrons have no effect on $f_{e1}$, we also consider a model where electrons have a minimal scattering rate which is agnostic of a physical explanation and purely motivated by our Fokker--Plank scattering frequency. We call this a semi-empirical model, and cover it in detail in \S\ref{sec:semi-empirical}. The heat flux implied by the model is plotted in figure~\ref{fig:IHF} with gray markers and labeled by $q_{\parallel{\rm model,SE}}$. This model matches simulation results well; however, the lack of a physics-based explanation for the scattering floor should give pause to a reader who wishes extrapolate this model to astrophysical scale separation.
Finally, another way around the $90$-degree scattering problem is to ignore the pitch-angle-dependence of the scattering frequency entirely, as in the model from \citet{Drake2021}, which we discuss in \S\ref{sec:model_CO_comp}. As shown in figure~\ref{fig:IHF} by the cyan markers, this model does reproduce the proper heat flux scaling and is reasonably consistent with our measurements from simulation. However, there are concerns with this operator in addition to the omission of any pitch-angle-dependence of the scattering frequency; we refer the reader to \S\ref{sec:model_CO_comp} for details on these concerns.

\subsection{Scattering across the gap}\label{sec:scattering_gap}

As discussed at the end of \S\ref{sec:IHF}, the heat flux implied by our model scattering frequency \eqref{eqn:nu_model_all} is unphysically large because the scattering rate approaches $0$ geometrically as ${|\vpar/\vthe|\rightarrow 0}$. This is a variation on the well-known `$90$-degree problem', which also afflicts theories of cosmic-ray transport in the Galactic and intracluster contexts. Just as in our formulation, the scattering rate for cosmic rays approaches $0$ for small parallel velocities because the limited power in resonant (in this case, Alfv{\'e}n) waves. It is argued that some nonlinear mechanism is in fact required to scatter cosmic rays across $\vpar=0$ and isotropize the distribution \citep[e.g.,][]{Salchi_2009,Holcomb_2019}. Options proposed in the literature include adiabatic mirroring \citep[i.e., trapping;][]{Jones1978,Felice_Kulsrud_2001} and resonance broadening \citep{Dupree1966}, among others. We see strong evidence for such trapping or mirroring in our simulations, which we detail in \S\ref{sec:evidence_trapped_electrons}. We then continue with a discussion of the repercussions of limiting our heat flux to only passing electrons (\S\ref{sec:q_passing}). Finally, we end \S\ref{sec:scattering_gap} with a discussion of how to improve our resonant scattering frequency calculations in \S\ref{sec:refined_scattering} and a discussion of resonance broadening in \S\ref{sec:resonance_broadening}.

\subsubsection{Evidence for trapped electrons}\label{sec:evidence_trapped_electrons}

\begin{figure}
	\centering
	\includegraphics[width=\textwidth]{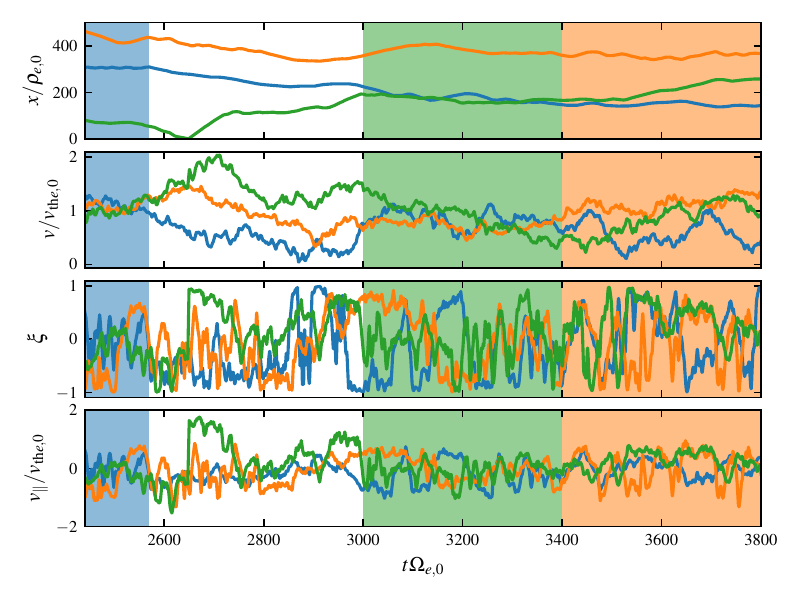}
	\caption{Parallel position (top), velocity (second from top), pitch angle (second from bottom), and parallel velocity (bottom) for three electrons from run~b40x4 that exhibit periods of extended wave trapping, denoted by the background of the corresponding colour. }
	\label{fig:e_trapped}
\end{figure}

One avenue to reduce the large contribution of low-pitch-angle electrons to the heat flux is to treat them as trapped particles that do not participate in diffusive transport. In figure~\ref{fig:e_trapped}, we show various quantities versus time for three selected electrons (blue, orange, and green) from run~b40x4 that exhibit periods of trapping (denoted by a background of the corresponding colour). We select time periods representative of trapping by eye, looking for large-amplitude, quasi-periodic oscillations in $\xi$. This method is by no means perfect, yet we can be confident in stating the selected electrons are largely trapped, rather than scattered: the tracks of velocity, which we have shown to be consistent with an Ornstein--Uhlenbeck process in \S\ref{sec:FP_v}, are clearly more reminiscent of a random walk than the tracks of $\xi(t)$ and $\vpar(t)$, which exhibit periods of coherent, quasi-periodic oscillations about $0$. By sampling and averaging the period of the largest-amplitude oscillations, we estimate that $\wb/\wceo\sim 0.2$. This value is approximately half what results from a na\"{i}ve calculation using \eqref{eqn:wb} after assuming that electrons with a thermal perpendicular velocity ($\vperp/\vthe\sim 1$) are trapped by $\kpar\rho_e\sim 0.5$ fluctuations with rms amplitude $\avg{\delta B^2}/B_0^2$. Our simple expression for the bounce frequency of an electron in a single mode, equation~\eqref{eqn:wb}, can therefore be reasonably applied to our turbulent simulations. It is more difficult to say the same of our expression for the trapped-passing boundary \eqref{eqn:xicrit}. For run~b40x4, $\xicrit\lesssim 0.5$; our tracked electrons in figure~\ref{fig:e_trapped} show trapped oscillations largely inside the boundary, but the electrons do occasionally veer outside it. That said, what really matters is how the boundary scales with fluctuation amplitude -- of which we can be confident -- and the conclusions of \S\ref{sec:q_passing} stand regardless of its precise value.

\subsubsection{Heat-flux contribution from passing electrons only}\label{sec:q_passing}

As an improvement to our model, we allow only passing particles to contribute to the diffusive part of the heat flux. The trapping/passing boundary occurs at a critical pitch angle $\xicrit$ given by
\begin{equation}\label{eqn:xicrit}
\xicrit\simeq \sqrt{\frac{\delta B/B_0}{1+\delta B/B_0}} ,
\end{equation}
where we have taken $\delta B/B_0=\sqrt{2}\sqrt{0.1\avg{\delta B^2}/B_0^2}$, i.e., the root mean square of the box-averaged fluctuation energy that is sampled by electrons at the $n=0$ resonance. For trapped electrons to have no contribution to the heat flux, their impact on the perturbed electron distribution function $f_{e1}$ must be neglected. We accomplish this by enforcing $f_{e1}={\rm const}$ for $|\xi|\le \xicrit$, requiring $f_{e1}$ to be continuous across the trapped-passing boundary $\xi=\pm\xicrit$ and choose ${\rm const}$ to satisfy the solvability conditions \eqref{eqn:f1_constraints}. We construct $f_{e1}$ subject to these conditions by simply altering the collision frequency to be infinite for trapped electrons, i.e.,
\begin{equation}\label{eqn:nuxicrit}
    \nu_{\xicrit}(v,\xi)=\begin{cases}
        \numodel(v,\xi), & (|\xi|> \xicrit)\\
        \infty, & (|\xi|\le \xicrit)
    \end{cases}.
\end{equation}
Our approach here is similar to that of \citet{Felice_Kulsrud_2001}, but much more simple. In that work, the authors find that adiabatic mirroring regularizes the singularity in $f_1$, effectively setting the distribution to a constant for trapped particles. Their treatment included a self-consistent solution to match the distribution function between the quasi-linear and adiabatic regions; our inner and outer solutions match automatically by construction of \eqref{eqn:nuxicrit}.

We evaluate the heat flux implied by \eqref{eqn:nuxicrit} according to \eqref{eqn:HF_integral} and plot the result in figure~\ref{fig:IHF} using green markers. Note that the heat flux implied by this method is at or below the results for $q_{v_{\rm w}}$, indicating that the diffusive contribution to the heat flux is zero or negative. The reason the diffusive heat flux is negative comes down to how setting $f_{e1}=\const$ inside $|\xi|<\xicrit$ affects the integrand of \eqref{eqn:HF_integral}. Without any modification to $f_{e1}$ calculated using \eqref{eqn:nu_model}, the integrand is dominated by values where $1/\nu$ is the largest; these regions are positive, hence the positive heat flux. When we set $f_{e1}=\const$ for $|\xi|<\xicrit$, the regions of largest $1/\nu$ no longer to contribute to the integrand; the positive contribution is therefore suppressed relative to the negative one and the overall diffusive heat flux becomes negative. Perhaps a more self-consistent treatment like \citet{Felice_Kulsrud_2001} is required when applying this method, as it clearly does not produce physical results for all physically motivated collision frequencies.

\subsubsection{Refined resonance condition at $|\vpar/v_{{\rm th}e}|\ll 1$}\label{sec:refined_scattering}
With the cyclotron resonance condition \eqref{eqn:kpar_res_n} requiring $\kpar\rho_e\rightarrow \infty$ as $|\vpar/\vthe|\rightarrow 0$, it ought to be checked whether or not the dispersion relation used to derive this result is in fact still appropriate as $\kpar\rho_e\rightarrow \infty$. As it turns out, \eqref{eqn:specific_res_condition} breaks down for ${|\vpar/\vthe|\lesssim \beta_e^{-1/2}}$. The breakdown occurs for two reasons: firstly, the term under the square root in \eqref{eqn:res_general_solution} is no longer small; secondly, and most importantly, $\kpar\rho_e\rightarrow \beta_e^{1/2}$, in which case $\omega(\kpar\rho_e)\rightarrow \wce$ under \eqref{eqn:cpwdr}. It is clear from \eqref{eqn:res_condition} that $\omega\rightarrow\wce$ as $|\vpar/\vthe|\rightarrow 0$, even without any assumptions on the size of $\beta_e$. However, the dispersion relation \eqref{eqn:cpwdr} used in \eqref{eqn:specific_res_condition} is only valid for $\kpar\rho_e\ll\beta_e^{1/2}$. One could use the full parallel cold-plasma dispersion relation \citep{Stix1992},
\begin{equation}\label{eqn:full_cpwdr}
    \frac{\omega(\kpar\rho_e)}{\wce}=\frac{(\kpar\rho_e)^2/\beta_e}{1+(\kpar\rho_e)^2/\beta_e},
\end{equation}
which would lead to a different resonance condition for $|\vpar/\vthe|\le \beta_e^{-1/2}$. However, even \eqref{eqn:full_cpwdr} is liable to be wrong in detail due to thermal corrections. Thus, a detailed understanding of the dispersion relation is necessary to understand the resonant wave vector at small parallel velocities.

\subsubsection{Resonance broadening}\label{sec:resonance_broadening}
One additional possibility for whistler waves to scatter electrons self-consistently across the gap is resonance broadening. The delta function that features in the quasi-linear operator is customarily replaced with a peaked function having finite width \citep{Berk1995}. Resonance broadening can result from particle trapping or deviation from linear orbits resulting from statistical variation in the integrated trajectories when calculating the quasi-linear operator. In the former case, resonance broadening function is often taken to be a table-top distribution of width $\Delta \Omega=4\wb$ \citep{Karimabadi1992}, where \citep{Roberg-Clark2016,Cai2020}
\begin{equation}\label{eqn:wb}
\frac{\wb(k\rho_e)}{\Omega_e}=\sqrt{k\rho_e\frac{\vperp}{\vthe}\frac{\delta B(k\rho_e)}{B_0}}.
\end{equation}
is the bounce frequency for a deeply trapped particle under the pendulum approximation with perpendicular velocity $v_\perp$ bouncing in a magnetic mirror of amplitude $\delta B$ and wavneumber $k$, {\em viz.}
\begin{equation}\label{eqn:R_K92}
R_{\rm K92}(x/\wce,\wb/\wce)=
\begin{cases}
\dfrac{\wce}{4\wb}, &|x|\le 2\wb\\
0, &|x|>2\wb
\end{cases}.
\end{equation}
Here, $x=\omega(\kpar)-\kpar\vpar+n\wce$ is the exact linear resonance condition. A recent numerical study of resonance broadening for a single wave mode \citep{Meng2018} has verified that the broadening width $\Delta \Omega=4\wb$ is valid at small wave amplitudes.

In the case of resonance broadening by statistical variation in the linear particle orbits, one obtains a resonance function \citep{Dupree1966}
\begin{equation}\label{eqn:R_D}
    R_{\rm D}=
    \frac{1}{\upi}\int_0^\infty\rmd\tau\,
    \rme^{\imag(\omega(\kpar)-\kpar\avg{\vpar}
    -n\Omega_e)\tau-\frac{1}{3}\kpar^2\nu(\vpar,\vperp)\vthe^2\tau^3}.
\end{equation}
Because the scattering frequency appears as a parameter in \eqref{eqn:R_D}, one is left with an implicit expression for $\nu$ when solving an expression like \eqref{eqn:nu_model} with the delta function replaced with \eqref{eqn:R_D}. In either the case of \eqref{eqn:R_K92} or \eqref{eqn:R_D}, because the scattering frequency or fluctuation energy controls the width of the broadened resonance, the resulting scattering frequency does not in general scale with the energy in the fluctuations. The heat flux implied by a resonance-broadened operator calculated from the magnetic energy spectrum observed in our simulations, therefore, does not equal the HWI threshold heat flux, except perhaps in special circumstances, casting doubt on whether current theories of resonance broadening can explain our results. This is not to say, however, that there could not be a physically relevant regime in which resonance broadening might control HWI saturation and in which the scattering rate scales non-diffusively with the magnetic energy -- we just have yet to see it.

Closely related to, and intertwined with the concept of resonance broadening, is the concept of resonance overlap. For a discrete wave mode of infinitesimal amplitude, resonant diffusion occurs along quasi-linear contours in velocity space within an infinitesimal neighborhood of the resonant parallel velocity. As the wave amplitude is increased, particles within a finite neighborhood of the resonance become trapped. Resonances are broadened by the trapping according to, e.g., equation~\eqref{eqn:R_K92} with \eqref{eqn:wb}, leading to diffusion with finite extent along the quasi-linear contours \citep{Karimabadi1992}. Particles can therefore only diffuse across discrete wave modes when the wave amplitudes are large enough that their resonances overlap \citep{Chirikov_1960}. Earlier work on the HWI by \citet{Roberg-Clark2016} concluded that the overlap of Landau ($n=0$) and cyclotron ($n=\pm 1$) resonances was necessary to explain the saturation of the HWI in 2D. Overlap was found to occur for wave amplitudes $\delta B^2/B_0^2\gtrsim 0.09$; all of our simulations  saturate in this regime. Whether or not the HWI saturates in the way we have observed in this work at wave amplitudes below this overlap threshold remains to be seen. Our analysis, however, suggests a more strict condition for HWI saturation than \citet{Roberg-Clark2016}, namely, that resonances not only need to overlap, but the scattering frequency where they overlap must also scale like the marginal stability criterion \eqref{eqn:nueff_sat}. We present a model that satisfies this condition in \S\ref{sec:semi-empirical}.

\subsubsection{Semi-empirical scattering frequency}\label{sec:semi-empirical}
While we are so far unable to affix an analytical theory to our quasi-linear model that satisfactorily scatters electrons across the $|\vpar/\vthe|\ll 1$ gap, our results clearly indicate that they are. Not only do all of our simulations saturate at the threshold heat flux (\S\ref{sec:sat_scaling}), but our Fokker--Planck analysis (\S\ref{sec:FP_xi}) directly reports a finite pitch-angle scattering rate across the gap that is also consistent with the threshold heat flux (\S\ref{sec:IHF}). Motivated by these observations and eschewing any detailed theoretical arguments, we propose a semi-empirical model that is identical to \eqref{eqn:nu_model_all} except that the scattering frequency \eqref{eqn:nu_model} has a floor that scales with $\beta_e\rho_e/L_T$, i.e.
\begin{equation}\label{eqn:nuse}
    \frac{\nu_{{\rm SE}}(v,\xi)}{\wce}=\max \bigg[\frac{\numodel(v,\xi)}{\wce}, \frac{1}{2}\frac{\beta_e\rho_e}{L_T}\bigg].
\end{equation}
The heat fluxes implied by \eqref{eqn:nuse} are shown by the gray dots in figure~\ref{fig:IHF} and closely agree with the heat fluxes measured in our simulations. While we cannot guarantee the asymptotic relevance of this operator, it is at least consistent out to the scale separations we were able to simulate. One can also argue that the scattering rate must be this in order to  saturate self-consistently under the diffusive scaling predicted by theory and observed in this paper. The fact that our simulations have extended out to larger scale separations numerically than previously published ones and yet the diffusive scaling still holds lends us a certain degree of confidence that the scaling might continue to astrophysical scale separations. We discuss the possibility that the asymptotic HWI does not saturate in this manner in \S\ref{sec:alt_transport}.

\subsection{Alternative transport regimes}\label{sec:alt_transport}
In the ICM, the saturated field amplitude $\delta B/B_0$ is expected to be ${\sim}3\times 10^{-7}$ at the largest temperature gradient length scales, leaving $\wb/\wce\sim 5\times 10^{-4}$ and $\xicrit\simeq 3\times 10^{-7}$. Thus, the effect of trapping is expected to be small. In reality, there are Coulomb collisions in the ICM that scatter electrons at a rate ${\sim}\nu_{ei}$, which may help regulate the heat flux. Let us explore this possibility. Coulomb collisions in the ICM correspond to a mean free path
\begin{equation}\label{eqn:ICM_mfpe}
    \lambda_{e,{\rm ICM}}\simeq 680~ \bigg(\frac{T_e}{6~{\rm keV}}\bigg)^2\bigg(\frac{10^{-2}~{\rm cm}^{-3}}{n_e}\bigg)~{\rm pc}.
\end{equation}
For temperature-gradient length scales shorter than $\beta_e\lambda_{e,{\rm ICM}}$, the Coulomb-collisional heat flux will exceed the HWI threshold and the instability will be active. If the HWI is only able to scatter resonant electrons, then there will be a range of $|\vpar/\vthe|\ll 1$ where scattering is dominated by the Coulomb collisions. However, by definition, the Coulomb scattering frequency is not fast enough to stabilize the HWI, so it is unlikely that background Coulomb collisions will cause the HWI to saturate in the manner described and observed in this work. Except perhaps in the case of a specific spectral index for magnetic field fluctuations or for temperature-gradient length scales near the Coulomb mean free path, the HWI fluctuations must continue to grow until the heat flux reaches the marginally stable value.

Another possibility is that large-scale turbulence traps electrons and isotropizes the distribution at small parallel velocities \citep[for a discussion of this effect in the cosmic-ray literature, see, e.g.][]{Yan_2008,Xu_2018}. Assuming the wavelengths of the turbulent fluctuations are significantly smaller than the temperature-gradient length scale, the problem here is in principle the same as for Coulomb collisions. Either the trapping is strong enough to suppress the heat flux below the HWI threshold, resulting in no instability, or the HWI is active and, if only able to scatter resonantly, must grow until the heat flux is regulated to the threshold. Whether it is realistic to expect the system to adapt to some hybrid state with Coulomb collisions or large-scale turbulence -- and even what such a state would look like -- remains to be seen. It is worth noting that our PIC simulations do have noise that is analogous to Coulomb collisions resulting from a finite number of PIC particles, which clearly affected the results of our quasi-linear model that was computed from the magnetic spectrum measured in our simulations. When we introduced a quasi-linear model derived from a model spectrum that does not incorporate PIC noise, the heat flux implied was significantly larger than that implied by the operator using the measured spectrum. However, because the PIC noise does not scale proportionally to $\beta_e\rho_e/L_T$, the quasi-linear model incorporating the noise did not imply heat fluxes that were consistent with HWI saturation.

\subsection{Comparison to model in Drake et al. 2021}\label{sec:model_CO_comp}
\citet{Drake2021} proposed a model collision operator for saturated for HWI similar to the one presented in \S\ref{sec:model_CO}, with the same differential form of the operator and a scattering frequency
\begin{equation}\label{eqn:nu_Drake}
    \nu_{\rm Drake21} = 0.1\wce \bigg(\frac{\delta B}{B_0}\bigg)^2\bigg(\frac{v}{\vthe}\bigg)^{4/3}.
\end{equation}
The power law index is similar to the pitch-angle-average results from \S\ref{sec:QL_res} and \S\ref{sec:FP_xi} for $v/\vthe\gtrsim 1$; however, this result is a coincidence. \citet{Drake2021} argues for a $4/3$ power law by invoking quasi-linear scattering off of an assumed $k^{-7/3}$ spectrum of electron-MHD turbulence; our results are direct calculations from the simulation. The model \eqref{eqn:nu_Drake} also contains no $\xi$-dependence, as \citet{Drake2021} argued the whistlers `should be able to scatter through the whole range of pitch angle\dots because of the multiple anomalous resonances' under a broad spectrum of oblique waves. While we have found that the anomalous cyclotron resonances play a significant part in the saturation of the instability, their contribution to the scattering rate -- according to quasi-linear theory and our Fokker--Planck measurements at the largest $L_T/\rhoeo$ -- is a factor ${\sim}10$ smaller than from the usual resonances. We also find that the `anomalous' resonances on their own cannot be responsible for scattering electrons across all pitch angles because of the issues mentioned in \S\ref{sec:scattering_gap}.

\section{Conclusion}\label{sec:conclusion}

Making use of an ensemble of electromagnetic PIC simulations across different scale separations $L_T/\rho_e$ and plasma $\beta_e$ parameters, we have demonstrated the ability of the HWI to limit the parallel electron heat flux to quasi-linear threshold levels ${\propto}\beta^{-1}_e$ in high-$\beta$, weakly collisional, magnetized plasma. We also show this result to be robust with respect to the initial hydrostatic equilibrium. To investigate in detail how the saturated state of the HWI regulates the heat flux, we have utilized three different methods to quantitatively characterize the effects of the HWI fluctuations on the velocity-space dynamics of the electrons.

The first method assumes that whistlers pitch-angle scatter electrons in a frame traveling with the whistler phase velocity and leverages a Chapman--Enskog expansion to solve for the scattering frequency given the electron distribution function measured in the simulations. Although we found this method to be too noisy to resolve the full pitch-angle dependence of the scattering frequency, we were able to resolve the speed dependence of its pitch-angle average. The second method is derived from the electromagnetic quasi-linear operator, which takes as its input the spectrum of magnetic-field fluctuations measured in our simulations. The resulting operator was rigorously shown to converge -- in the limit of slow whistler phase velocity -- to a pitch-angle scattering operator expressed in the frame of the phase velocity. This operator makes evident the previously known fact that oblique whistlers are necessary for heat-flux-limiting interactions to occur with electrons that travel down the temperature gradient. Using this operator, we also showed that the implied scattering frequency of these electrons is an order of magnitude less than for electrons that travel up the temperature gradient. This asymmetry in the scattering frequency directly follows from the ratio of energies in the left- and right-handed components of oblique whistler waves, which resonate with electrons moving down and up the temperature gradient, respectively. The third and final method relied on the calculation of Fokker--Planck jump moments in velocity space from tracked particle data. We showed not only how such jump moments are to be obtained in curvilinear coordinates, but also how to obtain drag and diffusion coefficients from said jump moments by using a jump interval that is self-consistently obtained from the particle statistics.

The pitch-angle scattering frequency from all of our methods show two power laws in $v$: one decreasing as a function of $v$ for subthermal electrons, due to transit-time damping at $|\vpar/\vthe|\ll 1$; and the other increasing for superthermal electrons, due to cyclotron-resonant scattering at $|\vpar/\vthe|\sim 1$. Unfortunately, the pitch-angle dependence of the method leveraging the Chapman--Enskog expansion was unable to be resolved due to finite-particle-number noise, and our quasi-linear and Fokker--Planck methods produced scattering frequencies that differed significantly from one another. The quasi-linear scattering frequency exhibits steep power laws in $|\vpar/\vthe|$, which results from the steep spectral slope of the magnetic energy. The Fokker--Planck scattering frequency is flat in comparison, only showing slight asymmetry in pitch-angle at the smallest $\betaeo\rhoeo/L_T$. The heat flux implied by the Fokker--Planck operator matches well with our simulations, although there is some random variation in the result that suggests our method could be improved. The quasi-linear operator, however, implies a heat flux that varies as a function of plasma $\beta$ in a way that is inconsistent with the HWI marginal stability criterion, due to transit-time damping for electrons with $|\vpar|/\vthe\ll 1$. We also introduce a simple model, which is motivated by the quasi-linear model but includes a model whistler spectrum. Our simple model implies heat fluxes which are qualitatively similar to the quasi-linear model, except that the implied heat flux is even higher in our simple model. Because a diffusive heat flux is proportional to the inverse of the scattering frequency, it is therefore sensitive to where $\nu$ is the smallest. The increased heat flux implied by the simple model suggests that the numerical quasi-linear result is sensitive to the level of PIC noise in the runs, which is not incorporated into our model spectrum. Even though our numerical quasi-linear operator does imply heat fluxes that are consistent with simulations producing small fluctuation amplitudes, these issues call into question whether a resonant quasi-linear approach can explain physically the saturation of the HWI.

Ultimately, though, it is the quasi-linear cyclotron resonance that explains why the marginal whistler scattering rate $\nu/\Omega_e\sim\beta_e\rho_e/L_T$ is also proportional to the square of the whistler fluctuation amplitude, i.e. $\nu/\Omega_e\sim \langle \delta B^2\rangle/B_0^2$, which we clearly show to be the case in our simulations. But there's the rub: electrons with $|\vpar/\vthe|\ll 1$ are cyclotron-resonant with $\kpar\rho_e\gg 1$ whistler waves, which have vanishingly little power at high parallel wavenumbers. The problem of how exactly these $|\vpar/\vthe|\ll 1$ electrons are scattered in a way that is consistent with the simple cyclotron-resonant quasilinear picture of the HWI is a vexing one. This is a variation on the well-known $90$-degree scattering problem in the cosmic-ray community, albeit with tighter constraints than is usually found in the literature. We commented on various ways to improve the quasi-linear scattering rate at low $|\vpar/\vthe|$, including using a more accurate treatment of the dispersion relation and incorporating various schemes for resonance broadening. The former would require an in-depth method to solve the plasma dispersion function for our distribution function and is beyond the scope of this work; the latter runs into the problem that resonances are broadened in a way that depends on the turbulent fluctuation amplitude. While detailed calculations of resonance-broadened scattering frequencies also lie beyond the scope of this work, we argue that, in general, such scattering frequencies will not be proportional to $\langle \delta B^2\rangle/B_0^2$ and will violate the basic picture of the HWI to which we are constrained by simulation results.

The best model we can currently produce is a simple quasi-linear scattering frequency derived from a model whistler spectrum, but with an empirically motivated floor that is proportional to the marginal whistler scattering frequency ${\propto}\beta_e\rho_e/L_T$. It is no surprise that this model does imply a heat flux consistent with our simulations; however, we stress that we do not have an explanation for the physical process that underlies such a floor, other than to say that whatever that process is, it must result in the same scaling. This conclusion limits the possibility that either Coulomb collisions or large-scale turbulence could play a role in HWI saturation. For either process to help saturate the instability effectively, they would need to scatter electrons at or near the instability threshold value; but, this also means that the plasma would be stable to the HWI in the first place and so the scheme is not logically consistent. Clearly, for the HWI to regulate the electron heat flux self-consistently, therefore, the effective collision operator must look closer to our relatively flat Fokker--Planck operator or to the model from \citet{Drake2021}. Unfortunately, a rigorous statement as to how this operator works in detail is beyond the scope of the current work.

While our three methods to obtain an effective HWI collision operator did not converge numerically to an unambiguous result, we hope the procedures outlined in this paper find utility in characterizing the phase-space dynamics of other kinetic instabilities in future studies. In fact, a concurrent work by some of the authors uses similar techniques to this work in the characterization of a pressure-anisotropy-driven firehose instability in hybrid-PIC simulations. That work saw more success in particular with the method that leverages the Chapman--Enskog expansion, owing to a much-better-resolved particle distribution function. That study also points out limitations with the Fokker--Planck method in regards to the locality of individual particle jumps; a curious reader is encouraged to refer to Bott {\em et al.} (2024, in preparation) for an in-depth discussion. We also plan to apply many of the lessons learned here in the study of a corresponding ion-heat-flux instability, discovered by \citet{Bott_2024}, using hybrid-PIC simulations. The utility of these methods to obtain effective collision operators lies in their distillation of complex kinetic physics into relatively simple expressions, which can then be used in a small-parameter expansion for an even simpler fluid-like closure of the kinetic equations. Such a closure would be much less computationally expensive to simulate and could be incorporated into Braginskii-MHD simulations of the ICM \citep[e.g.,][]{Berlok2021,Perrone2024} and simple models of the solar wind or low-luminosity black-hole accretion flows.

\section*{Acknowledgements}
The authors thank K.~Klein, E.~Quataert, F.~Rincon, A.~Schekochihin, A.~Vanthieghem, the participants of the 13th and 14th Plasma Kinetics Working Meetings at the Wolfgang Pauli Institute in Vienna, and especially B.~Chandran for many insightful conversations and suggestions that improved the approaches and analyses presented in this work. M.W.K.~additionally thanks the Institut de Plan\'etologie et d'Astrophysique de Grenoble (IPAG) for its hospitality and visitor support while this work was being completed.

This paper was prepared alongside the first author's PhD dissertation \citep{Yerger_2023}. Barring minor additions and changes, sections \ref{sec:background} through \ref{sec:results2} and Appendices \ref{apx:CEO} and \ref{apx:QL} are generally as they appear in \citet{Yerger_2023}, including the figures. Sections \ref{sec:intro}, \ref{sec:results3}, and \ref{sec:conclusion} contain passages and figures similar to \citet{Yerger_2023}, but have been reworked. The main conclusions and discussion contained within \S\ref{sec:results3}, however, are different from those in the dissertation.

\section*{Funding}
Support for E.L.Y.~and M.W.K.~was provided by the National Aeronautics and Space Administration (NASA) under grant number NNX17AK63G issued through the Astrophysics Theory Program and under Chandra award number G08-19110C issued by the Chandra X-ray Center, which is operated by the Smithsonian Astrophysical Observatory for and on behalf of NASA under contract NAS8-03060. Additional support was provided by U.S.~Department of Energy (DOE) Contract No.~DE-AC02-09CH11466 and NASA grants NNN06AA01C and 80NSSC24K0171. A.F.A.B.~was supported by DOE awards DE-SC0019046 and DE-SC0019047 through the NSF/DOE Partnership in Basic Plasma Science and Engineering and also by the UKRI (grant number MR/W006723/1). This research was also facilitated by Multimessenger Plasma Physics Center (MPPC), NSF grant PHY-2206607. High-performance computing resources supporting this work were provided by: the NASA High-End Computing (HEC) Program through the NASA Advanced Supercomputing (NAS) Division at Ames Research Center, and the Texas Advanced Computing Center (TACC) at The University of Texas at Austin under Stampede2 allocation TG-AST160068 and Frontera allocation AST20010. This work used the Extreme Science and Engineering Discovery Environment (XSEDE), which was supported by NSF grant number ACI-1548562. It also made extensive use of the {\em Perseus} and {\em Stellar} clusters at the PICSciE-OIT TIGRESS High Performance Computing Center and Visualization Laboratory at Princeton University.

\section*{Declarations of interests}
The authors report no conflict of interest.

\appendix

\section{Derivation of the Chapman--Enskog operator}\label{apx:CEO}

In this appendix, we detail the derivation of the Chapman--Enskog operator. This includes a statement of the ordering used to reduce the kinetic equation and a brief derivation of the resulting correction equation that determines the perturbed electron distribution function $f_{e1}$ in terms of the background distribution $f_{e0}$ and its phase-space gradients (\S\ref{apx:CEE}), the coordinate transform of the model collision operator to the frame of the whistler waves (\S\ref{apx:CEtransform}), and the solution of the correction equation for the gyro-averaged $f_{e1}$ (\S\ref{apx:CEsolution}).

\subsection{Chapman--Enskog ordering and the correction equation}\label{apx:CEE}

The Vlasov--Maxwell equation governing the evolution of the electron distribution function $f_e=f_e(t,\bb{r},\bb{w}\doteq\bb{v}-\bb{u}_e)$, written in the frame of the bulk electron velocity
\begin{equation}
	\bb{u}_e(\tr)\doteq\frac{1}{n_e(\tr)}\int \rmd^3\bb{v} \, \bb{v}f_e(\trv) ,
\end{equation}
is given by
\begin{equation}\label{eqn:vlasov-maxwell}
	\bigD{t}{f_e}+\bb{w}\bcdot \grad f_e -
	\bigg[\frac{e}{m_e}\bigg(\bb{E}+\frac{\bb{u}_e}{c}\btimes\bb{B}\bigg)+\bigD{t}{\bb{u}_e}+
	\bb{w}\bcdot\grad \bb{u}_e\bigg]
	\bcdot\pD{\bb{w}}{f_e}=	\Omega_e\pD{\phi}{f_e}+C[f_e] ,
\end{equation}
where ${\rm D}/{\rm D}t\doteq \partial/\partial t + \bb{u}_e\bcdot\grad$ is the material derivative and $\phi$ is the gyrophase. To tailor this equation for our problem, we introduce the dimensionless quantity
\begin{equation}
	\epsilon\sim\frac{\rho_e}{L_T}\sim \frac{\mfpe}{L_T} \sim \frac{u_e}{\vthe} \sim \frac{1}{\beta_e} \ll 1 ,
\end{equation}
which is designed to separate kinetic physics -- namely, Larmor gyrations occurring at frequency $\Omega_e=\rho_e/\vthe$ and particle scattering (whether Coulombic or anomalous) occurring at frequency $\nu=\mfpe/\vthe$ -- from any physics that occurs on the thermal crossing time $L_T/\vthe$ or the characteristic dynamical time scale $L_T/u_e$.\footnote{In the primary ordering used by \citet{Braginskii1965}, the plasma $\beta$ parameter and the Mach number relative to the ion thermal speed are both ordered unity. The electron bulk flow is then small relative to the electron thermal speed because S.I.~Braginskii also expands in the mass ratio $\sqrt{m_e/m_i}$, including this small number in the $\epsilon$ ordering. In our expansion, assuming that $\beta_e\gg 1$ is sufficient to guarantee that $u_e/\vthe\ll 1$, since $u_e \sim \vw \sim \vthe/\beta_e \ll \vthe$ for the whistler fluctuations with phase velocity $\vw$. This ordering is also consistent with our theoretical expectation (and numerical finding) that the effective mean free path is ${\sim}L_T/\beta$.} Note that spatial diffusion along the magnetic field occurs on a time scale ${\sim}L^2_T/(\vthe\mfpe)$ that is comparable to $L_T/u_e$. Expanding the electron distribution function in $\epsilon$ using
\begin{equation}\label{eqn:app:orderf}
	f_e=f_{e0}+\epsilon f_{e1} + \epsilon^2 f_{e2} + \dots
\end{equation}
and taking $\partial/\partial t \sim u_e/L_T \sim \epsilon \vthe/L_T \sim \epsilon^2 \Omega_e$, we can then determine $f_{e0}$ and then express $f_{en}$ in terms of the lower-order quantities $f_{e0},\,\dots,\,f_{e(n-1)}$ at progressively higher orders.

Equation~\eqref{eqn:vlasov-maxwell} at $O(\epsilon^0 f_{e0}\Omega_e)$ is
\begin{equation}\label{eqn:order0}
	0 = \Omega_e\pD{\phi}{f_{e0}}+C[f_{e0}] .
\end{equation}
The solution to this equation is a gyrotropic $f_{e0}$ that also annihilates the collision operator, $C[f_{e0}]=0$. For example, if the collision operator were to scatter only in pitch angle, then $f_{e0} = f_{e0}(w)$ would be a consistent solution of \eqref{eqn:order0}. If the collision operator were additionally required to satisfy Boltzmann's H theorem, then $f_{e0}$ would have the form of a Maxwellian. Regardless, in order for this solution to correspond to the `hydrodynamic' $f_{e0}$ that includes information only about the fluid quantities -- {\em viz.}, the number density, bulk velocity, and isotropic temperature of the electrons -- we must demand that $f_{e1}$ satisfies
\begin{equation*}
	\int \rmd^3\bb{w}\, (1,\bb{w},w^2)f_{e1}=0 ,
\end{equation*}
i.e., that it is purely kinetic. In what follows, we assume that the collision operator provides an isotropic $f_{e0}=f_{e0}(w)$.

At $O(\epsilon f_{e0}\Omega_e)$, equation~\eqref{eqn:vlasov-maxwell} provides the correction equation
\begin{equation}
	\bb{w}\bcdot\grad f_{e0}+\frac{\grad p_e}{m_en_e}\bcdot\frac{\bb{w}}{w}\D{w}{f_{e0}}=\Omega_e\pD{\phi}{f_{e1}}+C[f_{e1}] ,
\end{equation}
whose gyro-average is
\begin{equation}\label{eqn:app:correction}
    w_\parallel \nabla_\parallel f_{e0} + \frac{\nabla_\parallel p_e}{m_e n_e}\frac{w_\parallel}{w} \D{w}{f_{e0}} = C[\langle f_{e1}\rangle_\phi] .
\end{equation}
Provided we are able to invert the operator on the right-hand side, the solution to this equation provides $\langle f_{e1}\rangle_\phi$ in terms of $f_{e0}$ and its parallel gradients. From there, the corresponding parallel heat flux may be calculated. Alternatively, if $f_{e0}$ and $\langle f_{e1}\rangle_\phi$ are known, say, from a numerical simulation, the form of the collision operator might then be inferred. This latter option was the route taken in section~\ref{sec:CE}. In the remainder of this appendix, we provide additional calculations needed in that section to infer the effective collision operator of the HWI.

\subsection{Coordinate transformation for pitch-angle scattering operator in the whistler frame}\label{apx:CEtransform}

We assume that the unstable whistler waves act as sites of pitch-angle scattering in a frame moving with the whistler phase velocity $\bb{u}_e=\vw\eb$. The corresponding collision operator is then
\begin{equation}\label{eqn:app:pa_operator}
	\C[f_e]=\pD{\primed{\xi}}{} \biggl[ \frac{1-\xi^{\prime 2}}{2}\nu(\primed{w},\primed{\xi})\pD{\primed{\xi}}{f_e}\biggr],
\end{equation}
where the prime denotes quantities evaluated in the whistler frame and we have implicitly taken $f_e=f_e(w,\xi)$ to be gyrotropic. Writing $\primed{w}_\parallel = w'\xi' = w_\parallel - \vw$, we transform to the lab frame as follows. First, we expand $w'$ and $\xi'$ up to first order in $v_w/w$:
\begin{align*}
    w'^2 = (w_\parallel - \vw)^2 + w^2_\perp \quad\Longrightarrow\quad w' &= w\sqrt{1-2\xi \vw/w + (\vw/w)^2} \simeq w - \xi \vw , \\*
    \xi' &= \frac{w_\parallel - \vw}{w'} \simeq \xi - (1-\xi^2)\frac{\vw}{w} .
\end{align*}
The corresponding inverse transformations are
\[
w \simeq \primed{w}+\primed{\xi}\vw \quad{\rm and}\quad \xi \simeq \primed{\xi}+(1-\xi^{\prime 2})\frac{\vw}{\primed{w}} .
\]
The pitch-angle derivative in \eqref{eqn:app:pa_operator} is then
\begin{equation}\label{eqn:app:dxi}
\pD{\primed{\xi}}{} = \pD{\primed{\xi}}{\xi}\pD{\xi}{}+\pD{\primed{\xi}}{w}\pD{w}{} \simeq (1-2\xi \vw/w)\pD{\xi}{}+\vw\pD{w}{}.
\end{equation}
Using \eqref{eqn:app:dxi} in \eqref{eqn:app:pa_operator} then leads to
\begin{align}
    C[f_e] &\simeq \pD{\xi}{} \biggl[ \frac{1-\xi^2}{2} \nu(w',\xi') \biggl( \pD{\xi}{f_e} + \vw\pD{w}{f_e} \biggr) \biggr] + \frac{\vw}{w} \biggl( w \pD{w}{} - 2\xi\pD{\xi}{}\biggr) \biggl[ \frac{1-\xi^2}{2} \nu(w',\xi') \pD{\xi}{f_e} \biggr] ,
\end{align}
where
\begin{equation}
\begin{split}
\nu(\primed{w},\primed{\xi})&=\nu\big(w-\xi \vw,\xi-(1-\xi^2)\vw/w\big)\\
&\simeq \nu(w,\xi)-\xi \vw\pD{w}{\nu(w,\xi)}-(1-\xi^2)\frac{\vw}{w}\pD{\xi}{\nu(w,\xi)}.
\end{split}
\end{equation}
We now decompose $f_e$ according to \eqref{eqn:app:orderf} and use $C[f_{e0}]=0$ to find that
\begin{equation}\label{eqn:app:Cfe}
    C[f_e] \simeq \pD{\xi}{} \biggl[ \frac{1-\xi^2}{2} \nu(w,\xi) \biggl(\pD{\xi}{f_{e1}} + \vw \D{w}{f_{e0}}\biggr)\biggr] ,
\end{equation}
again dropping terms of order $(\vw/w)^2$.

\subsection{Chapman--Enskog expansion with the pitch-angle scattering operator \eqref{eqn:app:Cfe}}\label{apx:CEsolution}

Inserting \eqref{eqn:app:Cfe} into the correction equation~\eqref{eqn:app:correction} gives
\begin{equation}
    w_\parallel\nabla_\parallel f_{e0} + \frac{\nabla_\parallel p_e}{m_e n_e} \frac{w_\parallel}{w} \D{w}{f_{e0}} = \pD{\xi}{} \biggl[ \frac{1-\xi^2}{2} \nu(w,\xi) \biggl( \pD{\xi}{\langle f_{e1}\rangle_\phi} + \vw \D{w}{f_{e0}}\biggr)\biggr].
\end{equation}
Integrating both sides of this equation with respect to $\xi$, choosing the integration constant to keep ${\p \avg{f_{e1}}_\phi/\p \xi}$ finite, and re-arranging terms, we find that
\begin{equation}\label{eqn:CEnuw}
	\nu(w,\xi)=\left.
	-\bigg(w\nabla_\parallel f_{e0}+\frac{\nabla_{\parallel}	p_e}{m_en_e}\D{w}{f_{e0}}\bigg)
	\middle/ \bigg(\pD{\xi}{\avg{f_{e1}}_{\phi}} + v_w\D{w}{f_{e0}}\bigg) \right..
\end{equation}
This equation matches \eqref{eqn:CEnuw_CEsec}. Provided $f_{e0}$ and $\langle f_{e1}\rangle_\phi$, we may determine the effective collision frequency $\nu(w,\xi)$ within the assumptions of the model operator. Alternatively, given a model for $\nu(w,\xi)$, the value of $\langle f_{e1}\rangle_\phi$ may be calculated and its velocity-space dependence compared to that measured in the numerical simulations; this procedure yields
\begin{equation}\label{eqn:CEf1}
    \avg{f_{e1}}_{\phi} = -\int \rmd\xi\, \bigg[\frac{1}{\nu(w,\xi)}\bigg(w\nabla_\parallel f_{e0}+\frac{\nabla_{\parallel}	p_e}{m_en_e}\D{w}{f_{e0}}\bigg)-v_w\D{w}{f_{e0}}\bigg].
\end{equation}
Furthermore, if $f_{e0}$ were taken to be a Maxwellian, $f_{e0} = f_{\rm M}$, then \eqref{eqn:CEf1} would return the more familiar expression
\begin{equation}\label{eqn:CEf1M}
    \avg{f_{e1}}_{\phi} = -\bigg[\bigg(\frac{w^2}{\vthe^2}-\frac{5}{2}\bigg)w \nabla_\parallel\ln T_e\int \frac{\rmd\xi}{\nu(w,\xi)}+2\vw\frac{w\xi}{\vthe^2}\bigg] f_{\rm M}.
\end{equation}
The integrals over $\xi$ in the above equations are indefinite, with the constant chosen such that integral expression equals $0$ at $\xi=0$.

\section{Derivation of the quasi-linear collision operator}\label{apx:QL}

This appendix provides further information on the derivation of the quasi-linear collision operator \eqref{eqn:QL_op}. It begins by leveraging a particularly convenient form of the quasi-linear diffusion equation of \citet{Kennel&Engelmann1966} provided by \citet{Stix1992}:
\begin{equation}\label{eqn:fpstix}
    \pD{t}{f(t,v_\parallel,v_\perp)}=\lim_{V\rightarrow\infty}\upi \Omega^2\sum_{n=-\infty}^{\infty}\int \frac{\rmd^3\bb{k}}{V} \,\lop{L} \Biggl[v_\perp\delta \big(\omega-\kpar\vpar-n\Omega\big)\biggl|\frac{c\psi_{n,k}}{B_0}\biggr|^2 v_\perp \lop{L} f(t,v_\parallel,v_\perp) \Biggr].
\end{equation}
Here, $\Omega$ is the cyclotron frequency associated with the mean magnetic field $\bb{B}_0=B_0\ex$; $\omega=\omega(\bb{k})$ is the (real) wave frequency;
\begin{equation}\label{eqn:psink_apx}
    \psi_{n,k} \doteq \frac{E_{y,k}+\imag E_{z,k}}{2} \,\rme^{-\imag\phi}\besselj{n-1}(z)+\frac{E_{y,k}-\imag E_{z,k}}{2} \,\rme^{\imag\phi}\besselj{n+1}(z)+\frac{v_\parallel}{v_\perp}E_{x,k}\,\besselj{n}(z)
\end{equation}
is an effective electric field expressed in Fourier space, with $\bb{k}=k_\parallel\ex+k_\perp(\cos\phi\ey+\sin\phi\ez)$ and the argument of the Bessel functions $\besselj{n}$ being $z\doteq k_\perp v_\perp/\Omega$; and the differential operator
\begin{equation}\label{eqn:Lstix}
    \lop{L}\doteq \bigg(1-\frac{\kpar\vpar}{\omega}\bigg)\frac{1}{v_\perp}\pD{v_\perp}{}+
    \frac{\kpar}{\omega}\pD{v_\parallel}{}.
\end{equation}
For our 2D setup, $\phi=0$ and $V\rightarrow A$ where $A$ is the area of the domain, so that $\rmd^3\bb{k} \rightarrow \rmd^2\bb{k}$ with $k_\perp=k_y$. In this limit, we can express the circularly polarized electric fields in terms of the perturbed magnetic field using the Fourier-transformed version of Faraday's law:
\begin{equation}\label{eqn:Ek}
    E_k^\pm\doteq\frac{E_{y,k}\pm\imag E_{z,k}}{2}=\frac{\omega}{k_\parallel c}\frac{B_{z,k}\mp\imag B_{y,k}}{2} +\frac{1}{2}\frac{k_\perp}{k_\parallel}E_{x,k}
    \doteq \frac{\omega}{k_\parallel c}B_k^{\mp}+\frac{1}{2}\frac{k_\perp}{k_\parallel}E_{x,k}.
\end{equation}
Pulling out a factor of $\omega/k_\parallel c$ from the above expression for $\psi_{n,k}$, we then define
\begin{equation}
    \Psi_{n,k} \doteq B_k^-\,\besselj{n-1}(z) + B_k^+\,\besselj{n+1}(z) -\frac{k_\parallel c}{\omega}\bigg( \frac{v_\parallel}{v_\perp} + \frac{n\Omega}{k_\parallel v_\perp} \bigg) E_{x,k} \, \besselj{n}(z).
\end{equation}
Converting from $(v_\parallel,v_\perp)$ coordinates to $(v,\xi)$ coordinates and defining the dimensionless operator
\begin{equation}
    \mc{L}' \doteq \bigg(1-\frac{\omega}{k_\parallel}\frac{\xi}{v}\bigg)\pD{\xi}{}+\frac{\omega}{k_\parallel}\pD{v}{},
\end{equation}
equation~\eqref{eqn:fpstix} becomes
\begin{equation}\label{eqn:fpstix2}
    \pD{t}{f} = \lim_{A\rightarrow\infty} 2\upi \Omega^2_e \sum_{n=-\infty}^{\infty}\int \frac{\rmd^2\bb{k}}{A} \,  \frac{1}{v}\, \mc{L}' \biggl[ v \,\frac{1-\xi^2}{2} \delta \big(\omega-\kpar v\xi-n\Omega_e\big)\biggl|\frac{\Psi_{n,k}}{B_0}\biggr|^2 \, \mc{L}' f \biggr] ,
\end{equation}
where $f=f(\tvxi)$.

We now leverage the fact that, for the unstable whistler fluctuations, $\omega/k_\parallel \sim \vw \sim \vthe/\beta_e \ll \vthe$. We expand the electron distribution function in a power series, $f_e = f_{e0} + f_{e1} + \dots$, with the subscript denoting the order in $\omega/k_\parallel\vthe \sim \vw/\vthe \sim \beta^{-1}_e$ at which each contribution appears, and group terms in~\eqref{eqn:fpstix2} according to their size in $\beta^{-1}_e$. To leading order, we find that
\begin{equation}
    \pD{t}{f_{e0}} = \pD{\xi}{} \biggl[\frac{1-\xi^2}{2} \nu_{\rm QL}(v,\xi) \pD{\xi}{f_{e0}}\biggr] ,
\end{equation}
where
\begin{equation}\label{eqn:nuql_apx}
    \nu_{{\rm QL}}(v,\xi) = \lim_{A\rightarrow\infty} 2\upi \Omega^2_e \sum_{n=-\infty}^{\infty}\int \frac{\rmd^2\bb{k}}{A} \,  \delta \big(\omega-\kpar v\xi-n\Omega_e\big)\biggl|\frac{\Psi_{n,k}}{B_0}\biggr|^2
\end{equation}
 is the quasi-linear collision frequency. This equation states that $f_{e0}$ evolves under the action of the collision operator to become nearly independent of pitch angle, {\em viz.}~$\partial f_{e0}/\partial\xi \sim \beta^{-1}_e$. The next-order terms can be simplified considerably if we then anticipate $f_{e0}$ being isotropic, in which case
\begin{equation}\label{eqn:fpstix3}
    \pD{t}{f_{e1}} = \pD{\xi}{} \biggl[\frac{1-\xi^2}{2} \nu_{\rm QL}(v,\xi) \pD{\xi}{f_{e1}}\biggr] + \pD{\xi}{} \biggl[ \frac{1-\xi^2}{2} \nu_{\rm QL}(v,\xi) \, \vw \pD{v}{f_{e0}} \biggr] .
\end{equation}
Neglecting terms of order ${\sim}\beta^{-2}_e$ and smaller, the two terms in \eqref{eqn:fpstix3} may be combined to obtain the quasi-linear operator given by~\eqref{eqn:QL_op}, under whose influence $f_e$ becomes isotropic in a frame moving at speed $\vw$.

\end{document}